\documentclass[11pt,reqno]{article}
\pdfoutput=1

\usepackage{amsmath,amsthm,amssymb}
\usepackage{setspace}

\usepackage[letterpaper,total={15.5cm, 22.4cm}]{geometry}
\renewcommand{\baselinestretch}{1}

\usepackage{caption}
\renewcommand{\floatpagefraction}{.8}

\newcommand{\R}{\mathbb{R}}
\newcommand{\C}{\mathbb{C}}
\newcommand{\Z}{\mathbb{Z}}
\newcommand{\cZ}{\mathcal{Z}}
\newcommand{\SL}{\textup{SL}}
\newcommand{\PSL}{\textup{PSL}}
\newcommand{\vol}{\mathop{\textup{vol}}\nolimits}
\newcommand{\btau}{\bar{\tau}}
\newcommand{\DLPN}[2]{\Delta^{\textup{LP},#1}_{#2}}
\newcommand{\DLP}[1]{\Delta^{\textup{LP}}_{#1}}

\usepackage{microtype}
\usepackage{tikz}

\colorlet{darkblue}{blue!70!black}
\usepackage[pdftex,colorlinks=true,urlcolor=darkblue,linktocpage=true,linkcolor=darkblue,citecolor=darkblue]{hyperref}

\usepackage{amssymb}
\usepackage{booktabs}
\def\arXiv#1{arXiv:\href{http://arXiv.org/abs/#1}{#1}}
\usepackage{doi}

\numberwithin{equation}{section}
\newtheorem{theorem}{Theorem}[section]

\newtheorem{conjecture}[theorem]{Conjecture}

\begin{document}

\begin{spacing}{1.15}
\begin{titlepage}

\begin{center}
{\Large \bf 
High-dimensional sphere packing\\
and the modular bootstrap}

\vspace*{6mm}

Nima Afkhami-Jeddi,$^1$ Henry Cohn,$^2$ Thomas Hartman,$^3$\\
David de Laat,$^4$ and Amirhossein Tajdini$^3$

\vspace*{6mm}

$^1$\textit{Enrico Fermi Institute \& Kadanoff Center for Theoretical Physics,\\ University of Chicago, Chicago, Illinois, USA}

$^2$\textit{Microsoft Research New England, Cambridge, Massachusetts, USA \\}

$^3$\textit{Department of Physics, Cornell University, Ithaca, New York, USA\\}

$^4$\textit{Delft Institute of Applied Mathematics, Delft University of Technology,\\ Delft, The Netherlands\\}

\vspace{6mm}

{nimaaj@uchicago.edu, cohn@microsoft.com, hartman@cornell.edu,\\ d.delaat@tudelft.nl, at734@cornell.edu}

\vspace*{6mm}
\end{center}

\begin{abstract}
We carry out a numerical study of the spinless modular bootstrap for
conformal field theories with current algebra $\textup{U}(1)^c \times \textup{U}(1)^c$, or
equivalently the linear programming bound for sphere packing in $2c$
dimensions. We give a more detailed picture of the behavior for finite $c$
than was previously available, and we extrapolate as $c \to \infty$. Our
extrapolation indicates an exponential improvement for sphere packing
density bounds in high dimensions. Furthermore, we study when these bounds
can be tight. Besides the known cases $c=1/2$, $4$, and $12$ and the
conjectured case $c=1$, our calculations numerically rule out sharp bounds
for all other $c<90$, by combining the modular bootstrap with linear
programming bounds for spherical codes.
\end{abstract}

\setcounter{tocdepth}{1}
\tableofcontents

\end{titlepage}
\end{spacing}

\begin{spacing}{1.15}

\addtocounter{page}{1}

\section{Introduction}

To what extent do self-consistency principles constrain, or even determine,
the behavior of a system? This question underlies many topics in mathematics
and physics. One notable example is the conformal bootstrap program
\cite{Ferrara:1973yt,Polyakov:1974gs,Mack:1975jr,Rattazzi:2008pe} (see
\cite{Rychkov:2016iqz,Simmons-Duffin:2016gjk,Poland:2016chs} for reviews),
which seeks to map the space of possible conformal field theories (CFTs) and
to identify those on the boundary of theory space (i.e., those with extremal
properties, almost but not quite inconsistent). A more down to earth example
is the sphere packing problem, in which the goal is to maximize the fraction
of $\R^d$ covered by congruent spheres whose interiors are not allowed to
overlap. In low dimensions it is not hard to guess the optimal packings, but
even that remains mysterious in high dimensions. Proving upper bounds for the
packing density is particularly difficult, and in most cases the best bounds
currently known are obtained via the linear programming bound of Cohn and
Elkies \cite{CE}, which relies on harmonic analysis.

While these problems sound completely unrelated, Hartman, Maz\'a\v{c}, and
Rastelli \cite{HMR} discovered a surprising connection between them: the
spinless modular bootstrap for two-dimensional CFTs is very nearly the same
as the linear programming bound for sphere packing. The underlying
optimization problems are exactly equivalent when the current algebra
$\textup{U}(1)^c_\textup{left} \times \textup{U}(1)^{\bar{c}}_\textup{right}$ with total
central charge $c_\textup{total} = c + \bar{c}$ acts on the CFT and the
sphere packing dimension is given by $d=c_\textup{total}$, and they are
closely related (but not equivalent) under the Virasoro algebra. The
relationship between the modular bootstrap and linear programming bounds
seems to be specific to these particular techniques, rather than being based
on a direct connection between CFTs and sphere packings.

In this paper we will focus on the spinless modular bootstrap for
$\textup{U}(1)^c_\textup{left} \times \textup{U}(1)^{\bar{c}}_\textup{right}$ with
$c_\textup{total}$ large. The analysis depends only on $c_\textup{total}$, not
on the left and right central charges individually. To simplify the notation we
set $\bar{c} = c$ and refer simply to the $\textup{U}(1)^c$ modular bootstrap,
parameterizing our results by $c = c_\textup{total}/2$.

Neither the modular bootstrap nor the linear programming bound has been
completely analyzed, either theoretically or numerically. Each depends on
producing some additional information (namely, an auxiliary function or
linear functional satisfying certain inequalities), which must be chosen
carefully to optimize the resulting bound, and this optimization has proved
difficult. The equivalence between these problems adds to the motivation for
studying them, because any consequences will shed light on two seemingly
disparate topics. A third application is to generalizations of the
Bourgain-Clozel-Kahane uncertainty principle for signs of functions
\cite{BCK,CG}. Thus, these problems live at a particularly fruitful
intersection of several fields.

In this paper, we carry out the first large-scale numerical study of the
$\textup{U}(1)^c$ spinless modular bootstrap with $c$ large, or equivalently the
linear programming bound on sphere packing in high dimensions, by adapting
the numerical techniques introduced by Afkhami-Jeddi, Hartman, and Tajdini
for the Virasoro case \cite{AHT}. These techniques closely parallel the
approach independently taken by Cohn, Elkies, Kumar, and Gon\c calves
\cite{CE,CK2009,CG,CG2} in the sphere packing literature, but the paper
\cite{AHT} introduced better extrapolation techniques and achieved superior
performance.

In CFT terms, the spinless modular bootstrap corresponds to constraints on
the partition function at zero angular potential. A natural question is
whether the spinning modular bootstrap, i.e., including an angular potential,
also bounds the density of general sphere packings. The answer is that it
does not, as this would  contradict known packings. The spinning bootstrap
analysis for CFTs with $\textup{U}(1)^c_\textup{left} \times \textup{U}(1)^c_\textup{right}$
symmetry has interesting implications for holographic duality and will appear
in a separate paper \cite{ACHT}.

\subsection{Results from the spinless modular bootstrap for large $c$}

For sphere packing in high dimensions, the central question is the asymptotic
behavior of the packing density.  It is at least $2^{-d}$ in $\R^d$, with
only much lower-order improvements known \cite{Ball,V11,V13}, and it is at
most $2^{-(\kappa+o(1)) d}$ with $\kappa = 0.59905576\dots$. The latter bound
was found by Kabatyanskii and Levenshtein \cite{KL78} in 1978, and the
exponential decay rate has not been improved since then. Cohn and Zhao
\cite{CZ} showed how to obtain it via the linear programming bound, and a
fundamental open question is whether the linear programming bound is capable
of improving on this decay rate.

In terms of the $\textup{U}(1)^c$ spinless modular bootstrap, bounding the packing
density amounts to bounding the spectral gap of the CFT. Specifically, the
Kabatyanskii-Levenshtein bound says that the scaling dimension of the lowest
non-vacuum primary is at most $c/(K+o(1))$ as $c \to \infty$, where $K =
e\pi2^{2\kappa-1}=9.79674646\dots$. No better bound is known for the spectral
gap.

One of our primary results in this paper is a numerical estimate of the fully
optimized $\textup{U}(1)^c$ spinless modular bootstrap bound for the spectral gap
(Conjecture~\ref{conjecture:LP}). In sphere packing terms, it amounts to an
upper bound of $2^{-(\lambda+o(1)) d}$ for the sphere packing density in
$\R^d$ as $d \to \infty$ with $\lambda \approx 0.6044$; in modular bootstrap
terms, it amounts to an upper bound of  $c/(\Lambda+o(1))$ for the spectral
gap as $c \to \infty$ with $\Lambda \approx 9.869$. This bound is based on
numerical extrapolation, with no proof or even heuristic derivation, but we
give a careful accounting of the potential error from the extrapolation. We
furthermore guess that the exact value of $\Lambda$ is $\pi^2$
(Conjecture~\ref{conjecture:exact}), although that conjecture is much more
speculative.

Conceptually, what our computations indicate is that the
Kabatyanskii-Levenshtein upper bound can be decreased by an exponential
factor through optimizing the linear programming bound. If proved, this bound
would settle a longstanding open problem in discrete geometry. However, the
improvement in the decay rate will be small.

The analytical \cite{HMR} and numerical \cite{AHT} results for Virasoro
symmetry are quite a bit further away from each other (the analytical bound for
the spectral gap is $c/8.503$, while the numerical bound is $c/9.08$).
We have no conceptual explanation for why the Kabatyanskii-Levenshtein bound
should come rather close to optimizing the $\textup{U}(1)^c$ case, yet fall slightly
short. Perhaps generalizing this bound will offer new techniques for
optimizing the modular bootstrap more broadly, but we do not expect that it
will lead to an exact solution without some new idea.

Sphere packings are error-correcting codes for a continuous communication
channel, and they therefore play an important role in information theory.
Their discrete counterpart is error-correcting codes for a binary channel,
and these two theories are in many ways closely analogous \cite{SPLAG}, with
substantial interplay between them, both in results and in techniques. The
linear program bound originated in the discrete setting, in a fundamental
paper by Delsarte \cite{Del72}, before being generalized to sphere packing by
Cohn and Elkies \cite{CE}, and the Kabatyanskii-Levenshtein bound was
inspired by the MRRW bound, due to McEliece, Rodemich, Rumsey, and Welch
\cite{MRRW}.

Much like the case of sphere packing, the asymptotic rate in the MRRW bound
has not been beaten by any method, and it is an open problem whether it
optimizes the linear programming bound. Barg and Jaffe \cite{BargJaffe}
examined this issue numerically, and they conjectured that it is the optimal
rate in the linear programming bound. Their conjecture is widely believed,
but the evidence is not conclusive. While our results have no direct
implications for binary error-correcting codes, they suggest that the MRRW
bound may not be optimal, because it is the discrete analogue of the
Kabatyanskii-Levenshtein bound. It would be valuable to perform a more
extensive study than Barg and Jaffe were able to do in 2001, as well as to
compare the data with Section~3.2 of \cite{BDS}.

At the optimum, the linear programming approach provides not just a bound on
the spectral gap, but a candidate spectrum for a CFT that saturates it. In
sphere packing terms, this spectrum amounts to the pair correlation function
of the packing. We study the spectrum numerically in
Section~\ref{sec:properties} and find some intriguing structure. For
computational purposes, the infinite set of bootstrap constraints is
truncated to a finite system of $2N$ equations, with $N$ taken as large as
possible. The corresponding  spectrum has $N$ states other than the vacuum,
with scaling dimensions $\Delta_1 < \Delta_2 < \dots < \Delta_N$.  We
conjecture a formula for the ratio $\Delta_n/N$ in the limit $N \to \infty$
with $n/N$ held fixed. The formula is piecewise smooth, with an abrupt
transition from linear to nonlinear behavior at $n \sim (2/\pi) N$. We have
no analytic explanation for this transition. The linear portion of the
spectrum matches the spectrum of the generalized free fermion in one
dimension, which was used to construct analytic functionals for CFT in
\cite{M} and adapted to sphere packing in \cite{HMR}.

\subsection{New constraints on tight sphere packing bounds}

In addition to studying the asymptotic behavior of the modular bootstrap, we
also search for exceptional behavior at finite central charge.  Four
particular values are known to play a special role, namely $c=1/2$, $1$, $4$,
and $12$. In sphere packing terms, these cases correspond to exact solutions
of the sphere packing problem in dimensions $1$, $2$, $8$, and $24$.  While
$d=1$ is trivial, $d=8$ and $d=24$ are far deeper, and they are the only
cases in which the sphere packing problem has been solved above $d=3$ (which
was solved by Hales \cite{Hal05,Hplus2017}, with no connection to the modular
bootstrap). Dimension $8$ was a breakthrough due to Viazovska \cite{V}, and
dimension $24$ built on her techniques \cite{CKMRV}. The linear programming
bound seems to be exact when $d=2$ as well, but no proof is known, although
the two-dimensional sphere packing problem can be solved directly
\cite{T1892,HalesCH}.

These cases are more subtle for CFTs than they are for sphere packings. For
$(c,\bar{c})=(4,4)$, there is indeed a CFT invariant under
$\textup{U}(1)^c_\textup{left} \times \textup{U}(1)^{\bar{c}}_\textup{right}$ that achieves the
spinless modular bootstrap bound, namely eight free fermions \cite{Collier:2016cls}. No such CFT exists for $(c,\bar{c})=(12,12)$ (see
\cite{ACHT}), but there is a chiral CFT with $(c,\bar{c})=(24,0)$, namely
the $24$ chiral bosons compactified using the quotient of $\R^{24}$ by the
Leech lattice.\footnote{Note that we do not take an orbifold quotient, as in
the Monster CFT of Frenkel, Lepowsky, and Meurman \cite{FLM}, because we want
current algebra $\textup{U}(1)^{24}$. Similarly, $8$ chiral bosons compactified using
$E_8$ meet the bound with $(c,\bar{c})=(8,0)$, but the $(c,\bar{c})=(4,4)$
case is more noteworthy.} The case $c=1/2$ is not an integer, so
$\textup{U}(1)^c_\textup{left} \times \textup{U}(1)^c_\textup{right}$ symmetry does not even
make sense, but again we can use a chiral boson with $(c,\bar{c})=(1,0)$.
This time, however, it is not fully conformally invariant. Instead, it has a
nontrivial phase under the action of the generator $T$ of $\SL_2(\Z)$, but
the spinless modular bootstrap nevertheless applies. Finally, in the case
$c=1$ no CFT invariant under $\textup{U}(1)^c_\textup{left} \times
\textup{U}(1)^c_\textup{right}$ achieves the spinless bound (see \cite{ACHT}), but
it is achieved by two chiral bosons with $(c,\bar{c})=(2,0)$ and a nontrivial
phase under the $T$ transformation. Thus, the CFT picture encompasses all
four exceptional cases, provided we allow chiral CFTs and are willing to
relax conformal invariance.

Why should the exceptional solutions of the sphere packing problem be limited
to these specific dimensions?  It comes as no surprise to see sporadic
behavior tied to $E_8$ and the Leech lattice (for $c=4$ and $12$,
respectively), but it is difficult to explain why this behavior is not more
widespread. For a provocative example, why shouldn't the linear programming
bound solve the sphere packing problem in all sufficiently large dimensions?
We do not know how to rule out this possibility, although it is utterly
implausible.

To shed light on this problem, we examine the conditions that would have to
hold to obtain a sphere packing meeting the linear programming bound. To do
so, we incorporate additional constraints beyond the modular invariance of
the partition function. Specifically, we study the \emph{implied kissing
number}, the average number of tangencies between spheres in a hypothetical
packing with this property. In all dimensions up through $d=250$ other than
$1$, $2$, $8$, $24$, $180$, $181$, and $192$, we show that the implied
kissing number from our numerically optimized bound is impossibly large.
Thus, no sphere packing can attain the exact linear programming bound in
these dimensions.\footnote{Strictly speaking, our work does not amount to a
proof, because it leaves open the possibility that we have not fully
optimized the linear programming bound. However, we present strong numerical
evidence that we have optimized it.} We do not expect optimal solutions in
dimensions $180$, $181$, or $192$, and we see no sign of them,
but our bounds do not rule them out.

As the unexpected occurrence of dimensions such as $181$ indicates, this
problem has a surprisingly intricate structure. While certain aspects behave
in straightforward ways that are not hard to extrapolate, other aspects are
far more subtle. One feature of our numerical solutions for which we have no
conceptual explanation is a kind of periodicity: the degeneracies are
especially well described by a Cardy-like entropy formula when $c$ is a
multiple of $8$, and the scaling dimensions are especially close to those for
generalized one-dimensional free fermions when $c$ is $4$ more than a
multiple of $8$. In other words, multiples of $4$ behave particularly well,
but no value of $c$ looks equally simple from all perspectives. The reason
for this behavior remains mysterious.

\section{The spinless modular bootstrap}
\label{sec:prelim}

\subsection{Setting up the bootstrap}
\label{subsec:modularbootstrap}

In this section, we will briefly review the spinless modular bootstrap, which
is a technique for proving bounds on the possible scaling dimensions of
primary fields in a compact, unitary 2d CFT \cite{hellerman, Friedan:2013cba,
Collier:2016cls}.  Given such a CFT, let $Z(\tau,\btau)$ be its partition
function, i.e., the sum over all states of $q^{h-c/24}
\bar{q}^{\bar{h}-\bar{c}/24}$, where $h$ and $\bar{h}$ are the conformal
weights of the state, $c$ and $\bar{c}$ are the left and right central
charges, and $q=e^{2\pi i \tau}$ and $\bar{q} = e^{-2\pi i\btau}$ (with
$\tau$ and $-\btau$ in the upper half-plane).\footnote{Mathematicians should
note that the bars do not denote complex conjugates.} Because of conformal
invariance, the partition function satisfies modular invariance:
\begin{equation}
Z\mathopen{}\left(\frac{a\tau+b}{c\tau+d},\frac{a\btau+b}{c\btau+d}\right)\mathclose{} = Z(\tau,\btau)
\end{equation}
whenever
\begin{equation}
\begin{pmatrix} a & b\\c & d \end{pmatrix} \in \SL_2(\Z)
\end{equation}
(where in these formulas $c$ is of course not necessarily the left central
charge). In terms of the usual generators
\begin{equation}
S = \begin{pmatrix} 0 & -1\\1 & 0 \end{pmatrix}\qquad\text{and}\qquad T = \begin{pmatrix} 1 & 1\\0 & 1 \end{pmatrix}
\end{equation}
of $\SL_2(\Z)$, modular invariance amounts to
\begin{equation}
Z(-1/\tau,-1/\btau) = Z(\tau+1,\btau+1) = Z(\tau,\btau).
\end{equation}

For the spinless modular bootstrap, we specialize the partition function to
have zero angular potential. In other words, we set $\btau = -\tau$ (i.e.,
$\bar{q}=q$) and use the restricted partition function
\begin{equation}
\cZ(\tau) = Z(\tau,-\tau).
\end{equation}
The action of $S$ on $\tau$ and $\btau$ preserves the condition $\btau =
-\tau$, and thus
\begin{equation}
\cZ(-1/\tau) = \cZ(\tau),
\end{equation}
but the action of $T$ does not. Thus, we expect that usually $\cZ(\tau+1)\ne
\cZ(\tau)$.

The spinless modular bootstrap is based on the identity $\cZ(-1/\tau) =
\cZ(\tau)$. Because we make no use of the action of $T$, the bound applies
even to theories that are invariant only under $S$. A simple example is a
single chiral boson at the self-dual radius, for which $Z(\tau, \btau) =
\theta_3(\tau) / \eta(\tau)$ in terms of the Jacobi theta function and
Dedekind eta function. Such theories are not fully conformally invariant, but
the spinless modular bootstrap still applies.

The combined contribution of the descendants of a primary field of scaling
dimension $\Delta=h+\bar{h}$ to the partition function $\cZ(\tau)$ is a
character $\chi_\Delta(\tau)$ of a Verma module of the current algebra, and
thus the partition function is given by a sum
\begin{equation}
\cZ(\tau) = \sum_\Delta d_\Delta \chi_\Delta(\tau)
\end{equation}
over the scaling dimensions of the primary fields, each with multiplicity
given by the degeneracy $d_\Delta$. The vacuum corresponds to $\Delta=0$,
with degeneracy $d_0=1$, and the other scaling dimensions are positive
numbers $\Delta_1 < \Delta_2 < \dotsb$ that tend to infinity.

The precise form of the characters depends on the current algebra. Our main
interest in this paper will be the algebra $\textup{U}(1)^c_\textup{left} \times
\textup{U}(1)^{\bar{c}}_\textup{right}$ (more precisely, the corresponding affine Lie
algebra), in which case
\begin{equation}
\chi_\Delta(\tau) = \frac{e^{2\pi i \tau \Delta}}{\eta(\tau)^{c+\bar{c}}},
\end{equation}
where $\eta$ is again the Dedekind eta function. In particular, the only
dependence on the central charges is through their sum $c+\bar{c}$.

The spectral gap of the CFT is the lowest scaling dimension $\Delta_1$ of a
primary other than the vacuum. We can obtain an upper bound for the spectral
gap by producing a linear functional that acts in a certain way on functions
of $\tau$. The key observation is that if we set
\begin{equation}
\Phi_\Delta(\tau) = \chi_\Delta(\tau)-\chi_\Delta(-1/\tau),
\end{equation}
then we obtain the crossing equation
\begin{equation} \label{eq:crossing}
\sum_\Delta d_\Delta \Phi_\Delta(\tau) = 0
\end{equation}
by modular invariance.
Now suppose $\omega$ is a linear functional such that
\begin{equation}
\omega(\Phi_0) > 0
\end{equation}
and
\begin{equation}
\omega(\Phi_\Delta) \ge 0
\end{equation}
whenever $\Delta \ge \Delta_{\textup{gap}}$ for some constant
$\Delta_{\textup{gap}}$. If we apply $\omega$ to the crossing equation, we
find that
\begin{equation}
\omega(\Phi_0) + \sum_{\Delta > 0} d_\Delta \omega(\Phi_\Delta) = 0,
\end{equation}
which would be impossible if all the non-zero scaling dimensions were at
least $\Delta_{\textup{gap}}$, because the total would be positive. Thus, we
conclude that some primary must have a scaling dimension strictly between $0$
and $\Delta_{\textup{gap}}$. In other words, $\Delta_{\textup{gap}}$ is a
strict upper bound for the spectral gap. One can show that it is a weak upper
bound even if $\omega(\Phi_0) = 0$, as long as $\omega(\Phi_\Delta)$ is not
identically zero (see Appendix~\ref{appendix:vanishat0}), and that
$\omega(\Phi_0) = 0$ must hold for the optimal choice of $\omega$.

The optimal functional $\omega$ is not known, except in a handful of special
cases discussed below. In Sections~\ref{sec:numerics}
and~\ref{sec:properties}, we will give the most detailed numerical study so
far of how $\omega$ and $\Delta_{\textup{gap}}$ behave.

As noted earlier, because the spinless modular bootstrap for
$\textup{U}(1)^c_\textup{left} \times \textup{U}(1)^{\bar{c}}_\textup{right}$ depends only on
$c+\bar{c}$, we will set $\bar{c}=c$ and refer just to $c$. Strictly speaking
this notation is misleading when $c+\bar{c}$ is odd, because the current
algebra $\textup{U}(1)^c_\textup{left} \times \textup{U}(1)^{\bar{c}}_\textup{right}$ makes
sense only when $c$ and $\bar{c}$ are nonnegative integers. For example, the
only physically meaningful cases with $c+\bar{c}=1$ are $(c,\bar{c}) = (1,0)$
or $(0,1)$, and they are therefore what we mean when we refer to the $c=1/2$
case. More generally, the abstract problem of optimizing the bound makes
sense for any $c>0$, but there are consequences for CFTs only when $c$ is an
integer or half-integer.

\subsection{Uncertainty principle}
\label{subsec:uncertainty}

Hartman, Maz\'a\v{c}, and Rastelli \cite{HMR} reformulated the $\textup{U}(1)^c$
spinless modular bootstrap in terms of an uncertainty principle for
eigenfunctions of the Fourier transform as follows. Suppose $d = 2c$ is an
integer, which is the meaningful case for CFTs. Given a functional $\omega$
as above, we define a radial function $f_\omega \colon \R^d \to \R$ by
$f_\omega(x) = \omega(\Phi_{|x|^2/2})$. If we normalize the Fourier transform
in $\R^d$ by
\begin{equation}
\widehat{f}(y) = \int_{\R^d} dx\, f(x) e^{-2\pi i \langle x,y\rangle},
\end{equation}
then $f_\omega$ is an eigenfunction of the Fourier transform with eigenvalue
$-1$; in other words, $\widehat{f_\omega} = - f_\omega$. To see why, we start
with
\begin{equation} \begin{split}
\Phi_{|x|^2/2}(\tau) &= \chi_{|x|^2/2}(\tau)-\chi_{|x|^2/2}(-1/\tau)\\
&= \frac{e^{\pi i \tau |x|^2}}{\eta(\tau)^{d}} - \frac{e^{\pi i (-1/\tau) |x|^2}}{\eta(-1/\tau)^{d}}.\\
&= \frac{e^{\pi i \tau |x|^2} - (i/\tau)^{d/2} e^{\pi i (-1/\tau) |x|^2}}{\eta(\tau)^{d}},
\end{split} \end{equation}
because the Dedekind eta function satisfies the identity $\eta(-1/\tau) =
(\tau/i)^{1/2} \eta(\tau)$. The complex Gaussian $x \mapsto e^{\pi i \tau
|x|^2}$ on $\R^d$ has Fourier transform $y \mapsto (i/\tau)^{d/2} e^{\pi i
(-1/\tau) |y|^2}$. Thus, the function $x \mapsto e^{\pi i \tau |x|^2} -
(i/\tau)^{d/2} e^{\pi i (-1/\tau) |x|^2}$ in the numerator of
$\Phi_{|x|^2/2}(\tau)$ is a $-1$ eigenfunction of the Fourier transform for
each $\tau$, because it is the difference of a Gaussian and its Fourier
transform. We conclude that $f_\omega$ also satisfies $\widehat{f_\omega} = -
f_\omega$, by the linearity of $\omega$. The same holds even when $2c$ is not
an integer, if we interpret the radial Fourier transform in non-integral
dimension as a Hankel transform.

Conversely, every radial $-1$ eigenfunction of the Fourier transform in
$\R^d$ arises as $f_\omega$ for some $\omega$, which we can obtain as follows
by constructing a basis. If we let
\begin{equation}
\omega_k = \left.\frac{\partial^k}{\partial \tau^k}\right|_{\tau=i},
\end{equation}
then $\omega_k(\Phi_{\Delta})$ is the product of $e^{-2\pi \Delta}$ with a
polynomial in $\Delta$ of degree at most $k$, in which the coefficient of
$\Delta^k$ is
\begin{equation}
\left.\frac{(2\pi i)^k - (i/\tau)^{d/2} (2\pi i)^k (1/\tau^2)^k}{\eta(\tau)^d}\right|_{\tau=i} = \begin{cases}
0 & \text{if $k$ is even, and}\\
2(2\pi i)^k/\eta(i)^d & \text{if $k$ is odd.}
\end{cases}
\end{equation}
For comparison, the Laguerre polynomials $L_k^{(d/2-1)}$ give a basis for
radial functions on $\R^d$ as $x \mapsto L_k^{(d/2-1)}(2\pi |x|^2) e^{-\pi
|x|^2}$, with eigenvalues $(-1)^k$ under the Fourier transform. We conclude
that the functions $x \mapsto \omega_k(\Phi_{|x|^2/2})$ with $k = 1, 3, 5,
\dotsc, 2m-1$ must span the same space as the Laguerre eigenfunctions with
these values of $k$, and thus they span the entire $-1$ eigenspace as $m \to
\infty$.

We have seen that choosing the linear functional $\omega$ in the $\textup{U}(1)^c$
spinless modular bootstrap amounts to choosing an integrable, radial function
$f \colon \R^d \to \R$ with $\widehat{f}=-f$ such that $f$ is not identically
zero. The constraints on $\omega$ say that $f(0) \ge 0$ and $f(x) \ge 0$
whenever $|x| \ge r$ for some radius $r$. Then $\Delta_{\textup{gap}} =
r^2/2$, and optimizing the bound means minimizing $r$. This optimization
problem for signs of eigenfunctions was first studied by Cohn and Elkies
\cite{CE}, and it was placed in the context of more general uncertainty
principles for signs of functions by Cohn and Gon\c{c}alves \cite{CG}.

The corresponding problem for $+1$ eigenfunctions asks for an integrable,
radial function $g \colon \R^d \to \R$ with $\widehat{g}=g$ such that $g(0)
\le 0$ and $g(x) \ge 0$ for $|x| \ge r$. Again the goal is to minimize $r$,
without letting $g$ vanish identically. This problem does not arise in the
spinless modular bootstrap as set up above, but it would apply if the
partition function satisfied $\cZ(-1/\tau) = -\cZ(\tau)$ and $d_{\Delta}<0$
for $\Delta>0$ (see Section~2.1 of \cite{CG}). It behaves much like the $-1$
case. For example, Cohn and Gon\c{c}alves \cite{CG} obtained an exact
solution of the $+1$ problem for $c=6$, which is analogous to the solutions
of the $-1$ problem with $c=4$ or $12$. The partition function in this case
is given by $\mathcal{Z}(\tau) = \sqrt{j(\tau)-1728}$, which also arises as
the Norton series for a certain pair of elements in the Monster group (see
equation (7.3.4c) in \cite[p.~425]{G}). Although we do not have a direct
physical interpretation for this problem, generalized modular transformations
do arise in theories with discrete anomalies \cite{Lin:2019kpn} or fermions
\cite{Keller:2012mr,Benjamin:2020zbs}, including sectors that obey
$\cZ(-1/\tau) = -\cZ(\tau)$.

\subsection{Sphere packing}
\label{subsec:spherepacking}

The $-1$ eigenfunction uncertainty principle plays a fundamental role in
discrete geometry, where it underlies the linear programming bound for the
sphere packing density. Linear programming bounds are a powerful technique
for proving upper bounds for packing density or error-correcting code rates.
They were introduced for discrete error-correcting codes by Delsarte
\cite{Del72} in 1972, and extended to sphere packing in Euclidean space by
Cohn and Elkies \cite{CE} in 2003. The connection with the spinless modular
bootstrap for $\textup{U}(1)^c$ was derived by Hartman, Maz\'a\v{c}, and Rastelli
\cite{HMR} in 2019.

The linear programming bound for sphere packing in $\R^d$ converts an
auxiliary function satisfying certain inequalities into a sphere packing
density bound, as follows.\footnote{The original technical hypotheses in
\cite{CE} were slightly stronger; see \cite{CZ} and \cite{CT}.}

\begin{theorem}[Cohn and Elkies \cite{CE}] \label{theorem:LP}
Let $h \colon \R^d \to \R$ be an integrable, continuous, radial function such
that $\widehat{h}$ is integrable, and let $r$ be a positive real number. If
$h(0)=\widehat{h}(0)=1$, $h(x) \le 0$ whenever $|x| \ge r$, and
$\widehat{h}(y) \ge 0$ for all $y$, then every sphere packing in $\R^d$ has
density at most the volume of a sphere of radius $r/2$ in $\R^d$, i.e.,
\begin{equation}
\frac{\pi^{d/2}}{(d/2)!} \left(\frac{r}{2}\right)^d.
\end{equation}
\end{theorem}

The problem of choosing $h$ so as to minimize $r$ is clearly reminiscent of
the $-1$ eigenfunction uncertainty principle, but not obviously equivalent to
it. One direction is simple: if $h$ satisfies the hypotheses of
Theorem~\ref{theorem:LP}, then the function $f = \widehat{h}-h$ satisfies
$f(0)=0$, $\widehat{f}=-f$, and $f(x) \ge 0$ for $|x| \ge r$. Conversely,
Cohn and Elkies conjectured that an optimal solution $f$ of the $-1$
eigenfunction problem can always be lifted to a function $h$ for use in
Theorem~\ref{theorem:LP} with the same value of $r$, such that
$\widehat{h}-h=f$. In other words, the linear programming bound for sphere
packing should be identical to the spinless modular bootstrap. No proof is
known, but no counterexample has been found, either numerically or
analytically.

At first glance, it is not obvious that any auxiliary function satisfies the
hypotheses of the linear programming bound. For a first example, let $\chi
\colon \R^d \to \R$ be the characteristic function of a ball $B_{r/2}$
centered at the origin, with its radius $r/2$ chosen so that
$\vol(B_{r/2})=1$. Then the convolution $h := \chi * \chi$ has Fourier
transform $\widehat{h} = \widehat{\chi}^2$. By construction, $h(x) = 0$ for
$|x| \ge r$ and $\widehat{h}(y)\ge0$ for all $y$; furthermore, $h(0) =
\vol(B_{r/2})=1$ and $\widehat{h}(0) = \vol(B_{r/2})^2 =1$. Thus, the sphere
packing density in $\R^d$ is at most $\vol(B_{r/2})=1$. This bound is sharp
when $d=1$, but it is of course not an exciting packing bound. For $d>1$ the
linear programming bound is much better than this first attempt.

In fact, it is the best upper bound known for the sphere packing density in
high dimensions \cite{CZ}, but it is generally far from a tight bound
\cite{CT}. Only four cases seem to be sharp: $d=1$ (as shown above), $2$,
$8$, and $24$. The case $d=8$ was a breakthrough due to Viazovska \cite{V},
and the case $d=24$ extended her techniques \cite{CKMRV}. These are the only
cases in which the sphere packing problem has been solved above three
dimensions. The optimal auxiliary functions for $d=8$ and $24$ can also be
derived from analytic functionals constructed that same year by Maz\'a\v{c}
\cite{M} in the four-point function bootstrap for 1d CFTs, as shown by
Hartman, Maz\'a\v{c}, and Rastelli \cite{HMR}. Remarkably, the case $d=2$
remains unsolved analytically. There is no doubt that it matches the
two-dimensional packing density,\footnote{With the techniques from Appendix~A
of \cite{CK2009}, we can prove rigorously that they agree to more than a
thousand decimal places, by using $300$ forced double roots at the vector
lengths from the hexagonal lattice.} but no proof is known.

Linear programming bounds can be applied not just to sphere packing, but to
understand ground states under pair potential functions more broadly
\cite{CK07, CdCI, CKMRV2}. We will not address that topic in this paper,
except to note that our numerical results seem consistent with Conjecture~7.2
in \cite{CKMRV2}, which says that the linear programming bound for sphere
packing extends to the Gaussian core model and thereby proves a form of
universal optimality, despite the failure of the analogous statement for
binary error-correcting codes \cite{CZ2}.

\subsection{Numerics}
\label{subsec:numerics}

To obtain numerical bounds for the $\textup{U}(1)^c$ spinless modular bootstrap, we
must choose a finite-dimensional space of functionals $\omega$. We truncate
at derivative order $4N-1$; in other words, $\omega$ will be a linear
combination of $\omega_1,\omega_3,\dots,\omega_{4N-1}$, where as above
$\omega_k = \left.\partial^k/\partial \tau^k\right|_{\tau=i}$.

For convenience let $f(\Delta) = \omega(\Phi_\Delta)$, which differs from the
$-1$ Fourier eigenfunction in being a function of $\Delta$ rather than $x$
with $\Delta=|x|^2/2$. Then $f(\Delta)$ can be written in terms of the
Laguerre eigenfunctions as
\begin{equation}
f(\Delta) = \sum_{j=1}^{2N} \alpha_j f_j(\Delta),
\end{equation}
where
\begin{equation}
f_j(\Delta)  =  L_{2j-1}^{(c-1)}(4\pi \Delta) e^{-2\pi\Delta} .
\end{equation}
For fixed $\Delta_{\textup{gap}}$ and $N$, the question is whether $f$ can be
chosen to satisfy the positivity conditions $f(0) \ge 0$ and $f(\Delta)\ge0$
for $\Delta \ge \Delta_{\textup{gap}}$ without vanishing identically. This
problem can be solved using semidefinite programming, or approximated using
linear programming.

Let $\DLPN{N}{1}(c)$ be the best bound that can be obtained for a fixed
truncation order $N$, and let $\DLP{1}(c)$ be the best bound without
restriction on $\omega$. Increasing $N$ improves the bound, and we expect
that
\begin{equation}
\DLP{1}(c) = \lim_{N \to \infty}\DLPN{N}{1}(c).
\end{equation}
Numerical linear or semidefinite programming succeeds at small $N$, but has
been limited to $N \lesssim 100$ by the computational cost. It is much faster
to trade the linear program for a nonlinear optimization over the roots of
$f(\Delta)$.

At the optimum, $f(\Delta)$ is found empirically to have single roots at
$\Delta_0=0$ and $\Delta_1=\Delta_\textup{gap}$, and $N-1$ double roots
$\Delta_2, \Delta_3, \dots, \Delta_N$.  Assuming this to hold in general, we
can restate the optimization problem as follows: fix $\Delta_1$, and maximize
$f(0)$ over the parameters $\alpha_j$ for $1\leq j \leq 2N$ and $\Delta_n$
for $2 \leq n \leq N$, subject to the pattern of roots
\begin{equation} \begin{split}
f(\Delta_1) &= 0 \text{ and}\\
f(\Delta_n) &= f'(\Delta_n) = 0 \text{ for $2 \leq n \leq N$}.
\end{split} \end{equation}
If the optimized function has $f(0) > 0$, then this value of $\Delta_1$ is
excluded. The marginal bound has $f(0) = 0$, and the corresponding $\Delta_1$
gives $\DLPN{N}{1}(c)$.

If this problem can be solved with some mild non-degeneracy conditions, then
it provably gives $\DLPN{N}{1}(c)$ (see Section~5 of \cite{CG}). However,
there is no guarantee that the optimum must be of this form, and it fails for
$c=3/2$. Specifically, when $c=3/2$ it works for $1 \le N \le 21$ and $27 \le
N \le 32$, but for $22 \le N \le 26$ and $N=33$ there is a different pattern
of roots. We do not know what happens as $N \to \infty$.

Aside from $c=3/2$, this method has always worked in practice.\footnote{Our
primary interest is in large $c$, and the main role of $c=3/2$ is to dash any
hope of proving that the method works for all $c$. We have no conceptual
explanation for why this case seems to differ from all the others.} The
resulting bound can be made rigorous simply by proving that the optimal
functional satisfies the positivity conditions.

This is essentially the method used by Cohn and Elkies \cite{CE}. In the
conformal bootstrap, a similar approach was first discussed by El-Showk and
Paulos in the context of 1d correlation functions \cite{E-SP}. Recently,
their methods were improved by Afkhami-Jeddi, Hartman, and Tajdini and
applied to the modular bootstrap \cite{AHT}. The first step is to dualize the
optimization problem. The dual problem, together with the equation $f(0) =
0$, leads to the equations
\begin{equation}\label{truncprim}
f_k(0) + \sum_{n=1}^N d_n f_k(\Delta_n) = 0 \text{ for $1 \le k \le 2N$.}
\end{equation}
Here the unknowns are $d_1,\dots,d_N$ and $\Delta_1,\dots,\Delta_N$. These
equations are nothing but the original crossing equation \eqref{eq:crossing}
truncated to $N$ states. If we absorb the factors of $e^{-2\pi \Delta_n}$
from $f_k(\Delta_n)$ into the coefficients $d_n$, we are left with $2N$
polynomial equations in $2N$ unknowns.

Solutions to \eqref{truncprim} with $d_n > 0$ place a lower bound
$\DLPN{N}{1}(c) \geq \Delta_1$, almost by definition. What is more surprising
is that the solution saturating $\Delta_1 = \DLPN{N}{1}(c)$ has exactly $N$
states and can be found efficiently by Newton's method. The last ingredient
we need in the algorithm is a procedure to generate the initial guess for
Newton's method. We start at small $N$, where it is easy to find a guess by
hand, and then gradually increase $N$, using the results from lower $N$ to
generate the next guess as in Appendix~B of \cite{AHT}. This method allows
for large jumps in $N$, while still converging to the bound within a few
Newton steps.

\section{Numerical results}
\label{sec:numerics}

\subsection{Data and plots}

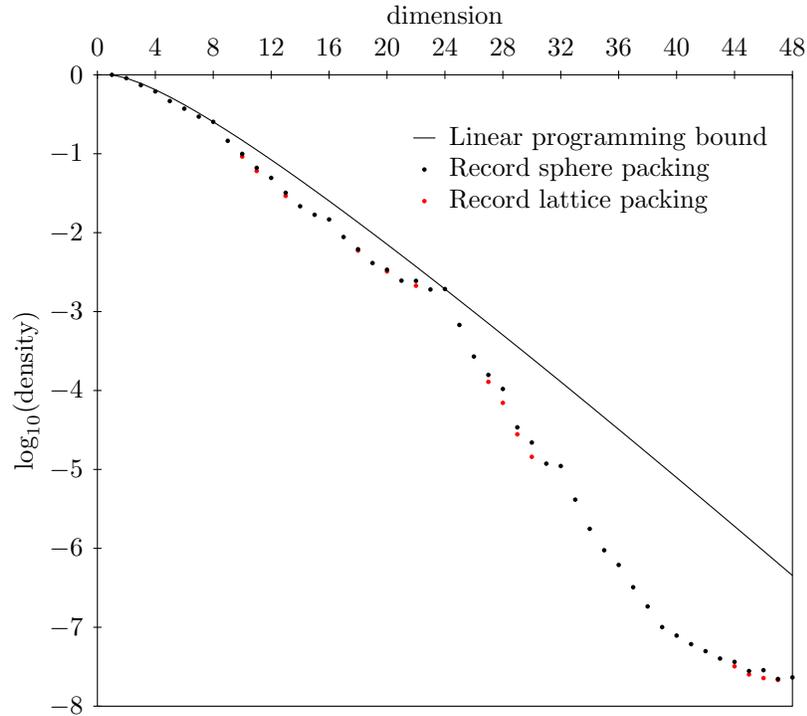
\begin{figure}
\begin{center}
\begin{tikzpicture}[x=0.525cm,y=0.84cm]
\draw (0,-10)--(17.6,-10)--(17.6,0);
\draw (8.65,-1) node[right] {\small Linear programming bound};
\draw (8.65,-1.5) node[right] {\small Record sphere packing};
\draw (8.65,-2) node[right] {\small Record lattice packing};
\draw (8,-1)--(8.56,-1);
\fill (8.28,-1.5) circle (0.03cm);
\fill[red] (8.28,-2) circle (0.03cm);
\foreach \x in {0,...,12}
\draw (1.46666*\x,-0.065)--(1.466666*\x,0.065);
\foreach \x in {0,4,...,48}
\draw ({1.46666*\x/4},0.15)  node[above] {\small \x};
\foreach \y in {-8,...,0} \draw (-0.104,{\y*10/8})--(0.104,{\y*10/8}); \draw (8.8,0.65)
node[above] {\small dimension}; \draw (-1.8,-5) node[rotate=90]
{\small $\log_{10}(\text{density})$};
\draw (19.1,-5) node[rotate=-90]
{\small $\phantom{\log_{10}(\text{density})}$};
\draw (-0.1,0) node[left] {\small $0$};
\draw (-0.1,-1.25) node[left] {\small $-1$};
\draw (-0.1,-2.5) node[left] {\small $-2$};
\draw (-0.1,-3.75) node[left] {\small $-3$};
\draw (-0.1,-5) node[left] {\small $-4$};
\draw (-0.1,-6.25) node[left] {\small $-5$};
\draw (-0.1,-7.5) node[left] {\small $-6$};
\draw (-0.1,-8.76) node[left] {\small $-7$};
\draw (-0.1,-10) node[left] {\small $-8$}; \draw(0,-10)--(0,0); \draw
(0,0)--(17.6,0);
\draw[join=round]  (0.3667,0.000)--(0.7333,-0.053)--(1.1000,-0.135)--(1.4667,-0.236)--(1.8333,-0.350)--(2.2000,-0.474)--(2.5667,-0.606)--(2.9333,-0.745)--(3.3000,-0.889)--(3.6667,-1.037)--(4.0333,-1.190)--(4.4000,-1.346)--(4.7667,-1.505)--(5.1333,-1.667)--(5.5000,-1.832)--(5.8667,-1.998)--(6.2333,-2.167)--(6.6000,-2.338)--(6.9667,-2.510)--(7.3333,-2.684)--(7.7000,-2.859)--(8.0667,-3.036)--(8.4333,-3.214)--(8.8000,-3.393)--(9.1667,-3.574)--(9.5333,-3.755)--(9.9000,-3.937)--(10.2667,-4.121)--(10.6333,-4.305)--(11.0000,-4.490)--(11.3667,-4.676)--(11.7333,-4.862)--(12.1000,-5.050)--(12.4667,-5.238)--(12.8333,-5.427)--(13.2000,-5.616)--(13.5667,-5.806)--(13.9333,-5.997)--(14.3000,-6.188)--(14.6667,-6.379)--(15.0333,-6.572)--(15.4000,-6.764)--(15.7667,-6.957)--(16.1333,-7.151)--(16.5000,-7.345)--(16.8667,-7.540)--(17.2333,-7.735)--(17.6000,-7.930);
\fill (0.36666667,0) circle (0.03cm);
\fill (0.73333333,-0.053050938) circle (0.03cm);
\fill (1.1000000,-0.16310797) circle (0.03cm);
\fill (1.4666667,-0.26227530) circle (0.03cm);
\fill (1.8333333,-0.41538314) circle (0.03cm);
\fill (2.2000000,-0.53544031) circle (0.03cm);
\fill (2.5666667,-0.66217460) circle (0.03cm);
\fill (2.9333333,-0.74466467) circle (0.03cm);
\fill (3.3000000,-1.0453966) circle (0.03cm);
\fill[red] (3.6666667,-1.2951406) circle (0.03cm);
\fill (3.6666667,-1.2520898) circle (0.03cm);
\fill[red] (4.0333333,-1.5234104) circle (0.03cm);
\fill (4.0333333,-1.4736158) circle (0.03cm);
\fill (4.4000000,-1.6322463) circle (0.03cm);
\fill[red] (4.7666667,-1.9181147) circle (0.03cm);
\fill (4.7666667,-1.8683201) circle (0.03cm);
\fill (5.1333333,-2.0813275) circle (0.03cm);
\fill (5.5000000,-2.2165063) circle (0.03cm);
\fill (5.8666667,-2.2905519) circle (0.03cm);
\fill (6.2333333,-2.5686988) circle (0.03cm);
\fill[red] (6.6000000,-2.7838310) circle (0.03cm);
\fill (6.6000000,-2.7623285) circle (0.03cm);
\fill (6.9666667,-2.9812723) circle (0.03cm);
\fill[red] (7.3333333,-3.1141929) circle (0.03cm);
\fill (7.3333333,-3.0865172) circle (0.03cm);
\fill (7.7000000,-3.2600340) circle (0.03cm);
\fill[red] (8.0666667,-3.3401222) circle (0.03cm);
\fill (8.0666667,-3.2633133) circle (0.03cm);
\fill (8.4333333,-3.4000378) circle (0.03cm);
\fill (8.8000000,-3.3931731) circle (0.03cm);
\fill (9.1666667,-3.9615942) circle (0.03cm);
\fill (9.5333333,-4.4623657) circle (0.03cm);
\fill[red] (9.9000000,-4.8631311) circle (0.03cm);
\fill (9.9000000,-4.7530741) circle (0.03cm);
\fill[red] (10.266667,-5.1955017) circle (0.03cm);
\fill (10.266667,-4.9753877) circle (0.03cm);
\fill[red] (10.633333,-5.6934038) circle (0.03cm);
\fill (10.633333,-5.5833467) circle (0.03cm);
\fill[red] (11.000000,-6.0509681) circle (0.03cm);
\fill (11.000000,-5.8240644) circle (0.03cm);
\fill (11.366667,-6.1584128) circle (0.03cm);
\fill (11.733333,-6.1962518) circle (0.03cm);
\fill (12.100000,-6.7286593) circle (0.03cm);
\fill (12.466667,-7.1909623) circle (0.03cm);
\fill (12.833333,-7.5300268) circle (0.03cm);
\fill (13.200000,-7.7628949) circle (0.03cm);
\fill (13.566667,-8.1162666) circle (0.03cm);
\fill (13.933333,-8.4201963) circle (0.03cm);
\fill (14.300000,-8.7477461) circle (0.03cm);
\fill (14.666667,-8.8815509) circle (0.03cm);
\fill (15.033333,-9.0177008) circle (0.03cm);
\fill (15.400000,-9.1283424) circle (0.03cm);
\fill (15.766667,-9.2452959) circle (0.03cm);
\fill[red] (16.133333,-9.3684177) circle (0.03cm);
\fill (16.133333,-9.2984913) circle (0.03cm);
\fill[red] (16.500000,-9.4975710) circle (0.03cm);
\fill (16.500000,-9.4431593) circle (0.03cm);
\fill[red] (16.866667,-9.5545379) circle (0.03cm);
\fill (16.866667,-9.4293607) circle (0.03cm);
\fill[red] (17.233333,-9.5853091) circle (0.03cm);
\fill (17.233333,-9.5684088) circle (0.03cm);
\fill (17.600000,-9.5436481) circle (0.03cm);
\end{tikzpicture}
\end{center}
\caption{The linear programming bound for the sphere packing density.}
\label{figure:LPbound}
\end{figure}

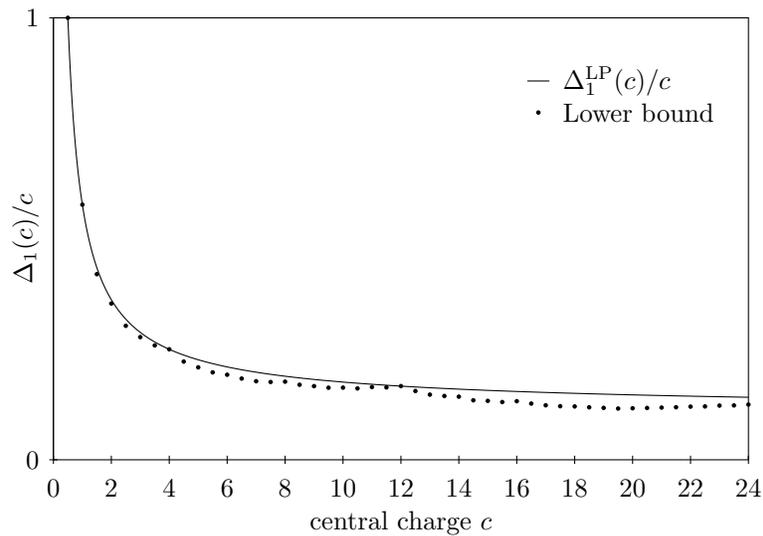
\begin{figure}
\begin{center}
\begin{tikzpicture}[x=0.525cm,y=0.84cm]
\draw (0,0)--(0,-7)--(17.6,-7)--(17.6,0)--(0,0);
\draw (12,-1)--(12.56,-1);
\fill (12.28,-1.5) circle (0.03cm);
\draw (12.65,-1) node[right] {\small $\Delta^\textup{LP}_1(c)/c$};
\draw (12.65,-1.5) node[right] {\small Lower bound};
\foreach \y in {-1,...,0} \draw (-0.104,{\y*7})--(0.104,{\y*7}); \draw (8.8,-7-0.65)
node[below] {\small central charge $c$}; \draw (-0.7,-3.5) node[rotate=90]
{\small $\Delta_1(c)/c$};
\draw (18,-3.5) node[rotate=-90]
{\small $\phantom{\Delta_1(c)/c}$};
\foreach \y in {0,...,1}
\draw (-0.1,-7+\y*7) node[left] {\small \y};
\draw (0,-7)--(0,0);
\draw (0,0)--(17.6,0);
\foreach \x in {0,...,12}
\draw (1.46666*\x,-7-0.065)--(1.466666*\x,-7+0.065);
\foreach \x in {0,2,...,24}
\draw ({1.46666*\x/2},-7-0.15)  node[below] {\small \x};
\draw[join=round]  (0.7333,-2.959)--(0.7150,-2.882)--(0.6967,-2.802)--(0.6783,-2.717)--(0.6600,-2.628)--(0.6417,-2.534)--(0.6233,-2.434)--(0.6050,-2.328)--(0.5867,-2.216)--(0.5683,-2.096)--(0.5500,-1.969)--(0.5317,-1.833)--(0.5133,-1.687)--(0.4950,-1.531)--(0.4767,-1.362)--(0.4583,-1.180)--(0.4400,-0.983)--(0.4217,-0.770)--(0.4033,-0.536)--(0.3850,-0.281)--(0.3667,0.000);
\draw[join=round]  (1.4667,-4.464)--(1.4483,-4.445)--(1.4300,-4.425)--(1.4117,-4.404)--(1.3933,-4.383)--(1.3750,-4.362)--(1.3567,-4.340)--(1.3383,-4.317)--(1.3200,-4.294)--(1.3017,-4.270)--(1.2833,-4.246)--(1.2650,-4.221)--(1.2467,-4.195)--(1.2283,-4.168)--(1.2100,-4.141)--(1.1917,-4.112)--(1.1733,-4.083)--(1.1550,-4.053)--(1.1367,-4.022)--(1.1183,-3.990)--(1.1000,-3.957)--(1.0817,-3.923)--(1.0633,-3.888)--(1.0450,-3.851)--(1.0267,-3.814)--(1.0083,-3.774)--(0.9900,-3.734)--(0.9717,-3.692)--(0.9533,-3.648)--(0.9350,-3.603)--(0.9167,-3.556)--(0.8983,-3.507)--(0.8800,-3.456)--(0.8617,-3.403)--(0.8433,-3.348)--(0.8250,-3.290)--(0.8067,-3.230)--(0.7883,-3.167)--(0.7700,-3.100)--(0.7517,-3.031)--(0.7333,-2.959);
\draw[join=round]  (2.9333,-5.250)--(2.9150,-5.245)--(2.8967,-5.240)--(2.8783,-5.234)--(2.8600,-5.229)--(2.8417,-5.224)--(2.8233,-5.218)--(2.8050,-5.213)--(2.7867,-5.207)--(2.7683,-5.201)--(2.7500,-5.196)--(2.7317,-5.190)--(2.7133,-5.184)--(2.6950,-5.178)--(2.6767,-5.172)--(2.6583,-5.166)--(2.6400,-5.160)--(2.6217,-5.154)--(2.6033,-5.147)--(2.5850,-5.141)--(2.5667,-5.134)--(2.5483,-5.128)--(2.5300,-5.121)--(2.5117,-5.114)--(2.4933,-5.107)--(2.4750,-5.100)--(2.4567,-5.093)--(2.4383,-5.086)--(2.4200,-5.079)--(2.4017,-5.071)--(2.3833,-5.064)--(2.3650,-5.056)--(2.3467,-5.048)--(2.3283,-5.041)--(2.3100,-5.033)--(2.2917,-5.025)--(2.2733,-5.016)--(2.2550,-5.008)--(2.2367,-5.000)--(2.2183,-4.991)--(2.2000,-4.982)--(2.1817,-4.973)--(2.1633,-4.964)--(2.1450,-4.955)--(2.1267,-4.946)--(2.1083,-4.937)--(2.0900,-4.927)--(2.0717,-4.917)--(2.0533,-4.907)--(2.0350,-4.897)--(2.0167,-4.887)--(1.9983,-4.876)--(1.9800,-4.866)--(1.9617,-4.855)--(1.9433,-4.844)--(1.9250,-4.832)--(1.9067,-4.821)--(1.8883,-4.809)--(1.8700,-4.797)--(1.8517,-4.785)--(1.8333,-4.773)--(1.8150,-4.760)--(1.7967,-4.748)--(1.7783,-4.734)--(1.7600,-4.721)--(1.7417,-4.707)--(1.7233,-4.694)--(1.7050,-4.679)--(1.6867,-4.665)--(1.6683,-4.650)--(1.6500,-4.635)--(1.6317,-4.620)--(1.6133,-4.604)--(1.5950,-4.588)--(1.5767,-4.571)--(1.5583,-4.554)--(1.5400,-4.537)--(1.5217,-4.519)--(1.5033,-4.501)--(1.4850,-4.483)--(1.4667,-4.464);
\draw[join=round]  (8.8000,-5.833)--(8.7817,-5.833)--(8.7633,-5.832)--(8.7450,-5.831)--(8.7267,-5.831)--(8.7083,-5.830)--(8.6900,-5.829)--(8.6717,-5.828)--(8.6533,-5.828)--(8.6350,-5.827)--(8.6167,-5.826)--(8.5983,-5.826)--(8.5800,-5.825)--(8.5617,-5.824)--(8.5433,-5.824)--(8.5250,-5.823)--(8.5067,-5.822)--(8.4883,-5.821)--(8.4700,-5.821)--(8.4517,-5.820)--(8.4333,-5.819)--(8.4150,-5.818)--(8.3967,-5.818)--(8.3783,-5.817)--(8.3600,-5.816)--(8.3417,-5.815)--(8.3233,-5.815)--(8.3050,-5.814)--(8.2867,-5.813)--(8.2683,-5.812)--(8.2500,-5.812)--(8.2317,-5.811)--(8.2133,-5.810)--(8.1950,-5.809)--(8.1767,-5.809)--(8.1583,-5.808)--(8.1400,-5.807)--(8.1217,-5.806)--(8.1033,-5.805)--(8.0850,-5.805)--(8.0667,-5.804)--(8.0483,-5.803)--(8.0300,-5.802)--(8.0117,-5.801)--(7.9933,-5.801)--(7.9750,-5.800)--(7.9567,-5.799)--(7.9383,-5.798)--(7.9200,-5.797)--(7.9017,-5.796)--(7.8833,-5.796)--(7.8650,-5.795)--(7.8467,-5.794)--(7.8283,-5.793)--(7.8100,-5.792)--(7.7917,-5.791)--(7.7733,-5.791)--(7.7550,-5.790)--(7.7367,-5.789)--(7.7183,-5.788)--(7.7000,-5.787)--(7.6817,-5.786)--(7.6633,-5.785)--(7.6450,-5.785)--(7.6267,-5.784)--(7.6083,-5.783)--(7.5900,-5.782)--(7.5717,-5.781)--(7.5533,-5.780)--(7.5350,-5.779)--(7.5167,-5.778)--(7.4983,-5.777)--(7.4800,-5.776)--(7.4617,-5.776)--(7.4433,-5.775)--(7.4250,-5.774)--(7.4067,-5.773)--(7.3883,-5.772)--(7.3700,-5.771)--(7.3517,-5.770)--(7.3333,-5.769)--(7.3150,-5.768)--(7.2967,-5.767)--(7.2783,-5.766)--(7.2600,-5.765)--(7.2417,-5.764)--(7.2233,-5.763)--(7.2050,-5.762)--(7.1867,-5.761)--(7.1683,-5.760)--(7.1500,-5.759)--(7.1317,-5.758)--(7.1133,-5.757)--(7.0950,-5.756)--(7.0767,-5.755)--(7.0583,-5.754)--(7.0400,-5.753)--(7.0217,-5.752)--(7.0033,-5.751)--(6.9850,-5.750)--(6.9667,-5.749)--(6.9483,-5.748)--(6.9300,-5.747)--(6.9117,-5.746)--(6.8933,-5.745)--(6.8750,-5.744)--(6.8567,-5.743)--(6.8383,-5.742)--(6.8200,-5.741)--(6.8017,-5.740)--(6.7833,-5.739)--(6.7650,-5.737)--(6.7467,-5.736)--(6.7283,-5.735)--(6.7100,-5.734)--(6.6917,-5.733)--(6.6733,-5.732)--(6.6550,-5.731)--(6.6367,-5.730)--(6.6183,-5.729)--(6.6000,-5.727)--(6.5817,-5.726)--(6.5633,-5.725)--(6.5450,-5.724)--(6.5267,-5.723)--(6.5083,-5.722)--(6.4900,-5.720)--(6.4717,-5.719)--(6.4533,-5.718)--(6.4350,-5.717)--(6.4167,-5.716)--(6.3983,-5.714)--(6.3800,-5.713)--(6.3617,-5.712)--(6.3433,-5.711)--(6.3250,-5.710)--(6.3067,-5.708)--(6.2883,-5.707)--(6.2700,-5.706)--(6.2517,-5.705)--(6.2333,-5.703)--(6.2150,-5.702)--(6.1967,-5.701)--(6.1783,-5.699)--(6.1600,-5.698)--(6.1417,-5.697)--(6.1233,-5.696)--(6.1050,-5.694)--(6.0867,-5.693)--(6.0683,-5.692)--(6.0500,-5.690)--(6.0317,-5.689)--(6.0133,-5.688)--(5.9950,-5.686)--(5.9767,-5.685)--(5.9583,-5.683)--(5.9400,-5.682)--(5.9217,-5.681)--(5.9033,-5.679)--(5.8850,-5.678)--(5.8667,-5.676)--(5.8483,-5.675)--(5.8300,-5.674)--(5.8117,-5.672)--(5.7933,-5.671)--(5.7750,-5.669)--(5.7567,-5.668)--(5.7383,-5.666)--(5.7200,-5.665)--(5.7017,-5.663)--(5.6833,-5.662)--(5.6650,-5.660)--(5.6467,-5.659)--(5.6283,-5.657)--(5.6100,-5.656)--(5.5917,-5.654)--(5.5733,-5.653)--(5.5550,-5.651)--(5.5367,-5.650)--(5.5183,-5.648)--(5.5000,-5.646)--(5.4817,-5.645)--(5.4633,-5.643)--(5.4450,-5.642)--(5.4267,-5.640)--(5.4083,-5.638)--(5.3900,-5.637)--(5.3717,-5.635)--(5.3533,-5.633)--(5.3350,-5.632)--(5.3167,-5.630)--(5.2983,-5.628)--(5.2800,-5.627)--(5.2617,-5.625)--(5.2433,-5.623)--(5.2250,-5.621)--(5.2067,-5.620)--(5.1883,-5.618)--(5.1700,-5.616)--(5.1517,-5.614)--(5.1333,-5.612)--(5.1150,-5.611)--(5.0967,-5.609)--(5.0783,-5.607)--(5.0600,-5.605)--(5.0417,-5.603)--(5.0233,-5.601)--(5.0050,-5.599)--(4.9867,-5.598)--(4.9683,-5.596)--(4.9500,-5.594)--(4.9317,-5.592)--(4.9133,-5.590)--(4.8950,-5.588)--(4.8767,-5.586)--(4.8583,-5.584)--(4.8400,-5.582)--(4.8217,-5.580)--(4.8033,-5.578)--(4.7850,-5.576)--(4.7667,-5.574)--(4.7483,-5.572)--(4.7300,-5.570)--(4.7117,-5.567)--(4.6933,-5.565)--(4.6750,-5.563)--(4.6567,-5.561)--(4.6383,-5.559)--(4.6200,-5.557)--(4.6017,-5.554)--(4.5833,-5.552)--(4.5650,-5.550)--(4.5467,-5.548)--(4.5283,-5.546)--(4.5100,-5.543)--(4.4917,-5.541)--(4.4733,-5.539)--(4.4550,-5.536)--(4.4367,-5.534)--(4.4183,-5.532)--(4.4000,-5.529)--(4.3817,-5.527)--(4.3633,-5.524)--(4.3450,-5.522)--(4.3267,-5.519)--(4.3083,-5.517)--(4.2900,-5.514)--(4.2717,-5.512)--(4.2533,-5.509)--(4.2350,-5.507)--(4.2167,-5.504)--(4.1983,-5.502)--(4.1800,-5.499)--(4.1617,-5.496)--(4.1433,-5.494)--(4.1250,-5.491)--(4.1067,-5.488)--(4.0883,-5.485)--(4.0700,-5.483)--(4.0517,-5.480)--(4.0333,-5.477)--(4.0150,-5.474)--(3.9967,-5.471)--(3.9783,-5.469)--(3.9600,-5.466)--(3.9417,-5.463)--(3.9233,-5.460)--(3.9050,-5.457)--(3.8867,-5.454)--(3.8683,-5.451)--(3.8500,-5.448)--(3.8317,-5.445)--(3.8133,-5.442)--(3.7950,-5.438)--(3.7767,-5.435)--(3.7583,-5.432)--(3.7400,-5.429)--(3.7217,-5.426)--(3.7033,-5.422)--(3.6850,-5.419)--(3.6667,-5.416)--(3.6483,-5.412)--(3.6300,-5.409)--(3.6117,-5.405)--(3.5933,-5.402)--(3.5750,-5.398)--(3.5567,-5.395)--(3.5383,-5.391)--(3.5200,-5.388)--(3.5017,-5.384)--(3.4833,-5.380)--(3.4650,-5.377)--(3.4467,-5.373)--(3.4283,-5.369)--(3.4100,-5.365)--(3.3917,-5.361)--(3.3733,-5.357)--(3.3550,-5.353)--(3.3367,-5.349)--(3.3183,-5.345)--(3.3000,-5.341)--(3.2817,-5.337)--(3.2633,-5.333)--(3.2450,-5.329)--(3.2267,-5.325)--(3.2083,-5.320)--(3.1900,-5.316)--(3.1717,-5.312)--(3.1533,-5.307)--(3.1350,-5.303)--(3.1167,-5.298)--(3.0983,-5.294)--(3.0800,-5.289)--(3.0617,-5.284)--(3.0433,-5.280)--(3.0250,-5.275)--(3.0067,-5.270)--(2.9883,-5.265)--(2.9700,-5.260)--(2.9517,-5.255)--(2.9333,-5.250);
\draw[join=round]  (17.6000,-6.010)--(17.5817,-6.009)--(17.5633,-6.009)--(17.5450,-6.009)--(17.5267,-6.009)--(17.5083,-6.009)--(17.4900,-6.008)--(17.4717,-6.008)--(17.4533,-6.008)--(17.4350,-6.008)--(17.4167,-6.008)--(17.3983,-6.007)--(17.3800,-6.007)--(17.3617,-6.007)--(17.3433,-6.007)--(17.3250,-6.007)--(17.3067,-6.006)--(17.2883,-6.006)--(17.2700,-6.006)--(17.2517,-6.006)--(17.2333,-6.006)--(17.2150,-6.005)--(17.1967,-6.005)--(17.1783,-6.005)--(17.1600,-6.005)--(17.1417,-6.004)--(17.1233,-6.004)--(17.1050,-6.004)--(17.0867,-6.004)--(17.0683,-6.004)--(17.0500,-6.003)--(17.0317,-6.003)--(17.0133,-6.003)--(16.9950,-6.003)--(16.9767,-6.003)--(16.9583,-6.002)--(16.9400,-6.002)--(16.9217,-6.002)--(16.9033,-6.002)--(16.8850,-6.001)--(16.8667,-6.001)--(16.8483,-6.001)--(16.8300,-6.001)--(16.8117,-6.001)--(16.7933,-6.000)--(16.7750,-6.000)--(16.7567,-6.000)--(16.7383,-6.000)--(16.7200,-5.999)--(16.7017,-5.999)--(16.6833,-5.999)--(16.6650,-5.999)--(16.6467,-5.999)--(16.6283,-5.998)--(16.6100,-5.998)--(16.5917,-5.998)--(16.5733,-5.998)--(16.5550,-5.997)--(16.5367,-5.997)--(16.5183,-5.997)--(16.5000,-5.997)--(16.4817,-5.997)--(16.4633,-5.996)--(16.4450,-5.996)--(16.4267,-5.996)--(16.4083,-5.996)--(16.3900,-5.995)--(16.3717,-5.995)--(16.3533,-5.995)--(16.3350,-5.995)--(16.3167,-5.995)--(16.2983,-5.994)--(16.2800,-5.994)--(16.2617,-5.994)--(16.2433,-5.994)--(16.2250,-5.993)--(16.2067,-5.993)--(16.1883,-5.993)--(16.1700,-5.993)--(16.1517,-5.992)--(16.1333,-5.992)--(16.1150,-5.992)--(16.0967,-5.992)--(16.0783,-5.991)--(16.0600,-5.991)--(16.0417,-5.991)--(16.0233,-5.991)--(16.0050,-5.991)--(15.9867,-5.990)--(15.9683,-5.990)--(15.9500,-5.990)--(15.9317,-5.990)--(15.9133,-5.989)--(15.8950,-5.989)--(15.8767,-5.989)--(15.8583,-5.989)--(15.8400,-5.988)--(15.8217,-5.988)--(15.8033,-5.988)--(15.7850,-5.988)--(15.7667,-5.987)--(15.7483,-5.987)--(15.7300,-5.987)--(15.7117,-5.987)--(15.6933,-5.986)--(15.6750,-5.986)--(15.6567,-5.986)--(15.6383,-5.986)--(15.6200,-5.985)--(15.6017,-5.985)--(15.5833,-5.985)--(15.5650,-5.985)--(15.5467,-5.984)--(15.5283,-5.984)--(15.5100,-5.984)--(15.4917,-5.984)--(15.4733,-5.983)--(15.4550,-5.983)--(15.4367,-5.983)--(15.4183,-5.983)--(15.4000,-5.982)--(15.3817,-5.982)--(15.3633,-5.982)--(15.3450,-5.982)--(15.3267,-5.981)--(15.3083,-5.981)--(15.2900,-5.981)--(15.2717,-5.981)--(15.2533,-5.980)--(15.2350,-5.980)--(15.2167,-5.980)--(15.1983,-5.980)--(15.1800,-5.979)--(15.1617,-5.979)--(15.1433,-5.979)--(15.1250,-5.979)--(15.1067,-5.978)--(15.0883,-5.978)--(15.0700,-5.978)--(15.0517,-5.977)--(15.0333,-5.977)--(15.0150,-5.977)--(14.9967,-5.977)--(14.9783,-5.976)--(14.9600,-5.976)--(14.9417,-5.976)--(14.9233,-5.976)--(14.9050,-5.975)--(14.8867,-5.975)--(14.8683,-5.975)--(14.8500,-5.975)--(14.8317,-5.974)--(14.8133,-5.974)--(14.7950,-5.974)--(14.7767,-5.973)--(14.7583,-5.973)--(14.7400,-5.973)--(14.7217,-5.973)--(14.7033,-5.972)--(14.6850,-5.972)--(14.6667,-5.972)--(14.6483,-5.972)--(14.6300,-5.971)--(14.6117,-5.971)--(14.5933,-5.971)--(14.5750,-5.970)--(14.5567,-5.970)--(14.5383,-5.970)--(14.5200,-5.970)--(14.5017,-5.969)--(14.4833,-5.969)--(14.4650,-5.969)--(14.4467,-5.968)--(14.4283,-5.968)--(14.4100,-5.968)--(14.3917,-5.968)--(14.3733,-5.967)--(14.3550,-5.967)--(14.3367,-5.967)--(14.3183,-5.966)--(14.3000,-5.966)--(14.2817,-5.966)--(14.2633,-5.966)--(14.2450,-5.965)--(14.2267,-5.965)--(14.2083,-5.965)--(14.1900,-5.964)--(14.1717,-5.964)--(14.1533,-5.964)--(14.1350,-5.963)--(14.1167,-5.963)--(14.0983,-5.963)--(14.0800,-5.963)--(14.0617,-5.962)--(14.0433,-5.962)--(14.0250,-5.962)--(14.0067,-5.961)--(13.9883,-5.961)--(13.9700,-5.961)--(13.9517,-5.960)--(13.9333,-5.960)--(13.9150,-5.960)--(13.8967,-5.960)--(13.8783,-5.959)--(13.8600,-5.959)--(13.8417,-5.959)--(13.8233,-5.958)--(13.8050,-5.958)--(13.7867,-5.958)--(13.7683,-5.957)--(13.7500,-5.957)--(13.7317,-5.957)--(13.7133,-5.956)--(13.6950,-5.956)--(13.6767,-5.956)--(13.6583,-5.956)--(13.6400,-5.955)--(13.6217,-5.955)--(13.6033,-5.955)--(13.5850,-5.954)--(13.5667,-5.954)--(13.5483,-5.954)--(13.5300,-5.953)--(13.5117,-5.953)--(13.4933,-5.953)--(13.4750,-5.952)--(13.4567,-5.952)--(13.4383,-5.952)--(13.4200,-5.951)--(13.4017,-5.951)--(13.3833,-5.951)--(13.3650,-5.950)--(13.3467,-5.950)--(13.3283,-5.950)--(13.3100,-5.949)--(13.2917,-5.949)--(13.2733,-5.949)--(13.2550,-5.948)--(13.2367,-5.948)--(13.2183,-5.948)--(13.2000,-5.947)--(13.1817,-5.947)--(13.1633,-5.947)--(13.1450,-5.946)--(13.1267,-5.946)--(13.1083,-5.946)--(13.0900,-5.945)--(13.0717,-5.945)--(13.0533,-5.945)--(13.0350,-5.944)--(13.0167,-5.944)--(12.9983,-5.944)--(12.9800,-5.943)--(12.9617,-5.943)--(12.9433,-5.943)--(12.9250,-5.942)--(12.9067,-5.942)--(12.8883,-5.942)--(12.8700,-5.941)--(12.8517,-5.941)--(12.8333,-5.941)--(12.8150,-5.940)--(12.7967,-5.940)--(12.7783,-5.940)--(12.7600,-5.939)--(12.7417,-5.939)--(12.7233,-5.939)--(12.7050,-5.938)--(12.6867,-5.938)--(12.6683,-5.937)--(12.6500,-5.937)--(12.6317,-5.937)--(12.6133,-5.936)--(12.5950,-5.936)--(12.5767,-5.936)--(12.5583,-5.935)--(12.5400,-5.935)--(12.5217,-5.935)--(12.5033,-5.934)--(12.4850,-5.934)--(12.4667,-5.933)--(12.4483,-5.933)--(12.4300,-5.933)--(12.4117,-5.932)--(12.3933,-5.932)--(12.3750,-5.932)--(12.3567,-5.931)--(12.3383,-5.931)--(12.3200,-5.930)--(12.3017,-5.930)--(12.2833,-5.930)--(12.2650,-5.929)--(12.2467,-5.929)--(12.2283,-5.929)--(12.2100,-5.928)--(12.1917,-5.928)--(12.1733,-5.927)--(12.1550,-5.927)--(12.1367,-5.927)--(12.1183,-5.926)--(12.1000,-5.926)--(12.0817,-5.925)--(12.0633,-5.925)--(12.0450,-5.925)--(12.0267,-5.924)--(12.0083,-5.924)--(11.9900,-5.924)--(11.9717,-5.923)--(11.9533,-5.923)--(11.9350,-5.922)--(11.9167,-5.922)--(11.8983,-5.922)--(11.8800,-5.921)--(11.8617,-5.921)--(11.8433,-5.920)--(11.8250,-5.920)--(11.8067,-5.920)--(11.7883,-5.919)--(11.7700,-5.919)--(11.7517,-5.918)--(11.7333,-5.918)--(11.7150,-5.917)--(11.6967,-5.917)--(11.6783,-5.917)--(11.6600,-5.916)--(11.6417,-5.916)--(11.6233,-5.915)--(11.6050,-5.915)--(11.5867,-5.915)--(11.5683,-5.914)--(11.5500,-5.914)--(11.5317,-5.913)--(11.5133,-5.913)--(11.4950,-5.912)--(11.4767,-5.912)--(11.4583,-5.912)--(11.4400,-5.911)--(11.4217,-5.911)--(11.4033,-5.910)--(11.3850,-5.910)--(11.3667,-5.909)--(11.3483,-5.909)--(11.3300,-5.909)--(11.3117,-5.908)--(11.2933,-5.908)--(11.2750,-5.907)--(11.2567,-5.907)--(11.2383,-5.906)--(11.2200,-5.906)--(11.2017,-5.905)--(11.1833,-5.905)--(11.1650,-5.905)--(11.1467,-5.904)--(11.1283,-5.904)--(11.1100,-5.903)--(11.0917,-5.903)--(11.0733,-5.902)--(11.0550,-5.902)--(11.0367,-5.901)--(11.0183,-5.901)--(11.0000,-5.901)--(10.9817,-5.900)--(10.9633,-5.900)--(10.9450,-5.899)--(10.9267,-5.899)--(10.9083,-5.898)--(10.8900,-5.898)--(10.8717,-5.897)--(10.8533,-5.897)--(10.8350,-5.896)--(10.8167,-5.896)--(10.7983,-5.895)--(10.7800,-5.895)--(10.7617,-5.894)--(10.7433,-5.894)--(10.7250,-5.893)--(10.7067,-5.893)--(10.6883,-5.892)--(10.6700,-5.892)--(10.6517,-5.892)--(10.6333,-5.891)--(10.6150,-5.891)--(10.5967,-5.890)--(10.5783,-5.890)--(10.5600,-5.889)--(10.5417,-5.889)--(10.5233,-5.888)--(10.5050,-5.888)--(10.4867,-5.887)--(10.4683,-5.887)--(10.4500,-5.886)--(10.4317,-5.886)--(10.4133,-5.885)--(10.3950,-5.885)--(10.3767,-5.884)--(10.3583,-5.884)--(10.3400,-5.883)--(10.3217,-5.883)--(10.3033,-5.882)--(10.2850,-5.881)--(10.2667,-5.881)--(10.2483,-5.880)--(10.2300,-5.880)--(10.2117,-5.879)--(10.1933,-5.879)--(10.1750,-5.878)--(10.1567,-5.878)--(10.1383,-5.877)--(10.1200,-5.877)--(10.1017,-5.876)--(10.0833,-5.876)--(10.0650,-5.875)--(10.0467,-5.875)--(10.0283,-5.874)--(10.0100,-5.874)--(9.9917,-5.873)--(9.9733,-5.872)--(9.9550,-5.872)--(9.9367,-5.871)--(9.9183,-5.871)--(9.9000,-5.870)--(9.8817,-5.870)--(9.8633,-5.869)--(9.8450,-5.869)--(9.8267,-5.868)--(9.8083,-5.867)--(9.7900,-5.867)--(9.7717,-5.866)--(9.7533,-5.866)--(9.7350,-5.865)--(9.7167,-5.865)--(9.6983,-5.864)--(9.6800,-5.863)--(9.6617,-5.863)--(9.6433,-5.862)--(9.6250,-5.862)--(9.6067,-5.861)--(9.5883,-5.861)--(9.5700,-5.860)--(9.5517,-5.859)--(9.5333,-5.859)--(9.5150,-5.858)--(9.4967,-5.858)--(9.4783,-5.857)--(9.4600,-5.856)--(9.4417,-5.856)--(9.4233,-5.855)--(9.4050,-5.855)--(9.3867,-5.854)--(9.3683,-5.853)--(9.3500,-5.853)--(9.3317,-5.852)--(9.3133,-5.852)--(9.2950,-5.851)--(9.2767,-5.850)--(9.2583,-5.850)--(9.2400,-5.849)--(9.2217,-5.848)--(9.2033,-5.848)--(9.1850,-5.847)--(9.1667,-5.847)--(9.1483,-5.846)--(9.1300,-5.845)--(9.1117,-5.845)--(9.0933,-5.844)--(9.0750,-5.843)--(9.0567,-5.843)--(9.0383,-5.842)--(9.0200,-5.841)--(9.0017,-5.841)--(8.9833,-5.840)--(8.9650,-5.839)--(8.9467,-5.839)--(8.9283,-5.838)--(8.9100,-5.837)--(8.8917,-5.837)--(8.8733,-5.836)--(8.8550,-5.835)--(8.8367,-5.835)--(8.8183,-5.834)--(8.8000,-5.833);
\fill (0.36666667,0) circle (0.03cm);
\fill (0.73333333,-2.9585481) circle (0.03cm);
\fill (1.1000000,-4.0601842) circle (0.03cm);
\fill (1.4666667,-4.5251263) circle (0.03cm);
\fill (1.8333333,-4.8779968) circle (0.03cm);
\fill (2.2000000,-5.0570726) circle (0.03cm);
\fill (2.5666667,-5.1885527) circle (0.03cm);
\fill (2.9333333,-5.2500000) circle (0.03cm);
\fill (3.3000000,-5.4444444) circle (0.03cm);
\fill (3.6666667,-5.5361046) circle (0.03cm);
\fill (4.0333333,-5.6151518) circle (0.03cm);
\fill (4.4000000,-5.6528494) circle (0.03cm);
\fill (4.7666667,-5.7131653) circle (0.03cm);
\fill (5.1333333,-5.7556780) circle (0.03cm);
\fill (5.5000000,-5.7684593) circle (0.03cm);
\fill (5.8666667,-5.7625631) circle (0.03cm);
\fill (6.2333333,-5.8113666) circle (0.03cm);
\fill (6.6000000,-5.8333333) circle (0.03cm);
\fill (6.9666667,-5.8584403) circle (0.03cm);
\fill (7.3333333,-5.8570346) circle (0.03cm);
\fill (7.7000000,-5.8695147) circle (0.03cm);
\fill (8.0666667,-5.8484899) circle (0.03cm);
\fill (8.4333333,-5.8538178) circle (0.03cm);
\fill (8.8000000,-5.8333333) circle (0.03cm);
\fill (9.1666667,-5.9106265) circle (0.03cm);
\fill (9.5333333,-5.9676336) circle (0.03cm);
\fill (9.9000000,-5.9892471) circle (0.03cm);
\fill (10.266667,-6.0000000) circle (0.03cm);
\fill (10.633333,-6.0572866) circle (0.03cm);
\fill (11.000000,-6.0666667) circle (0.03cm);
\fill (11.366667,-6.0856209) circle (0.03cm);
\fill (11.733333,-6.0719223) circle (0.03cm);
\fill (12.100000,-6.1094480) circle (0.03cm);
\fill (12.466667,-6.1368700) circle (0.03cm);
\fill (12.833333,-6.1510296) circle (0.03cm);
\fill (13.200000,-6.1550750) circle (0.03cm);
\fill (13.566667,-6.1689330) circle (0.03cm);
\fill (13.933333,-6.1779363) circle (0.03cm);
\fill (14.300000,-6.1882152) circle (0.03cm);
\fill (14.666667,-6.1834320) circle (0.03cm);
\fill (15.033333,-6.1790413) circle (0.03cm);
\fill (15.400000,-6.1730009) circle (0.03cm);
\fill (15.766667,-6.1676610) circle (0.03cm);
\fill (16.133333,-6.1580592) circle (0.03cm);
\fill (16.500000,-6.1551410) circle (0.03cm);
\fill (16.866667,-6.1415226) circle (0.03cm);
\fill (17.233333,-6.1386685) circle (0.03cm);
\fill (17.600000,-6.1250000) circle (0.03cm);
\end{tikzpicture}
\end{center}
\caption{The ratio $\Delta^{\textrm{LP}}_1(c)/c$, together with lower
bounds from sphere packings.}
\label{figure:ratioDeltac}
\end{figure}

The linear programming bound for the sphere packing density in $\R^d$ is
shown in Figure~\ref{figure:LPbound} for $d \le 48$, along with the record
packing densities from \cite[Table~I.1, pp.~xix--xx]{SPLAG}, and
Table~\ref{table:KLLP}, while Figure~\ref{figure:ratioDeltac} shows
$\Delta_1^\textup{LP}(c)/c$. We believe that the truncation order in our
calculations is high enough that these plots and table are indistinguishable
from the fully optimized bounds; see Appendix~\ref{appendix:convergence} for
how we calibrated the convergence rate. All of our data is available from
\url{https://hdl.handle.net/1721.1/125646}, including truncation orders, scaling dimensions, and
degeneracies.

The linear programming bound is sharp when $d=1$, $2$ (conjecturally), $8$,
or $24$, and seemingly nowhere else. From the perspective of discrete
geometry, one of the most mysterious aspects is the role of these special
dimensions. Unlike some other cases of the conformal bootstrap (see, for
example, \cite[Section~V.B.4]{PRV}), the bound itself shows no sign of kinks
or other non-analytic behavior at these points in
Figure~\ref{figure:LPbound}. Instead, the upper bound looks much the same
there as elsewhere, and it is the packing densities that show anomalous
behavior.

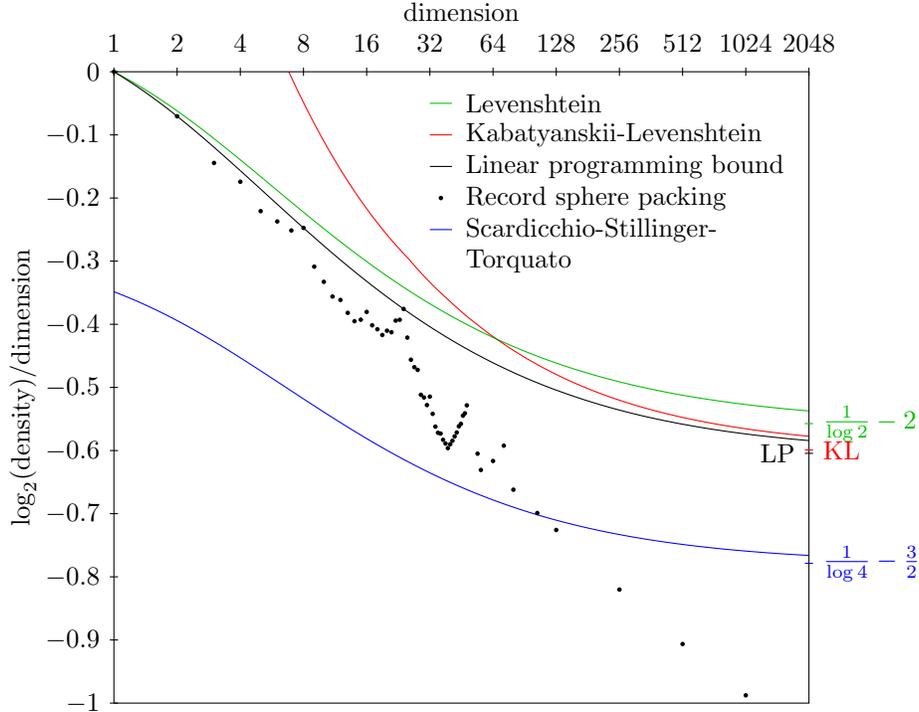
\begin{figure}
\begin{center}
\begin{tikzpicture}[x=0.525cm,y=0.84cm]
\draw[blue,blend mode=overlay,join=round] (0.0, -3.48504)--(0.133333, -3.5172)--(0.266667, -3.55045)--(0.4, -3.58477)--(0.533333, -3.62018)--(0.666667, -3.65666)--(0.8, -3.69422)--(0.933333, -3.73284)--(1.06667, -3.77251)--(1.2, -3.81323)--(1.33333, -3.85497)--(1.46667, -3.8977)--(1.6, -3.94142)--(1.73333, -3.98609)--(1.86667, -4.03168)--(2.0, -4.07816)--(2.13333, -4.1255)--(2.26667, -4.17365)--(2.4, -4.22259)--(2.53333, -4.27226)--(2.66667, -4.32262)--(2.8, -4.37363)--(2.93333, -4.42523)--(3.06667, -4.47739)--(3.2, -4.53004)--(3.33333, -4.58314)--(3.46667, -4.63664)--(3.6, -4.69047)--(3.73333, -4.7446)--(3.86667, -4.79896)--(4.0, -4.8535)--(4.13333, -4.90817)--(4.26667, -4.96292)--(4.4, -5.01769)--(4.53333, -5.07243)--(4.66667, -5.12709)--(4.8, -5.18162)--(4.93333, -5.23598)--(5.06667, -5.29011)--(5.2, -5.34398)--(5.33333, -5.39754)--(5.46667, -5.45074)--(5.6, -5.50355)--(5.73333, -5.55593)--(5.86667, -5.60784)--(6.0, -5.65924)--(6.13333, -5.71012)--(6.26667, -5.76042)--(6.4, -5.81014)--(6.53333, -5.85923)--(6.66667, -5.90767)--(6.8, -5.95545)--(6.93333, -6.00254)--(7.06667, -6.04892)--(7.2, -6.09457)--(7.33333, -6.13948)--(7.46667, -6.18364)--(7.6, -6.22704)--(7.73333, -6.26965)--(7.86667, -6.31149)--(8.0, -6.35253)--(8.13333, -6.39277)--(8.26667, -6.43221)--(8.4, -6.47085)--(8.53333, -6.50868)--(8.66667, -6.54571)--(8.8, -6.58193)--(8.93333, -6.61734)--(9.06667, -6.65196)--(9.2, -6.68578)--(9.33333, -6.71881)--(9.46667, -6.75106)--(9.6, -6.78253)--(9.73333, -6.81322)--(9.86667, -6.84316)--(10.0, -6.87234)--(10.1333, -6.90077)--(10.2667, -6.92847)--(10.4, -6.95544)--(10.5333, -6.9817)--(10.6667, -7.00725)--(10.8, -7.03212)--(10.9333, -7.0563)--(11.0667, -7.07981)--(11.2, -7.10267)--(11.3333, -7.12488)--(11.4667, -7.14646)--(11.6, -7.16742)--(11.7333, -7.18777)--(11.8667, -7.20753)--(12.0, -7.22671)--(12.1333, -7.24532)--(12.2667, -7.26338)--(12.4, -7.28089)--(12.5333, -7.29787)--(12.6667, -7.31434)--(12.8, -7.3303)--(12.9333, -7.34577)--(13.0667, -7.36076)--(13.2, -7.37528)--(13.3333, -7.38934)--(13.4667, -7.40296)--(13.6, -7.41615)--(13.7333, -7.42892)--(13.8667, -7.44128)--(14.0, -7.45324)--(14.1333, -7.46481)--(14.2667, -7.47601)--(14.4, -7.48684)--(14.5333, -7.49732)--(14.6667, -7.50745)--(14.8, -7.51724)--(14.9333, -7.52672)--(15.0667, -7.53587)--(15.2, -7.54472)--(15.3333, -7.55328)--(15.4667, -7.56155)--(15.6, -7.56953)--(15.7333, -7.57725)--(15.8667, -7.5847)--(16.0, -7.59191)--(16.1333, -7.59886)--(16.2667, -7.60558)--(16.4, -7.61206)--(16.5333, -7.61832)--(16.6667, -7.62437)--(16.8, -7.63021)--(16.9333, -7.63584)--(17.0667, -7.64128)--(17.2, -7.64653)--(17.3333, -7.65159)--(17.4667, -7.65648)--(17.6, -7.6612);
\draw[green!75!black,blend mode=overlay,join=round] (0, 0)--(0.133333, -0.0443171)--(0.266667, -0.0900193)--(0.4, -0.137101)--(0.533333, -0.185554)--(0.666667, -0.235365)--(0.8, -0.286517)--(0.933333, -0.338989)--(1.06667, -0.392755)--(1.2, -0.447788)--(1.33333, -0.504053)--(1.46667, -0.561515)--(1.6, -0.620131)--(1.73333, -0.679859)--(1.86667, -0.740651)--(2.0, -0.802455)--(2.13333, -0.865217)--(2.26667, -0.928883)--(2.4, -0.993391)--(2.53333, -1.05868)--(2.66667, -1.12469)--(2.8, -1.19135)--(2.93333, -1.25861)--(3.06667, -1.32638)--(3.2, -1.39461)--(3.33333, -1.46322)--(3.46667, -1.53216)--(3.6, -1.60134)--(3.73333, -1.67071)--(3.86667, -1.74019)--(4.0, -1.80973)--(4.13333, -1.87925)--(4.26667, -1.9487)--(4.4, -2.01802)--(4.53333, -2.08713)--(4.66667, -2.156)--(4.8, -2.22456)--(4.93333, -2.29276)--(5.06667, -2.36054)--(5.2, -2.42787)--(5.33333, -2.49469)--(5.46667, -2.56095)--(5.6, -2.62663)--(5.73333, -2.69168)--(5.86667, -2.75606)--(6.0, -2.81974)--(6.13333, -2.88269)--(6.26667, -2.94488)--(6.4, -3.00628)--(6.53333, -3.06686)--(6.66667, -3.12661)--(6.8, -3.18551)--(6.93333, -3.24354)--(7.06667, -3.30067)--(7.2, -3.35691)--(7.33333, -3.41222)--(7.46667, -3.46662)--(7.6, -3.52008)--(7.73333, -3.57259)--(7.86667, -3.62416)--(8.0, -3.67479)--(8.13333, -3.72446)--(8.26667, -3.77317)--(8.4, -3.82093)--(8.53333, -3.86774)--(8.66667, -3.9136)--(8.8, -3.95852)--(8.93333, -4.00249)--(9.06667, -4.04553)--(9.2, -4.08764)--(9.33333, -4.12882)--(9.46667, -4.1691)--(9.6, -4.20846)--(9.73333, -4.24693)--(9.86667, -4.28451)--(10.0, -4.32122)--(10.1333, -4.35705)--(10.2667, -4.39204)--(10.4, -4.42618)--(10.5333, -4.45949)--(10.6667, -4.49198)--(10.8, -4.52367)--(10.9333, -4.55456)--(11.0667, -4.58467)--(11.2, -4.61402)--(11.3333, -4.64261)--(11.4667, -4.67046)--(11.6, -4.69758)--(11.7333, -4.724)--(11.8667, -4.74971)--(12.0, -4.77473)--(12.1333, -4.79908)--(12.2667, -4.82278)--(12.4, -4.84583)--(12.5333, -4.86824)--(12.6667, -4.89004)--(12.8, -4.91124)--(12.9333, -4.93184)--(13.0667, -4.95187)--(13.2, -4.97133)--(13.3333, -4.99023)--(13.4667, -5.0086)--(13.6, -5.02644)--(13.7333, -5.04377)--(13.8667, -5.0606)--(14.0, -5.07693)--(14.1333, -5.09279)--(14.2667, -5.10818)--(14.4, -5.12312)--(14.5333, -5.13762)--(14.6667, -5.15168)--(14.8, -5.16532)--(14.9333, -5.17856)--(15.0667, -5.19139)--(15.2, -5.20384)--(15.3333, -5.21591)--(15.4667, -5.22761)--(15.6, -5.23895)--(15.7333, -5.24994)--(15.8667, -5.2606)--(16.0, -5.27093)--(16.1333, -5.28093)--(16.2667, -5.29063)--(16.4, -5.30002)--(16.5333, -5.30911)--(16.6667, -5.31793)--(16.8, -5.32646)--(16.9333, -5.33472)--(17.0667, -5.34272)--(17.2, -5.35047)--(17.3333, -5.35797)--(17.4667, -5.36523)--(17.6, -5.37226);
\draw[red,blend mode=overlay,join=round] (4.42723,0)--(4.5222,-0.1280)--(4.6444,-0.2890)--(4.7667,-0.4459)--(4.8889,-0.5986)--(5.0111,-0.7473)--(5.1333,-0.8920)--(5.2556,-1.0327)--(5.3778,-1.1696)--(5.5000,-1.3025)--(5.6222,-1.4317)--(5.7444,-1.5572)--(5.8667,-1.6790)--(5.9889,-1.7972)--(6.1111,-1.9118)--(6.2333,-2.0230)--(6.3556,-2.1308)--(6.4778,-2.2353)--(6.6000,-2.3365)--(6.7222,-2.4345)--(6.8444,-2.5293)--(6.9667,-2.6212)--(7.0889,-2.7100)--(7.2111,-2.7959)--(7.3333,-2.8790)--(7.4556,-2.9593)--(7.5778,-3.0484)--(7.7000,-3.1347)--(7.8222,-3.2181)--(7.9444,-3.2985)--(8.0667,-3.3762)--(8.1889,-3.4511)--(8.3111,-3.5232)--(8.4333,-3.5928)--(8.5556,-3.6636)--(8.6778,-3.7336)--(8.8000,-3.8010)--(8.9222,-3.8659)--(9.0444,-3.9284)--(9.1667,-3.9885)--(9.2889,-4.0487)--(9.4111,-4.1078)--(9.5333,-4.1646)--(9.6556,-4.2192)--(9.7778,-4.2717)--(9.9000,-4.3247)--(10.0222,-4.3756)--(10.1444,-4.4244)--(10.2667,-4.4718)--(10.3889,-4.5188)--(10.5111,-4.5639)--(10.6333,-4.6071)--(10.7556,-4.6502)--(10.8778,-4.6914)--(11.0000,-4.7314)--(11.1222,-4.7706)--(11.2444,-4.8081)--(11.3667,-4.8451)--(11.4889,-4.8806)--(11.6111,-4.9153)--(11.7333,-4.9487)--(11.8556,-4.9814)--(11.9778,-5.0128)--(12.1000,-5.0437)--(12.2222,-5.0734)--(12.3444,-5.1021)--(12.4667,-5.1302)--(12.5889,-5.1574)--(12.7111,-5.1837)--(12.8333,-5.2091)--(12.9556,-5.2339)--(13.0778,-5.2579)--(13.2000,-5.2812)--(13.3222,-5.3037)--(13.4444,-5.3256)--(13.5667,-5.3468)--(13.6889,-5.3674)--(13.8111,-5.3873)--(13.9333,-5.4066)--(14.0556,-5.4253)--(14.1778,-5.4434)--(14.3000,-5.4610)--(14.4222,-5.4780)--(14.5444,-5.4945)--(14.6667,-5.5104)--(14.7889,-5.5259)--(14.9111,-5.5409)--(15.0333,-5.5554)--(15.1556,-5.5695)--(15.2778,-5.5831)--(15.4000,-5.5963)--(15.5222,-5.6091)--(15.6444,-5.6215)--(15.7667,-5.6335)--(15.8889,-5.6451)--(16.0111,-5.6563)--(16.1333,-5.6672)--(16.2556,-5.6777)--(16.3778,-5.6879)--(16.5000,-5.6978)--(16.6222,-5.7074)--(16.7444,-5.7166)--(16.8667,-5.7256)--(16.9889,-5.7343)--(17.1111,-5.7427)--(17.2333,-5.7508)--(17.3556,-5.7587)--(17.4778,-5.7663)--(17.6000,-5.7737);
\draw (0,-10)--(17.6,-10)--(17.6,0);
\draw (8.65,-0.5) node[right] {\small Levenshtein};
\draw (8.65,-1) node[right] {\small Kabatyanskii-Levenshtein};
\draw (8.65,-1.5) node[right] {\small Linear programming bound};
\draw[green!75!black] (8,-0.5)--(8.56,-0.5);
\draw[red] (8,-1)--(8.56,-1);
\draw (8,-1.5)--(8.56,-1.5);
\fill (8.28,-2) circle (0.03cm);
\draw (8.65,-2) node[right] {\small Record sphere packing};
\draw (8.65,-2.5) node[right] {\small Scardicchio-Stillinger-};
\draw (8.65,-3) node[right] {\small Torquato};
\draw[blue] (8,-2.5)--(8.56,-2.5);
\foreach \x in {0,...,11}
\draw (1.6*\x,-0.065)--(1.6*\x,0.065); \draw (0,0.15) node[above] {\small $1$}; \draw
(1.6,0.15) node[above] {\small $2$}; \draw (3.2,0.15) node[above] {\small $4$}; \draw
(4.8,0.15) node[above] {\small $8$}; \draw (6.4,0.15) node[above] {\small $16$}; \draw
(8.0,0.15) node[above] {\small $32$}; \draw (9.6,0.15) node[above] {\small $64$}; \draw
(11.2,0.15) node[above] {\small $128$}; \draw (12.8,0.15) node[above] {\small $256$}; \draw
(14.4,0.15) node[above] {\small $512$}; \draw (16,0.15) node[above] {\small $1024$};  \draw (17.6,0.15) node[above] {\small $2048$};
\draw[red] (17.7,-5.9906) node[right] {\small KL};
\draw[red] (17.496,-5.9906)--(17.704,-5.9906);
\draw (17.5,-6.044) node[left] {\small LP};
\draw (17.496,-6.044)--(17.704,-6.044);
\draw[green!75!black] (17.7,-5.57303) node[right] {\small $\frac{1}{\log 2} - 2$};
\draw[green!75!black] (17.496,-5.57305)--(17.704,-5.57305);
\draw[blue] (17.7,-7.7865) node[right] {\small $\frac{1}{\log 4} - \frac{3}{2}$};
\draw[blue] (17.496,-7.7865)--(17.704,-7.7865);
\foreach \y in {-10,...,0} \draw (-0.104,{\y})--(0.104,{\y}); \draw (8.8,0.65)
node[above] {\small dimension}; \draw (-2.3,-5) node[rotate=90]
{\small $\log_2(\text{density})/\text{dimension}$};
\draw (19.6,-5) node[rotate=-90]
{\small $\phantom{\log_2(\text{density})/\text{dimension}}$};
\draw (-0.1,0) node[left] {\small $0$};
\draw (-0.1,-1) node[left] {\small $-0.1$};
\draw (-0.1,-2) node[left] {\small $-0.2$};
\draw (-0.1,-3) node[left] {\small $-0.3$};
\draw (-0.1,-4) node[left] {\small $-0.4$};
\draw (-0.1,-5) node[left] {\small $-0.5$};
\draw (-0.1,-6) node[left] {\small $-0.6$};
\draw (-0.1,-7) node[left] {\small $-0.7$};
\draw (-0.1,-8) node[left] {\small $-0.8$};
\draw (-0.1,-9) node[left] {\small $-0.9$};
\draw
(-0.1,-10) node[left] {\small $-1$}; \draw(0,-10)--(0,0); \draw
(0,0)--(17.6,0); \draw[join=round]  (3.2000,-1.566)--(3.7151,-1.859)--(4.1359,-2.099)--(4.4918,-2.301)--(4.8000,-2.474)--(5.0719,-2.624)--(5.3151,-2.757)--(5.5351,-2.875)--(5.7359,-2.981)--(5.9207,-3.077)--(6.0918,-3.165)--(6.2510,-3.245)--(6.4000,-3.319)--(6.5399,-3.388)--(6.6719,-3.451)--(6.7967,-3.511)--(6.9151,-3.566)--(7.0277,-3.618)--(7.1351,-3.667)--(7.2377,-3.714)--(7.3359,-3.757)--(7.4302,-3.799)--(7.5207,-3.838)--(7.6078,-3.875)--(7.6918,-3.911)--(7.7728,-3.945)--(7.8510,-3.977)--(7.9267,-4.008)--(8.0000,-4.038)--(8.0710,-4.067)--(8.1399,-4.094)--(8.2069,-4.120)--(8.2719,-4.146)--(8.3351,-4.170)--(8.3967,-4.194)--(8.4566,-4.216)--(8.5151,-4.238)--(8.5721,-4.260)--(8.6277,-4.280)--(8.6820,-4.300)--(8.7351,-4.319)--(8.7870,-4.338)--(8.8377,-4.356)--(8.8873,-4.373)--(8.9359,-4.390)--(8.9835,-4.407)--(9.0302,-4.423)--(9.0759,-4.439)--(9.1207,-4.454)--(9.1647,-4.469)--(9.2078,-4.483)--(9.2502,-4.497)--(9.2918,-4.511)--(9.3326,-4.525)--(9.3728,-4.538)--(9.4122,-4.550)--(9.4510,-4.563)--(9.4892,-4.575)--(9.5267,-4.587)--(9.5636,-4.599)--(9.6000,-4.610)--(9.6358,-4.621)--(9.6710,-4.632)--(9.7057,-4.643)--(9.7399,-4.653)--(9.7736,-4.663)--(9.8069,-4.674)--(9.8396,-4.683)--(9.8719,-4.693)--(9.9037,-4.702)--(9.9351,-4.712)--(9.9661,-4.721)--(9.9967,-4.730)--(10.0269,-4.738)--(10.0566,-4.747)--(10.0860,-4.755)--(10.1151,-4.764)--(10.1438,-4.772)--(10.1721,-4.780)--(10.2001,-4.788)--(10.2277,-4.796)--(10.2550,-4.803)--(10.2820,-4.811)--(10.3087,-4.818)--(10.3351,-4.825)--(10.3612,-4.832)--(10.3870,-4.839)--(10.4125,-4.846)--(10.4377,-4.853)--(10.4627,-4.860)--(10.4873,-4.866)--(10.5118,-4.873)--(10.5359,-4.879)--(10.5599,-4.885)--(10.5835,-4.892)--(10.6070,-4.898)--(10.6302,-4.904)--(10.6531,-4.910)--(10.6759,-4.915)--(10.6984,-4.921)--(10.7207,-4.927)--(10.7428,-4.932)--(10.7647,-4.938)--(10.7863,-4.943)--(10.8078,-4.949)--(10.8291,-4.954)--(10.8502,-4.959)--(10.8711,-4.965)--(10.8918,-4.970)--(10.9123,-4.975)--(10.9326,-4.980)--(10.9528,-4.985)--(10.9728,-4.989)--(10.9926,-4.994)--(11.0122,-4.999)--(11.0317,-5.004)--(11.0510,-5.008)--(11.0702,-5.013)--(11.0892,-5.017)--(11.1080,-5.022)--(11.1267,-5.026)--(11.1453,-5.030)--(11.1636,-5.035)--(11.1819,-5.039)--(11.2000,-5.043)--(11.4719,-5.104)--(11.7151,-5.156)--(11.9351,-5.201)--(12.1359,-5.240)--(12.3207,-5.275)--(12.4918,-5.306)--(12.6510,-5.334)--(12.8000,-5.359)--(13.0719,-5.403)--(13.3151,-5.440)--(13.5351,-5.472)--(13.7359,-5.499)--(13.9207,-5.524)--(14.0918,-5.545)--(14.2510,-5.565)--(14.4000,-5.582)--(14.6719,-5.613)--(14.9151,-5.639)--(15.1351,-5.661)--(15.3359,-5.680)--(15.5207,-5.697)--(15.6918,-5.712)--(15.8510,-5.725)--(16.0000,-5.737)--(16.2719,-5.758)--(16.5151,-5.775)--(16.7351,-5.790)--(16.9359,-5.803)--(17.1207,-5.815)--(17.2918,-5.825)--(17.4510,-5.834)--(17.6000,-5.842);
\draw[join=round]
(1.6000,-0.705)--(1.4816,-0.646)--(1.3568,-0.585)--(1.2249,-0.522)--(1.0849,-0.457)--(0.9359,-0.388)--(0.7767,-0.317)--(0.6056,-0.243)--(0.4209,-0.166)--(0.2200,-0.085)--(0,0);
\draw[join=round]
(3.2000,-1.566)--(3.1416,-1.5335)--(3.0816,-1.4997)--(3.0200,-1.4651)--(2.9568,-1.4296)--(2.8918,-1.3932)--(2.8249,-1.3559)--(2.7559,-1.3176)--(2.6849,-1.2783)--(2.6116,-1.2379)--(2.5359,-1.1964)--(2.4577,-1.1537)--(2.3767,-1.1098)--(2.2927,-1.0646)--(2.2056,-1.0180)--(2.1151,-0.9699)--(2.0209,-0.9203)--(1.9226,-0.8691)--(1.8200,-0.8163)--(1.7126,-0.7615)--(1.6000,-0.7049);
\fill (0,0) circle (0.03cm); \fill (1.6000000,-0.70480403) circle (0.03cm);
\fill (2.5359400,-1.4447980) circle (0.03cm); \fill (3.2000000,-1.7425194)
circle (0.03cm); \fill (3.7150850,-2.2077427) circle (0.03cm); \fill
(4.1359400,-2.3715517) circle (0.03cm); \fill (4.4917679,-2.5139388) circle
(0.03cm); \fill (4.8000000,-2.4737225) circle (0.03cm); \fill
(5.0718800,-3.0870247) circle (0.03cm); \fill (5.3150850,-3.3275742) circle
(0.03cm); \fill (5.5350906,-3.5600387) circle (0.03cm); \fill
(5.7359400,-3.6147070) circle (0.03cm); \fill (5.9207035,-3.8192202) circle
(0.03cm); \fill (6.0917679,-3.9509943) circle (0.03cm); \fill
(6.2510250,-3.9270639) circle (0.03cm); \fill (6.4000000,-3.8045244) circle
(0.03cm); \fill (6.5399405,-4.0155449) circle (0.03cm); \fill
(6.6718800,-4.0783863) circle (0.03cm); \fill (6.7966840,-4.1699109) circle
(0.03cm); \fill (6.9150850,-4.1012632) circle (0.03cm); \fill
(7.0277079,-4.1255486) circle (0.03cm); \fill (7.1350906,-3.9420161) circle
(0.03cm); \fill (7.2376991,-3.9285847) circle (0.03cm); \fill
(7.3359400,-3.7572924) circle (0.03cm); \fill (7.4301699,-4.2112393) circle
(0.03cm); \fill (7.5207035,-4.5611258) circle (0.03cm); \fill
(7.6078200,-4.6783295) circle (0.03cm); \fill (7.6917679,-4.7222515) circle
(0.03cm); \fill (7.7727696,-5.1165429) circle (0.03cm); \fill
(7.8510250,-5.1592328) circle (0.03cm); \fill (7.9267141,-5.2794344) circle
(0.03cm); \fill (8.0000000,-5.1458765) circle (0.03cm); \fill
(8.0710306,-5.4186973) circle (0.03cm); \fill (8.1399405,-5.6206738) circle
(0.03cm); \fill (8.2068528,-5.7175328) circle (0.03cm); \fill
(8.2718800,-5.7306176) circle (0.03cm); \fill (8.3351254,-5.8295471) circle
(0.03cm); \fill (8.3966840,-5.8886919) circle (0.03cm); \fill
(8.4566436,-5.9608977) circle (0.03cm); \fill (8.5150850,-5.9007748) circle
(0.03cm); \fill (8.5720832,-5.8451032) circle (0.03cm); \fill
(8.6277079,-5.7759424) circle (0.03cm); \fill (8.6820236,-5.7138985) circle
(0.03cm); \fill (8.7350906,-5.6161670) circle (0.03cm); \fill
(8.7869650,-5.5767992) circle (0.03cm); \fill (8.8376991,-5.4475924) circle
(0.03cm); \fill (8.8873422,-5.4103091) circle (0.03cm); \fill
(8.9359400,-5.2838857) circle (0.03cm); \fill (9.2078200,-6.0493417) circle
(0.03cm); \fill (9.2917679,-6.3075543) circle (0.03cm); \fill
(9.6000000,-6.1649043) circle (0.03cm); \fill (9.8718800,-5.9222872) circle
(0.03cm); \fill (10.115085,-6.6198993) circle (0.03cm); \fill
(10.720704,-6.9897312) circle (0.03cm); \fill (11.200000,-7.2577650) circle
(0.03cm); \fill (12.800000,-8.2019616) circle (0.03cm); \fill
(14.400000,-9.0643218) circle (0.03cm); \fill (16.000000,-9.8769541) circle
(0.03cm);
\end{tikzpicture}
\end{center}
\caption{A comparison of the linear programming bound with other bounds.
The annotations on the right show the limits of the colored curves in high dimensions,
including our conjectured limit for the linear programming bound from Conjecture~\ref{conjecture:LP}.}
\label{figure:loglog}
\end{figure}

Figures~\ref{figure:LPbound} and~\ref{figure:ratioDeltac} faithfully reflect
the qualitative properties of these bounds, but they are of limited use in
extrapolating to high dimensions. For that purpose, a log-log plot is more
effective, as in Figure~\ref{figure:loglog}. This figure shows the linear
programming bound and the record sphere packing densities from \cite{SPLAG}
in black. The green and red curves show the best bounds that have been
analytically derived: the green curve is Levenshtein's bound \cite{Lev79},
and the red curve is the Kabatyanskii-Levenshtein bound \cite{KL78}, computed
using Levenshtein's universal bound for spherical codes \cite{Lev92,Lev98}
and the approach of Cohn and Zhao \cite{CZ} (we review these bounds in
Sections~\ref{ss:lub} and~\ref{ss:kl}). It is known that the linear
programming bound is at least as strong as these bounds \cite{CE,CZ}, but no
further analytic results are known. As shown in Figure~\ref{figure:loglog},
our numerical calculations indicate that the linear programming bound is not
much stronger than the better of these two bounds.

The blue curve in Figure~\ref{figure:loglog} is a lower bound for the linear
programming bound due to Scardicchio, Stillinger, and Torquato
\cite{SST2008}, which is a variant of a bound obtained by Torquato and
Stillinger \cite{TS2006}. No better lower bound is known for the linear
programming bound in the limit as $d \to \infty$, and the known lower bounds
for the sphere packing density are much worse. For example, the
Minkowski-Hlawka bound says that there exist sphere packings of density at
least $2^{-d}$ in $\R^d$, which amounts to the bottom edge of
Figure~\ref{figure:loglog}; this exponential decay rate is the same as in the
ideal glass phase of the hard sphere model (see \cite[p.~247]{PUZ}), and no
known lower bound improves on this rate. These lower bounds are obtained from
probabilistic or averaging arguments, and as far as we are aware, no explicit
construction in dimension $2048$ or greater has been shown to achieve the
Minkowski-Hlawka bound.

\subsection{Extrapolation}

The annotations on the right side of Figure~\ref{figure:loglog} show the
limits of the various curves as $d \to \infty$. The limits of the green, red,
and blue curves are known explicitly, and one can see that even $d=2048$ is
not especially close to the asymptotic limit as $d \to \infty$. We know that
the black curve always lies below the red curve, and it appears to be getting
steadily closer. Thus, we expect the linear programming bound to be at most
slightly better than the Kabatyanskii-Levenshtein bound in the limit as $d
\to \infty$. What Figure~\ref{figure:loglog} does not reveal is whether the
gap in fact tends to zero.

\begin{table}
\caption{Numerical comparison of the linear programming bound in $\R^d$ (denoted LP) with the
Kabatyanskii-Levenshtein bound (denoted KL).  The bounds are in the form
$(\log_2 \textup{density})/d$, and are rounded
rather than truncated.}
\label{table:KLLP}
\begin{center}
\begin{tabular}{rccccc}
\toprule
$d$ & $\phantom{-}\textup{LP}$ & $\phantom{-}\textup{KL}$ & $\phantom{-}\textup{KL}-\textup{LP}$ & Difference & Ratio\\
\midrule
$1$ & $\phantom{-}0.00000$ & $\phantom{-}0.00000$ & $\phantom{-}0.00000$\\
$2$ & $-0.07049$ & $\phantom{-}0.29248$ & $\phantom{-}0.36297$\\
$4$ & $-0.15665$ & $\phantom{-}0.17511$ & $\phantom{-}0.33176$ & \raisebox{1.4ex}[0ex]{$0.03122$}\\
$8$ & $-0.24737$ & $-0.04879$ & $\phantom{-}0.19858$ & \raisebox{1.4ex}[0ex]{$0.13318$} & \raisebox{2.8ex}[0ex]{$0.23$}\\
$16$ & $-0.33192$ & $-0.21692$ & $\phantom{-}0.11501$ & \raisebox{1.4ex}[0ex]{$0.08357$} & \raisebox{2.8ex}[0ex]{$1.59$}\\
$32$ & $-0.40382$ & $-0.33342$ & $\phantom{-}0.07040$ & \raisebox{1.4ex}[0ex]{$0.04461$} & \raisebox{2.8ex}[0ex]{$1.87$}\\
$64$ & $-0.46101$ & $-0.41947$ & $\phantom{-}0.04154$ & \raisebox{1.4ex}[0ex]{$0.02885$} & \raisebox{2.8ex}[0ex]{$1.55$}\\
$128$ & $-0.50432$ & $-0.47947$ & $\phantom{-}0.02485$ & \raisebox{1.4ex}[0ex]{$0.01669$} & \raisebox{2.8ex}[0ex]{$1.73$}\\
$256$ & $-0.53589$ & $-0.52023$ & $\phantom{-}0.01566$ & \raisebox{1.4ex}[0ex]{$0.00919$} & \raisebox{2.8ex}[0ex]{$1.82$}\\
$512$ & $-0.55824$ & $-0.54749$ & $\phantom{-}0.01075$ & \raisebox{1.4ex}[0ex]{$0.00491$} & \raisebox{2.8ex}[0ex]{$1.87$}\\
$1024$ & $-0.57370$ & $-0.56553$ & $\phantom{-}0.00816$ & \raisebox{1.4ex}[0ex]{$0.00259$} & \raisebox{2.8ex}[0ex]{$1.90$}\\
$2048$ & $-0.58418$ & $-0.57737$ & $\phantom{-}0.00682$ & \raisebox{1.4ex}[0ex]{$0.00135$} & \raisebox{2.8ex}[0ex]{$1.92$}\\
$4096$ & $-0.59120$ & $-0.58508$ & $\phantom{-}0.00611$ & \raisebox{1.4ex}[0ex]{$0.00070$} & \raisebox{2.8ex}[0ex]{$1.92$}\\
\midrule
$\infty$ & $\phantom{-}?$ & $-0.59906$ & $\phantom{-}?$\\
\bottomrule
\end{tabular}
\end{center}
\end{table}

To estimate the asymptotic gap between the red and black curves, it is
helpful to examine numerical data. Table~\ref{table:KLLP} shows the
difference $\textup{KL}-\textup{LP}$ between these two bounds, as well as the
differences between consecutive values of $\textup{KL}-\textup{LP}$ and their
ratios. The $\textup{KL}-\textup{LP}$ column does indeed seem to decrease for
$d \ge 2$, and we expect that it is converging towards its limit at a rate
proportional to $1/d$. In that case, the difference column to its right
should be decreasing towards zero at the same rate $1/d$, and so the ratios
in the last column should tend to $2$. The behavior of the ratio column is
not absolutely clear, but it does look like it may be increasing towards $2$
beyond $d=32$.

We therefore hypothesize that the difference column will continue to decrease
by a factor between $1.92$ and $2$ in each additional row. In that case, the
sum of all the entries below $0.00070$ in the difference column must lie
between
\begin{equation}
\sum_{n=1}^\infty \frac{0.00070}{2^n} = 0.00070
\qquad \text{and} \qquad
\sum_{n=1}^\infty \frac{0.00070}{1.92^n} < 0.00077.
\end{equation}
In other words, the $\textup{KL}-\textup{LP}$ column will fall below
$0.00611$ by an amount between $0.00070$ and $0.00077$, and so its limiting
value will be between $0.00534$ and $0.00541$. We conclude that the limit of
the LP column must be $-0.6044 \pm 0.0001$ (i.e., between $-0.6045$ and
$-0.6043$). In other words, the linear programming bound will be
approximately $2^{-0.6044 d}$ when $d$ is large.

\begin{conjecture} \label{conjecture:LP}
There exists a constant $\lambda$ with $0.604 < \lambda < 0.605$ such that
the linear programming bound for the sphere packing density in $\R^d$ is
$2^{-(\lambda+o(1))d}$ as $d \to \infty$ when the auxiliary function is fully
optimized.
\end{conjecture}

Of course we have no proof of this conjecture, or even a heuristic
derivation. It is possible that the numbers could behave entirely differently
when $d$ is much larger, but that does not seem plausible. We are very
confident in the first three decimal places of the estimate $0.6044$ for
$\lambda$, and fairly confident in the fourth. In fact, we have a proposal
for the exact constant:

\begin{conjecture} \label{conjecture:exact}
The constant $\lambda$ in Conjecture~\ref{conjecture:LP} is given by
$2^{-\lambda} =  \sqrt{e/(2\pi)}$. Equivalently,
\begin{equation}
\lim_{c \to \infty} \frac{\Delta_1^\textup{LP}(c)}{c} = \frac{1}{\pi^2}
\end{equation}
and
\begin{equation}
\lim_{d \to \infty} \frac{\textup{A}_-(d)}{\sqrt{d}} = \frac{1}{\pi},
\end{equation}
where $\textup{A}_-(d)$ denotes the optimal radius for the $-1$ eigenfunction
uncertainty principle in $\R^d$.
\end{conjecture}

Of course Conjecture~\ref{conjecture:exact} is speculative, and four digits
of accuracy is far from enough to make a definitive argument for this value.
The equivalent limits are much more appealing than the formula for $\lambda$,
and their simplicity justifies going out on a limb. We have no great faith in
this conjecture, but it is worth noting that a simple formula fits the data
beautifully. For comparison, $\lambda = 0.6044 \pm 0.0001$ would amount to
\begin{equation}
0.318287 < \lim_{d \to \infty} \frac{\textup{A}_-(d)}{\sqrt{d}} < 0.318333,
\end{equation}
and
$1/\pi = 0.318309\dots.$

\begin{table}
\caption{The ratio of the $-1$ eigenfunction uncertainty principle
radius to the $+1$ radius.}
\label{table:plusoneratio}
\begin{center}
\begin{tabular}{rccc}
\toprule
$d$ & $-1$ radius & $+1$ radius & Ratio\\
\midrule
   $4$ & $1.2038$ & $0.9660$ & $1.2462$\\
   $8$ & $1.4142$ & $1.2173$ & $1.1618$\\
  $16$ & $1.7393$ & $1.5812$ & $1.1000$\\
  $32$ & $2.2241$ & $2.1000$ & $1.0591$\\
  $64$ & $2.9317$ & $2.8359$ & $1.0338$\\
 $128$ & $3.9515$ & $3.8787$ & $1.0188$\\
 $256$ & $5.4109$ & $5.3565$ & $1.0102$\\
 $512$ & $7.4905$ & $7.4504$ & $1.0054$\\
$1024$ & $10.446$ & $10.417$ & $1.0028$\\
$2048$ & $14.640$ & $14.619$ & $1.0014$\\
\bottomrule
\end{tabular}
\end{center}
\end{table}

Our calculations also support Conjecture~1.5 from \cite{CG}, which says that
the sign change radii for the $+1$ and $-1$ eigenfunction uncertainty
principles in $\R^d$ are the same asymptotically as $d \to \infty$. See
Table~\ref{table:plusoneratio} and \url{https://hdl.handle.net/1721.1/125646} for the numerical
data.\footnote{The reason why this table omits $d=1$ and $2$ is that we do
not have good numerical data in these dimensions. See the end of Section~4 in
\cite{CG}.} In the notation of \cite{CG},
\begin{equation}
\lim_{d \to \infty} \frac{\textup{A}_+(d)}{\sqrt{d}} = \lim_{d \to \infty} \frac{\textup{A}_-(d)}{\sqrt{d}}.
\end{equation}
Specifically, the ratio $\textup{A}_+(d)/\textup{A}_-(d)$ seems to be
$1+O(1/d)$.

\section{Properties of the spectrum and degeneracies}
\label{sec:properties}

\begin{figure}
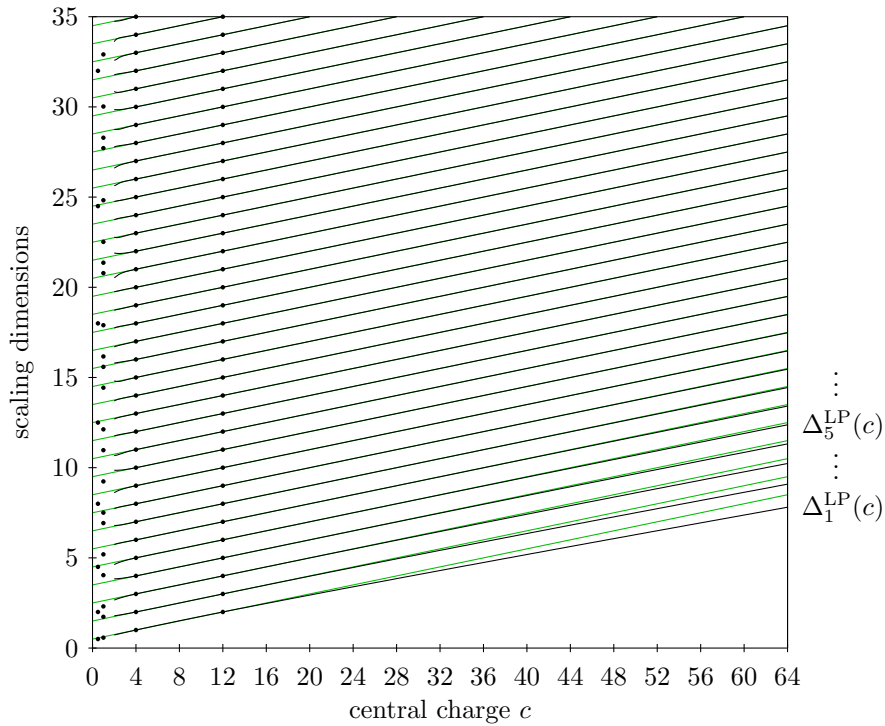

\begin{center}

\end{center}
\caption{The spectra $\Delta_n^\textup{LP}(c)$ for $2 \le c \le 64$ and $c \in \{1/2,1\}$, drawn in black.
The dots highlight the spectra for $c \in \{1/2,1,4,12\}$, and the green lines
show the 1d generalized free fermion spectra.}
\label{figure:limitingspectra}
\end{figure}

The spectra of our numerically optimized solutions of the spinless modular
bootstrap behave remarkably regularly when the central charge $c$ is large.
Figure~\ref{figure:limitingspectra} shows the scaling dimensions for $c \le
64$, with the sharp cases $c=1/2$, $1$, $4$, and $12$ highlighted. When $c=4$
or $12$, the scaling dimensions are positive integers, excluding $1$ when
$c=12$. The green lines in the figure extrapolate these arithmetic
progressions to other values of $c$. In other words, the $n$-th green line
from the bottom amounts to $\Delta_n = n + (c-4)/8$. While this equation
never holds exactly except when $c=4$ or $12$, it is an excellent
approximation when $n$ is large and $c \ge 4$. Later in this section we will
examine how close this approximation is.

The equation $\Delta_n = n + (c-4)/8$ amounts to the 1d generalized free
fermion spectrum. This spectrum arose in analytic functionals for the 1d
conformal bootstrap constructed by Maz\'a\v{c} \cite{M}, which were
generalized to a basis by Maz\'a\v{c}  and Paulos \cite{MP}. Hartman,
Maz\'a\v{c}, and Rastelli \cite{HMR} discovered that these functionals could
be adapted to the 2d modular bootstrap with $\textup{U}(1)^c$ or Virasoro symmetry,
and special cases were independently constructed by Rolen and Wagner
\cite{RW} and by Feigenbaum, Grabner, and Hardin \cite{FGH}.

\begin{figure}
\begin{center}
\begin{tikzpicture}[x=0.525cm,y=0.84cm]
\draw[green!75!black,join=round] (1.40681,0)--(1.46667, -0.0408176)--(1.6, -0.132879)--(1.73333, -0.226435)--(1.86667, -0.321376)--(2, -0.417581)--(2.13333, -0.514925)--(2.26667, -0.613275)--(2.4, -0.712494)--(2.53333, -0.812436)--(2.66667, -0.912955)--(2.8, -1.0139)--(2.93333, -1.11511)--(3.06667, -1.21643)--(3.2, -1.31771)--(3.33333, -1.41878)--(3.46667, -1.51948)--(3.6, -1.61966)--(3.73333, -1.71917)--(3.86667, -1.81785)--(4, -1.91555)--(4.13333, -2.01214)--(4.26667, -2.10747)--(4.4, -2.20141)--(4.53333, -2.29385)--(4.66667, -2.38465)--(4.8, -2.47372)--(4.93333, -2.56096)--(5.06667, -2.64626)--(5.2, -2.72956)--(5.33333, -2.81077)--(5.46667, -2.88984)--(5.6, -2.9667)--(5.73333, -3.04132)--(5.86667, -3.11365)--(6, -3.18367)--(6.13333, -3.25136)--(6.26667, -3.31671)--(6.4, -3.37971)--(6.53333, -3.44038)--(6.66667, -3.49872)--(6.8, -3.55475)--(6.93333, -3.6085)--(7.06667, -3.66)--(7.2, -3.70928)--(7.33333, -3.75639)--(7.46667, -3.80138)--(7.6, -3.84428)--(7.73333, -3.88516)--(7.86667, -3.92406)--(8, -3.96105)--(8.13333, -3.99618)--(8.26667, -4.02952)--(8.4, -4.06112)--(8.53333, -4.09105)--(8.66667, -4.11938)--(8.8, -4.14615)--(8.93333, -4.17144)--(9.06667, -4.19531)--(9.2, -4.21782)--(9.33333, -4.23903)--(9.46667, -4.259)--(9.6, -4.27778)--(9.73333, -4.29544)--(9.86667, -4.31202)--(10, -4.32758)--(10.1333, -4.34218)--(10.2667, -4.35587)--(10.4, -4.36868)--(10.5333, -4.38067)--(10.6667, -4.39189)--(10.8, -4.40237)--(10.9333, -4.41217)--(11.0667, -4.4213)--(11.2, -4.42983)--(11.3333, -4.43777)--(11.4667, -4.44517)--(11.6, -4.45205)--(11.7333, -4.45846)--(11.8667, -4.46441)--(12, -4.46994)--(12.1333, -4.47508)--(12.2667, -4.47984)--(12.4, -4.48425)--(12.5333, -4.48834)--(12.6667, -4.49212)--(12.8, -4.49562)--(12.9333, -4.49885)--(13.0667, -4.50183)--(13.2, -4.50458)--(13.3333, -4.50712)--(13.4667, -4.50945)--(13.6, -4.5116)--(13.7333, -4.51357)--(13.8667, -4.51539)--(14, -4.51705)--(14.1333, -4.51857)--(14.2667, -4.51996)--(14.4, -4.52123)--(14.5333, -4.52239)--(14.6667, -4.52345)--(14.8, -4.52441)--(14.9333, -4.52528)--(15.0667, -4.52608)--(15.2, -4.52679)--(15.3333, -4.52744)--(15.4667, -4.52802)--(15.6, -4.52854)--(15.7333, -4.52901)--(15.8667, -4.52943)--(16, -4.5298)--(16.1333, -4.53013)--(16.2667, -4.53042)--(16.4, -4.53068)--(16.5333, -4.5309)--(16.6667, -4.53109)--(16.8, -4.53126)--(16.9333, -4.5314)--(17.0667, -4.53152)--(17.2, -4.53162)--(17.3333, -4.5317)--(17.4667, -4.53176)--(17.6, -4.53181);
\draw (8.65,-1) node[right] {\small Linear programming bound};
\draw (8.65,-1.5) node[right] {\small Hypothetical bound};
\draw (8.65,-2) node[right] {\small Record sphere packing};
\draw (8,-1)--(8.56,-1);
\draw[green!75!black] (8,-1.5)--(8.56,-1.5);
\fill (8.28,-2) circle (0.03cm);
\draw (0,-7)--(17.6,-7);
\draw (-0.1,-1) node[left] {\small $-0.1$};
\draw (-0.1,-2) node[left] {\small $-0.2$};
\draw (-0.1,-3) node[left] {\small $-0.3$};
\draw (-0.1,-4) node[left] {\small $-0.4$};
\draw (-0.1,-5) node[left] {\small $-0.5$};
\draw (-0.1,-6) node[left] {\small $-0.6$};
\foreach \x in {0,...,11}
\draw (1.6*\x,-0.065)--(1.6*\x,0.065); \draw (0,0.15) node[above] {\small $1$}; \draw
(1.6,0.15) node[above] {\small $2$}; \draw (3.2,0.15) node[above] {\small $4$}; \draw
(4.8,0.15) node[above] {\small $8$}; \draw (6.4,0.15) node[above] {\small $16$}; \draw
(8.0,0.15) node[above] {\small $32$}; \draw (9.6,0.15) node[above] {\small $64$}; \draw
(11.2,0.15) node[above] {\small $128$}; \draw (12.8,0.15) node[above] {\small $256$}; \draw
(14.4,0.15) node[above] {\small $512$}; \draw (16,0.15) node[above] {\small $1024$};  \draw (17.6,0.15) node[above] {\small $2048$};
\draw (17.6,0)--(17.6,-7)--(0,-7);
\draw (17.7,-4.529044145) node[right] {\small $\frac{1+\log \pi}{\log 4}-2$};
\draw (17.496,-4.529044145)--(17.704,-4.529044145);
\foreach \y in {-7,...,0} \draw (-0.104,{\y})--(0.104,{\y}); \draw (8.8,0.65)
node[above] {\small dimension}; \draw (-2.3,-3.5) node[rotate=90]
{\small $\log_2(\text{density})/\text{dimension}$};
\draw (19.6,-3.5) node[rotate=-90]
{\small $\phantom{\log_2(\text{density})/\text{dimension}}$};
\draw (-0.1,0) node[left] {\small $0$}; \draw
(-0.1,-7) node[left] {\small $-0.7$}; \draw(0,-7)--(0,0); \draw
(0,0)--(17.6,0);
\draw[join=round]
(3.2000,-1.566)--(3.7151,-1.859)--(4.1359,-2.099)--(4.4918,-2.301)--(4.8000,-2.474)--(5.0719,-2.624)--(5.3151,-2.757)--(5.5351,-2.875)--(5.7359,-2.981)--(5.9207,-3.077)--(6.0918,-3.165)--(6.2510,-3.245)--(6.4000,-3.319)--(6.5399,-3.388)--(6.6719,-3.451)--(6.7967,-3.511)--(6.9151,-3.566)--(7.0277,-3.618)--(7.1351,-3.667)--(7.2377,-3.714)--(7.3359,-3.757)--(7.4302,-3.799)--(7.5207,-3.838)--(7.6078,-3.875)--(7.6918,-3.911)--(7.7728,-3.945)--(7.8510,-3.977)--(7.9267,-4.008)--(8.0000,-4.038)--(8.0710,-4.067)--(8.1399,-4.094)--(8.2069,-4.120)--(8.2719,-4.146)--(8.3351,-4.170)--(8.3967,-4.194)--(8.4566,-4.216)--(8.5151,-4.238)--(8.5721,-4.260)--(8.6277,-4.280)--(8.6820,-4.300)--(8.7351,-4.319)--(8.7870,-4.338)--(8.8377,-4.356)--(8.8873,-4.373)--(8.9359,-4.390)--(8.9835,-4.407)--(9.0302,-4.423)--(9.0759,-4.439)--(9.1207,-4.454)--(9.1647,-4.469)--(9.2078,-4.483)--(9.2502,-4.497)--(9.2918,-4.511)--(9.3326,-4.525)--(9.3728,-4.538)--(9.4122,-4.550)--(9.4510,-4.563)--(9.4892,-4.575)--(9.5267,-4.587)--(9.5636,-4.599)--(9.6000,-4.610)--(9.6358,-4.621)--(9.6710,-4.632)--(9.7057,-4.643)--(9.7399,-4.653)--(9.7736,-4.663)--(9.8069,-4.674)--(9.8396,-4.683)--(9.8719,-4.693)--(9.9037,-4.702)--(9.9351,-4.712)--(9.9661,-4.721)--(9.9967,-4.730)--(10.0269,-4.738)--(10.0566,-4.747)--(10.0860,-4.755)--(10.1151,-4.764)--(10.1438,-4.772)--(10.1721,-4.780)--(10.2001,-4.788)--(10.2277,-4.796)--(10.2550,-4.803)--(10.2820,-4.811)--(10.3087,-4.818)--(10.3351,-4.825)--(10.3612,-4.832)--(10.3870,-4.839)--(10.4125,-4.846)--(10.4377,-4.853)--(10.4627,-4.860)--(10.4873,-4.866)--(10.5118,-4.873)--(10.5359,-4.879)--(10.5599,-4.885)--(10.5835,-4.892)--(10.6070,-4.898)--(10.6302,-4.904)--(10.6531,-4.910)--(10.6759,-4.915)--(10.6984,-4.921)--(10.7207,-4.927)--(10.7428,-4.932)--(10.7647,-4.938)--(10.7863,-4.943)--(10.8078,-4.949)--(10.8291,-4.954)--(10.8502,-4.959)--(10.8711,-4.964)--(10.8918,-4.970)--(10.9123,-4.975)--(10.9326,-4.980)--(10.9528,-4.985)--(10.9728,-4.989)--(10.9926,-4.994)--(11.0122,-4.999)--(11.0317,-5.004)--(11.0510,-5.008)--(11.0702,-5.013)--(11.0892,-5.017)--(11.1080,-5.022)--(11.1267,-5.026)--(11.1453,-5.030)--(11.1636,-5.035)--(11.1819,-5.039)--(11.2000,-5.043)--(11.4719,-5.104)--(11.7151,-5.156)--(11.9351,-5.201)--(12.1359,-5.240)--(12.3207,-5.275)--(12.4918,-5.306)--(12.6510,-5.334)--(12.8000,-5.359)--(13.0719,-5.403)--(13.3151,-5.440)--(13.5351,-5.472)--(13.7359,-5.499)--(13.9207,-5.524)--(14.0918,-5.545)--(14.2510,-5.565)--(14.4000,-5.582)--(14.6719,-5.613)--(14.9151,-5.639)--(15.1351,-5.661)--(15.3359,-5.680)--(15.5207,-5.697)--(15.6918,-5.712)--(15.8510,-5.725)--(16.0000,-5.737)--(16.2719,-5.758)--(16.5151,-5.775)--(16.7351,-5.790)--(16.9359,-5.803)--(17.1207,-5.815)--(17.2918,-5.825)--(17.4510,-5.834)--(17.6000,-5.842);
\draw[join=round]
(1.6000,-0.705)--(1.4816,-0.646)--(1.3568,-0.585)--(1.2249,-0.522)--(1.0849,-0.457)--(0.9359,-0.388)--(0.7767,-0.317)--(0.6056,-0.243)--(0.4209,-0.166)--(0.2200,-0.085)--(0,0);
\draw[join=round]
(3.2000,-1.566)--(3.1416,-1.5335)--(3.0816,-1.4997)--(3.0200,-1.4651)--(2.9568,-1.4296)--(2.8918,-1.3932)--(2.8249,-1.3559)--(2.7559,-1.3176)--(2.6849,-1.2783)--(2.6116,-1.2379)--(2.5359,-1.1964);
\draw[join=round]
(2.5359,-1.1964)--(2.4577,-1.1537)--(2.3767,-1.1098)--(2.2927,-1.0646)--(2.2056,-1.0180)--(2.1151,-0.9699)--(2.0209,-0.9203)--(1.9226,-0.8691)--(1.8200,-0.8163)--(1.7126,-0.7615)--(1.6000,-0.7049);
\fill (0,0) circle (0.03cm); \fill (1.6000000,-0.70480403) circle (0.03cm);
\fill (2.5359400,-1.4447980) circle (0.03cm); \fill (3.2000000,-1.7425194)
circle (0.03cm); \fill (3.7150850,-2.2077427) circle (0.03cm); \fill
(4.1359400,-2.3715517) circle (0.03cm); \fill (4.4917679,-2.5139388) circle
(0.03cm); \fill (4.8000000,-2.4737225) circle (0.03cm); \fill
(5.0718800,-3.0870247) circle (0.03cm); \fill (5.3150850,-3.3275742) circle
(0.03cm); \fill (5.5350906,-3.5600387) circle (0.03cm); \fill
(5.7359400,-3.6147070) circle (0.03cm); \fill (5.9207035,-3.8192202) circle
(0.03cm); \fill (6.0917679,-3.9509943) circle (0.03cm); \fill
(6.2510250,-3.9270639) circle (0.03cm); \fill (6.4000000,-3.8045244) circle
(0.03cm); \fill (6.5399405,-4.0155449) circle (0.03cm); \fill
(6.6718800,-4.0783863) circle (0.03cm); \fill (6.7966840,-4.1699109) circle
(0.03cm); \fill (6.9150850,-4.1012632) circle (0.03cm); \fill
(7.0277079,-4.1255486) circle (0.03cm); \fill (7.1350906,-3.9420161) circle
(0.03cm); \fill (7.2376991,-3.9285847) circle (0.03cm); \fill
(7.3359400,-3.7572924) circle (0.03cm); \fill (7.4301699,-4.2112393) circle
(0.03cm); \fill (7.5207035,-4.5611258) circle (0.03cm); \fill
(7.6078200,-4.6783295) circle (0.03cm); \fill (7.6917679,-4.7222515) circle
(0.03cm); \fill (7.7727696,-5.1165429) circle (0.03cm); \fill
(7.8510250,-5.1592328) circle (0.03cm); \fill (7.9267141,-5.2794344) circle
(0.03cm); \fill (8.0000000,-5.1458765) circle (0.03cm); \fill
(8.0710306,-5.4186973) circle (0.03cm); \fill (8.1399405,-5.6206738) circle
(0.03cm); \fill (8.2068528,-5.7175328) circle (0.03cm); \fill
(8.2718800,-5.7306176) circle (0.03cm); \fill (8.3351254,-5.8295471) circle
(0.03cm); \fill (8.3966840,-5.8886919) circle (0.03cm); \fill
(8.4566436,-5.9608977) circle (0.03cm); \fill (8.5150850,-5.9007748) circle
(0.03cm); \fill (8.5720832,-5.8451032) circle (0.03cm); \fill
(8.6277079,-5.7759424) circle (0.03cm); \fill (8.6820236,-5.7138985) circle
(0.03cm); \fill (8.7350906,-5.6161670) circle (0.03cm); \fill
(8.7869650,-5.5767992) circle (0.03cm); \fill (8.8376991,-5.4475924) circle
(0.03cm); \fill (8.8873422,-5.4103091) circle (0.03cm); \fill
(8.9359400,-5.2838857) circle (0.03cm); \fill (9.2078200,-6.0493417) circle
(0.03cm); \fill (9.2917679,-6.3075543) circle (0.03cm); \fill
(9.6000000,-6.1649043) circle (0.03cm); \fill (9.8718800,-5.9222872) circle
(0.03cm); \fill (10.115085,-6.6198993) circle (0.03cm); \fill
(10.720704,-6.9897312) circle (0.03cm);
\end{tikzpicture}
\end{center}
\caption{The linear programming bound and the hypothetical bound based on the 1d generalized free
fermion spectrum $\Delta_n(c) = n + (c-4)/8$.}
\label{figure:hypotheticalbound}
\end{figure}
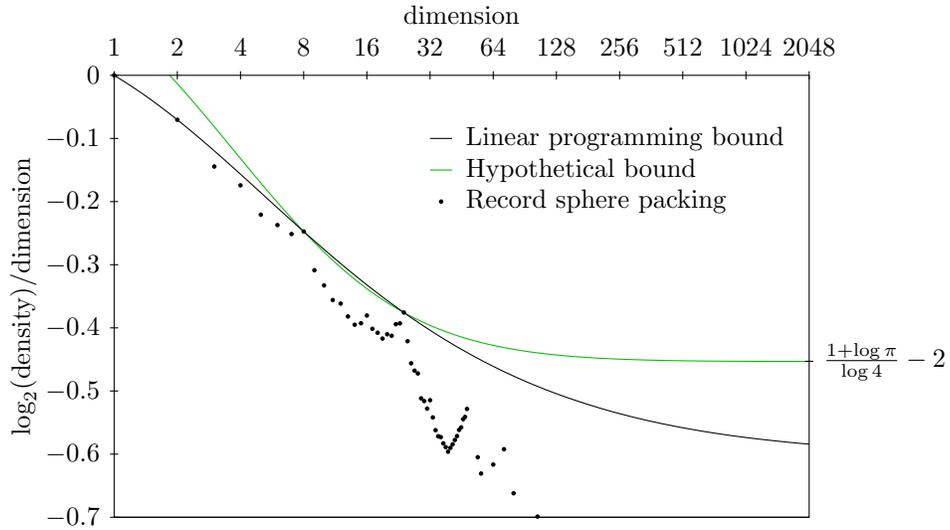

Figure~\ref{figure:hypotheticalbound} shows the sphere packing bounds
obtained from these functionals. No proof is known that the functionals
satisfy the required inequalities, and indeed they do not for $8 < d < 24$;
in particular, they would prove an impossibly good linear programming bound
for $d=16$ (see the dual bounds in \cite{CT}). However, the inequalities seem
to hold for other values of $d$. Unfortunately, the resulting bounds are
disappointing. The fact that the $\Delta_1^\textup{LP}(c)$ curve in
Figure~\ref{figure:limitingspectra} bends below the free fermion line
$\Delta_1 = 1 + (c-4)/8$ is crucial for obtaining a strong bound, and the
quality of the bound depends on the degree of deflection.

In contrast to the behavior for large $c$, the spectra for small $c$ are much
less regularly spaced. As we decrease $c$ below $4$, the scaling dimensions
in Figure~\ref{figure:limitingspectra} start to diverge unpredictably from
the green lines, and the behavior for $c \in \{1/2,1\}$ is entirely
different. Our numerical techniques break down at $c=3/2$, and it is
presumably not a coincidence that this failure occurs at the transition
between different regimes. It would be interesting to explore this transition
for $1 < c < 2$.

\subsection{Convergence to the free fermion spectrum}\label{ss:betacon}

\begin{figure}
\begin{center}
\begin{tikzpicture}[x=0.525cm,y=0.84cm]
\draw (1.4,-0.5) node[right] {\small $\Delta_n^{\textup{LP}, 1}(128)$};
\draw (1.57,-0.875) node[right] {\small $\vdots$};
\draw (1.4,-1.5) node[right] {\small $\Delta_n^{\textup{LP}, 128}(128)$};
\draw (1.4,-2) node[right] {\small 1d generalized free fermions};
\fill[red] (1.03,-0.5) circle (0.03cm);
\fill (1.03,-1.5) circle (0.03cm);
\fill[green!75!black] (1.03,-2) circle (0.03cm);
\draw (-1.8,-5) node[rotate=90]
{\small $\phantom{\text{spectrum}}$};
\draw (19.1,-5) node[rotate=-90]
{\small $\phantom{\text{spectrum}}$};
\draw (0,-10)--(17.6,-10)--(17.6,0)--(0,0);
\foreach \y in {-8,...,0} \draw (-0.104,{\y*10/8})--(0.104,{\y*10/8});
\foreach \y in {0,25,...,200}
\draw (-0.1,-10+\y/20) node[left] {\small \y};
\foreach \x in {0,32,...,256}
\draw ({17.6*\x/256},-10-0.065)--({17.6*\x/256},-10+0.065);
\foreach \x in {0,16,...,128}
\draw ({17.6*\x/128},-10-0.15)  node[below] {\small \x};
\draw (8.8,-10-0.75)
node[below] {\small $n$};
\draw (0,-10)--(0,0); \draw
(0,0)--(17.6,0);
\fill[red!100.00!black] (0.1375,-8.958) circle (0.03cm);
\fill[red!85.71!black] (0.1375,-9.026) circle (0.03cm); \fill[red!85.71!black] (0.2750,-8.842) circle (0.03cm);
\fill[red!71.43!black] (0.1375,-9.100) circle (0.03cm); \fill[red!71.43!black] (0.2750,-8.963) circle (0.03cm); \fill[red!71.43!black] (0.4125,-8.823) circle (0.03cm); \fill[red!71.43!black] (0.5500,-8.660) circle (0.03cm);
\fill[red!57.14!black] (0.1375,-9.172) circle (0.03cm); \fill[red!57.14!black] (0.2750,-9.064) circle (0.03cm); \fill[red!57.14!black] (0.4125,-8.963) circle (0.03cm); \fill[red!57.14!black] (0.5500,-8.861) circle (0.03cm); \fill[red!57.14!black] (0.6875,-8.756) circle (0.03cm); \fill[red!57.14!black] (0.8250,-8.642) circle (0.03cm); \fill[red!57.14!black] (0.9625,-8.513) circle (0.03cm); \fill[red!57.14!black] (1.1000,-8.356) circle (0.03cm);
\fill[red!42.86!black] (0.1375,-9.229) circle (0.03cm); \fill[red!42.86!black] (0.2750,-9.141) circle (0.03cm); \fill[red!42.86!black] (0.4125,-9.061) circle (0.03cm); \fill[red!42.86!black] (0.5500,-8.985) circle (0.03cm); \fill[red!42.86!black] (0.6875,-8.909) circle (0.03cm); \fill[red!42.86!black] (0.8250,-8.834) circle (0.03cm); \fill[red!42.86!black] (0.9625,-8.757) circle (0.03cm); \fill[red!42.86!black] (1.1000,-8.678) circle (0.03cm); \fill[red!42.86!black] (1.2375,-8.596) circle (0.03cm); \fill[red!42.86!black] (1.3750,-8.511) circle (0.03cm); \fill[red!42.86!black] (1.5125,-8.421) circle (0.03cm); \fill[red!42.86!black] (1.6500,-8.325) circle (0.03cm); \fill[red!42.86!black] (1.7875,-8.222) circle (0.03cm); \fill[red!42.86!black] (1.9250,-8.108) circle (0.03cm); \fill[red!42.86!black] (2.0625,-7.978) circle (0.03cm); \fill[red!42.86!black] (2.2000,-7.817) circle (0.03cm);
\fill[red!28.57!black] (0.1375,-9.260) circle (0.03cm); \fill[red!28.57!black] (0.2750,-9.182) circle (0.03cm); \fill[red!28.57!black] (0.4125,-9.114) circle (0.03cm); \fill[red!28.57!black] (0.5500,-9.050) circle (0.03cm); \fill[red!28.57!black] (0.6875,-8.988) circle (0.03cm); \fill[red!28.57!black] (0.8250,-8.927) circle (0.03cm); \fill[red!28.57!black] (0.9625,-8.868) circle (0.03cm); \fill[red!28.57!black] (1.1000,-8.808) circle (0.03cm); \fill[red!28.57!black] (1.2375,-8.749) circle (0.03cm); \fill[red!28.57!black] (1.3750,-8.689) circle (0.03cm); \fill[red!28.57!black] (1.5125,-8.629) circle (0.03cm); \fill[red!28.57!black] (1.6500,-8.568) circle (0.03cm); \fill[red!28.57!black] (1.7875,-8.506) circle (0.03cm); \fill[red!28.57!black] (1.9250,-8.444) circle (0.03cm); \fill[red!28.57!black] (2.0625,-8.380) circle (0.03cm); \fill[red!28.57!black] (2.2000,-8.316) circle (0.03cm); \fill[red!28.57!black] (2.3375,-8.249) circle (0.03cm); \fill[red!28.57!black] (2.4750,-8.181) circle (0.03cm); \fill[red!28.57!black] (2.6125,-8.112) circle (0.03cm); \fill[red!28.57!black] (2.7500,-8.040) circle (0.03cm); \fill[red!28.57!black] (2.8875,-7.966) circle (0.03cm); \fill[red!28.57!black] (3.0250,-7.889) circle (0.03cm); \fill[red!28.57!black] (3.1625,-7.810) circle (0.03cm); \fill[red!28.57!black] (3.3000,-7.727) circle (0.03cm); \fill[red!28.57!black] (3.4375,-7.641) circle (0.03cm); \fill[red!28.57!black] (3.5750,-7.550) circle (0.03cm); \fill[red!28.57!black] (3.7125,-7.453) circle (0.03cm); \fill[red!28.57!black] (3.8500,-7.350) circle (0.03cm); \fill[red!28.57!black] (3.9875,-7.238) circle (0.03cm); \fill[red!28.57!black] (4.1250,-7.115) circle (0.03cm); \fill[red!28.57!black] (4.2625,-6.973) circle (0.03cm); \fill[red!28.57!black] (4.4000,-6.798) circle (0.03cm);
\fill[red!14.29!black] (0.1375,-9.268) circle (0.03cm); \fill[red!14.29!black] (0.2750,-9.192) circle (0.03cm); \fill[red!14.29!black] (0.4125,-9.127) circle (0.03cm); \fill[red!14.29!black] (0.5500,-9.066) circle (0.03cm); \fill[red!14.29!black] (0.6875,-9.007) circle (0.03cm); \fill[red!14.29!black] (0.8250,-8.951) circle (0.03cm); \fill[red!14.29!black] (0.9625,-8.896) circle (0.03cm); \fill[red!14.29!black] (1.1000,-8.842) circle (0.03cm); \fill[red!14.29!black] (1.2375,-8.788) circle (0.03cm); \fill[red!14.29!black] (1.3750,-8.735) circle (0.03cm); \fill[red!14.29!black] (1.5125,-8.683) circle (0.03cm); \fill[red!14.29!black] (1.6500,-8.630) circle (0.03cm); \fill[red!14.29!black] (1.7875,-8.578) circle (0.03cm); \fill[red!14.29!black] (1.9250,-8.526) circle (0.03cm); \fill[red!14.29!black] (2.0625,-8.475) circle (0.03cm); \fill[red!14.29!black] (2.2000,-8.423) circle (0.03cm); \fill[red!14.29!black] (2.3375,-8.371) circle (0.03cm); \fill[red!14.29!black] (2.4750,-8.319) circle (0.03cm); \fill[red!14.29!black] (2.6125,-8.267) circle (0.03cm); \fill[red!14.29!black] (2.7500,-8.215) circle (0.03cm); \fill[red!14.29!black] (2.8875,-8.163) circle (0.03cm); \fill[red!14.29!black] (3.0250,-8.110) circle (0.03cm); \fill[red!14.29!black] (3.1625,-8.058) circle (0.03cm); \fill[red!14.29!black] (3.3000,-8.005) circle (0.03cm); \fill[red!14.29!black] (3.4375,-7.952) circle (0.03cm); \fill[red!14.29!black] (3.5750,-7.898) circle (0.03cm); \fill[red!14.29!black] (3.7125,-7.844) circle (0.03cm); \fill[red!14.29!black] (3.8500,-7.789) circle (0.03cm); \fill[red!14.29!black] (3.9875,-7.735) circle (0.03cm); \fill[red!14.29!black] (4.1250,-7.679) circle (0.03cm); \fill[red!14.29!black] (4.2625,-7.623) circle (0.03cm); \fill[red!14.29!black] (4.4000,-7.567) circle (0.03cm); \fill[red!14.29!black] (4.5375,-7.509) circle (0.03cm); \fill[red!14.29!black] (4.6750,-7.452) circle (0.03cm); \fill[red!14.29!black] (4.8125,-7.393) circle (0.03cm); \fill[red!14.29!black] (4.9500,-7.333) circle (0.03cm); \fill[red!14.29!black] (5.0875,-7.273) circle (0.03cm); \fill[red!14.29!black] (5.2250,-7.212) circle (0.03cm); \fill[red!14.29!black] (5.3625,-7.150) circle (0.03cm); \fill[red!14.29!black] (5.5000,-7.087) circle (0.03cm); \fill[red!14.29!black] (5.6375,-7.023) circle (0.03cm); \fill[red!14.29!black] (5.7750,-6.957) circle (0.03cm); \fill[red!14.29!black] (5.9125,-6.891) circle (0.03cm); \fill[red!14.29!black] (6.0500,-6.823) circle (0.03cm); \fill[red!14.29!black] (6.1875,-6.753) circle (0.03cm); \fill[red!14.29!black] (6.3250,-6.683) circle (0.03cm); \fill[red!14.29!black] (6.4625,-6.610) circle (0.03cm); \fill[red!14.29!black] (6.6000,-6.536) circle (0.03cm); \fill[red!14.29!black] (6.7375,-6.460) circle (0.03cm); \fill[red!14.29!black] (6.8750,-6.382) circle (0.03cm); \fill[red!14.29!black] (7.0125,-6.302) circle (0.03cm); \fill[red!14.29!black] (7.1500,-6.219) circle (0.03cm); \fill[red!14.29!black] (7.2875,-6.133) circle (0.03cm); \fill[red!14.29!black] (7.4250,-6.045) circle (0.03cm); \fill[red!14.29!black] (7.5625,-5.953) circle (0.03cm); \fill[red!14.29!black] (7.7000,-5.857) circle (0.03cm); \fill[red!14.29!black] (7.8375,-5.757) circle (0.03cm); \fill[red!14.29!black] (7.9750,-5.652) circle (0.03cm); \fill[red!14.29!black] (8.1125,-5.541) circle (0.03cm); \fill[red!14.29!black] (8.2500,-5.422) circle (0.03cm); \fill[red!14.29!black] (8.3875,-5.294) circle (0.03cm); \fill[red!14.29!black] (8.5250,-5.152) circle (0.03cm); \fill[red!14.29!black] (8.6625,-4.989) circle (0.03cm); \fill[red!14.29!black] (8.8000,-4.789) circle (0.03cm);
\fill[red!0.00!black] (0.1375,-9.268) circle (0.03cm); \fill[red!0.00!black] (0.2750,-9.193) circle (0.03cm); \fill[red!0.00!black] (0.4125,-9.127) circle (0.03cm); \fill[red!0.00!black] (0.5500,-9.066) circle (0.03cm); \fill[red!0.00!black] (0.6875,-9.008) circle (0.03cm); \fill[red!0.00!black] (0.8250,-8.952) circle (0.03cm); \fill[red!0.00!black] (0.9625,-8.897) circle (0.03cm); \fill[red!0.00!black] (1.1000,-8.844) circle (0.03cm); \fill[red!0.00!black] (1.2375,-8.790) circle (0.03cm); \fill[red!0.00!black] (1.3750,-8.738) circle (0.03cm); \fill[red!0.00!black] (1.5125,-8.686) circle (0.03cm); \fill[red!0.00!black] (1.6500,-8.634) circle (0.03cm); \fill[red!0.00!black] (1.7875,-8.583) circle (0.03cm); \fill[red!0.00!black] (1.9250,-8.531) circle (0.03cm); \fill[red!0.00!black] (2.0625,-8.480) circle (0.03cm); \fill[red!0.00!black] (2.2000,-8.430) circle (0.03cm); \fill[red!0.00!black] (2.3375,-8.379) circle (0.03cm); \fill[red!0.00!black] (2.4750,-8.328) circle (0.03cm); \fill[red!0.00!black] (2.6125,-8.278) circle (0.03cm); \fill[red!0.00!black] (2.7500,-8.227) circle (0.03cm); \fill[red!0.00!black] (2.8875,-8.177) circle (0.03cm); \fill[red!0.00!black] (3.0250,-8.127) circle (0.03cm); \fill[red!0.00!black] (3.1625,-8.076) circle (0.03cm); \fill[red!0.00!black] (3.3000,-8.026) circle (0.03cm); \fill[red!0.00!black] (3.4375,-7.976) circle (0.03cm); \fill[red!0.00!black] (3.5750,-7.926) circle (0.03cm); \fill[red!0.00!black] (3.7125,-7.876) circle (0.03cm); \fill[red!0.00!black] (3.8500,-7.826) circle (0.03cm); \fill[red!0.00!black] (3.9875,-7.775) circle (0.03cm); \fill[red!0.00!black] (4.1250,-7.725) circle (0.03cm); \fill[red!0.00!black] (4.2625,-7.675) circle (0.03cm); \fill[red!0.00!black] (4.4000,-7.625) circle (0.03cm); \fill[red!0.00!black] (4.5375,-7.575) circle (0.03cm); \fill[red!0.00!black] (4.6750,-7.525) circle (0.03cm); \fill[red!0.00!black] (4.8125,-7.475) circle (0.03cm); \fill[red!0.00!black] (4.9500,-7.425) circle (0.03cm); \fill[red!0.00!black] (5.0875,-7.375) circle (0.03cm); \fill[red!0.00!black] (5.2250,-7.325) circle (0.03cm); \fill[red!0.00!black] (5.3625,-7.274) circle (0.03cm); \fill[red!0.00!black] (5.5000,-7.224) circle (0.03cm); \fill[red!0.00!black] (5.6375,-7.174) circle (0.03cm); \fill[red!0.00!black] (5.7750,-7.124) circle (0.03cm); \fill[red!0.00!black] (5.9125,-7.074) circle (0.03cm); \fill[red!0.00!black] (6.0500,-7.024) circle (0.03cm); \fill[red!0.00!black] (6.1875,-6.974) circle (0.03cm); \fill[red!0.00!black] (6.3250,-6.923) circle (0.03cm); \fill[red!0.00!black] (6.4625,-6.873) circle (0.03cm); \fill[red!0.00!black] (6.6000,-6.823) circle (0.03cm); \fill[red!0.00!black] (6.7375,-6.772) circle (0.03cm); \fill[red!0.00!black] (6.8750,-6.722) circle (0.03cm); \fill[red!0.00!black] (7.0125,-6.672) circle (0.03cm); \fill[red!0.00!black] (7.1500,-6.621) circle (0.03cm); \fill[red!0.00!black] (7.2875,-6.571) circle (0.03cm); \fill[red!0.00!black] (7.4250,-6.520) circle (0.03cm); \fill[red!0.00!black] (7.5625,-6.469) circle (0.03cm); \fill[red!0.00!black] (7.7000,-6.418) circle (0.03cm); \fill[red!0.00!black] (7.8375,-6.367) circle (0.03cm); \fill[red!0.00!black] (7.9750,-6.316) circle (0.03cm); \fill[red!0.00!black] (8.1125,-6.265) circle (0.03cm); \fill[red!0.00!black] (8.2500,-6.214) circle (0.03cm); \fill[red!0.00!black] (8.3875,-6.162) circle (0.03cm); \fill[red!0.00!black] (8.5250,-6.111) circle (0.03cm); \fill[red!0.00!black] (8.6625,-6.059) circle (0.03cm); \fill[red!0.00!black] (8.8000,-6.007) circle (0.03cm); \fill[red!0.00!black] (8.9375,-5.954) circle (0.03cm); \fill[red!0.00!black] (9.0750,-5.902) circle (0.03cm); \fill[red!0.00!black] (9.2125,-5.849) circle (0.03cm); \fill[red!0.00!black] (9.3500,-5.796) circle (0.03cm); \fill[red!0.00!black] (9.4875,-5.743) circle (0.03cm); \fill[red!0.00!black] (9.6250,-5.689) circle (0.03cm); \fill[red!0.00!black] (9.7625,-5.635) circle (0.03cm); \fill[red!0.00!black] (9.9000,-5.581) circle (0.03cm); \fill[red!0.00!black] (10.0375,-5.526) circle (0.03cm); \fill[red!0.00!black] (10.1750,-5.471) circle (0.03cm); \fill[red!0.00!black] (10.3125,-5.416) circle (0.03cm); \fill[red!0.00!black] (10.4500,-5.360) circle (0.03cm); \fill[red!0.00!black] (10.5875,-5.303) circle (0.03cm); \fill[red!0.00!black] (10.7250,-5.247) circle (0.03cm); \fill[red!0.00!black] (10.8625,-5.189) circle (0.03cm); \fill[red!0.00!black] (11.0000,-5.132) circle (0.03cm); \fill[red!0.00!black] (11.1375,-5.073) circle (0.03cm); \fill[red!0.00!black] (11.2750,-5.015) circle (0.03cm); \fill[red!0.00!black] (11.4125,-4.955) circle (0.03cm); \fill[red!0.00!black] (11.5500,-4.895) circle (0.03cm); \fill[red!0.00!black] (11.6875,-4.835) circle (0.03cm); \fill[red!0.00!black] (11.8250,-4.774) circle (0.03cm); \fill[red!0.00!black] (11.9625,-4.712) circle (0.03cm); \fill[red!0.00!black] (12.1000,-4.650) circle (0.03cm); \fill[red!0.00!black] (12.2375,-4.587) circle (0.03cm); \fill[red!0.00!black] (12.3750,-4.523) circle (0.03cm); \fill[red!0.00!black] (12.5125,-4.459) circle (0.03cm); \fill[red!0.00!black] (12.6500,-4.393) circle (0.03cm); \fill[red!0.00!black] (12.7875,-4.327) circle (0.03cm); \fill[red!0.00!black] (12.9250,-4.260) circle (0.03cm); \fill[red!0.00!black] (13.0625,-4.193) circle (0.03cm); \fill[red!0.00!black] (13.2000,-4.124) circle (0.03cm); \fill[red!0.00!black] (13.3375,-4.054) circle (0.03cm); \fill[red!0.00!black] (13.4750,-3.984) circle (0.03cm); \fill[red!0.00!black] (13.6125,-3.912) circle (0.03cm); \fill[red!0.00!black] (13.7500,-3.840) circle (0.03cm); \fill[red!0.00!black] (13.8875,-3.766) circle (0.03cm); \fill[red!0.00!black] (14.0250,-3.691) circle (0.03cm); \fill[red!0.00!black] (14.1625,-3.615) circle (0.03cm); \fill[red!0.00!black] (14.3000,-3.538) circle (0.03cm); \fill[red!0.00!black] (14.4375,-3.460) circle (0.03cm); \fill[red!0.00!black] (14.5750,-3.379) circle (0.03cm); \fill[red!0.00!black] (14.7125,-3.298) circle (0.03cm); \fill[red!0.00!black] (14.8500,-3.215) circle (0.03cm); \fill[red!0.00!black] (14.9875,-3.130) circle (0.03cm); \fill[red!0.00!black] (15.1250,-3.044) circle (0.03cm); \fill[red!0.00!black] (15.2625,-2.955) circle (0.03cm); \fill[red!0.00!black] (15.4000,-2.865) circle (0.03cm); \fill[red!0.00!black] (15.5375,-2.772) circle (0.03cm); \fill[red!0.00!black] (15.6750,-2.677) circle (0.03cm); \fill[red!0.00!black] (15.8125,-2.579) circle (0.03cm); \fill[red!0.00!black] (15.9500,-2.479) circle (0.03cm); \fill[red!0.00!black] (16.0875,-2.375) circle (0.03cm); \fill[red!0.00!black] (16.2250,-2.268) circle (0.03cm); \fill[red!0.00!black] (16.3625,-2.157) circle (0.03cm); \fill[red!0.00!black] (16.5000,-2.041) circle (0.03cm); \fill[red!0.00!black] (16.6375,-1.921) circle (0.03cm); \fill[red!0.00!black] (16.7750,-1.794) circle (0.03cm); \fill[red!0.00!black] (16.9125,-1.660) circle (0.03cm); \fill[red!0.00!black] (17.0500,-1.518) circle (0.03cm); \fill[red!0.00!black] (17.1875,-1.364) circle (0.03cm); \fill[red!0.00!black] (17.3250,-1.195) circle (0.03cm); \fill[red!0.00!black] (17.4625,-1.001) circle (0.03cm); \fill[red!0.00!black] (17.6000,-0.764) circle (0.03cm);
\foreach \x in {1,2,...,128}
\fill[green!75!black] (17.6*\x/128,-10+10/200*\x+10/200*15.5) circle (0.03cm);
\end{tikzpicture}
\end{center}
\caption{The spectrum for $c=128$ computed using truncation order $N=1$, $2$, $4$, \dots, $128$ (red
through black)
and the free fermion spectrum $\Delta_n(c) = n + (c-4)/8$ (green).}
\label{figure:spectracomparison}
\end{figure}
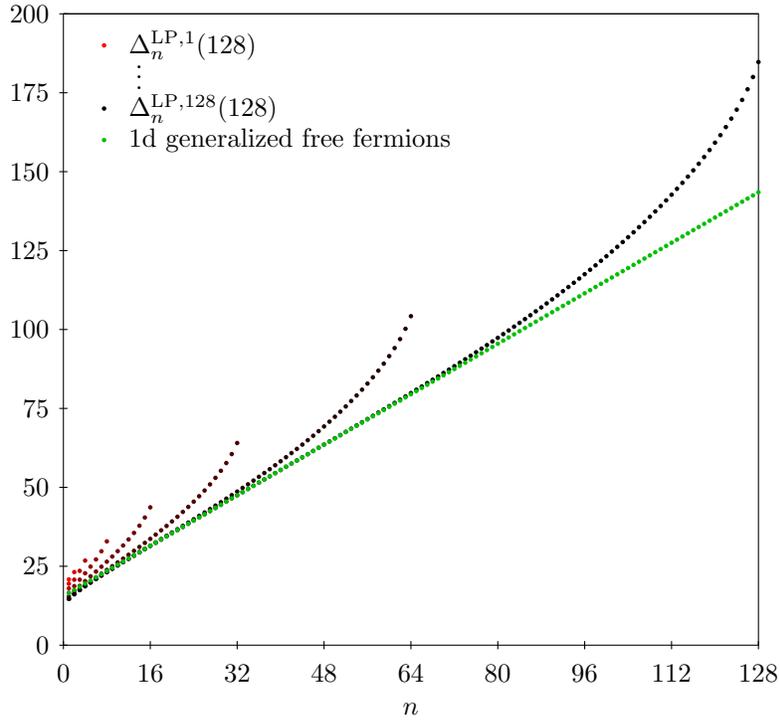

Figure~\ref{figure:spectracomparison} shows how the spectrum converges as we
increase the truncation order $N$. The limiting values for
$\Delta_n^{\textup{LP},N}(c)$ as $N \to \infty$ are quite close to the 1d
generalized free fermion values, but there is a substantial divergence when
$n$ is large enough relative to $N$, as noted in \cite[Section~3]{CM} and
\cite[Figure~2]{AHT}. Although we have no proof, we can predict the form of
this divergence. It occurs starting at $n \sim (2/\pi)N$, with shape
determined by the following conjecture:

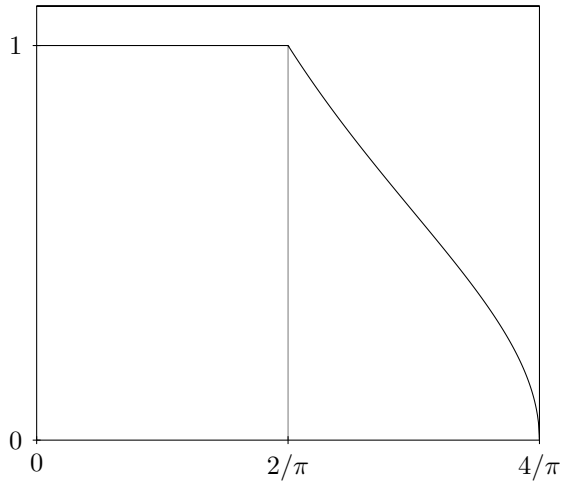
\begin{figure}
\begin{center}
\begin{tikzpicture}[x=0.525cm,y=0.525cm]
\draw[black!50] (6.36620,0)--(6.36620,10);
\draw (0,11)--(12.7324,11)--(12.7324,0)--(0,0)--(0,11)--(12.7324,11);
\draw (-0.1,10)--(0.1,10); \draw (-0.1,10) node[left] {\small $1$};
\draw (-0.1,0)--(0.1,0); \draw (-0.1,0) node[left] {\small $0$};
\draw (12.7324,0.1)--(12.7324,-0.1); \draw (12.7324,-0.1) node[below] {\small $4/\pi$};
\draw (6.3662,0.1)--(6.3662,-0.1); \draw (6.3662,-0.1) node[below] {\small $2/\pi$};
\draw (0,-0.1)--(0,0.1); \draw(0,-0.1) node[below] {\small $0$};
\draw (0,10)--(6.36620,10)--(6.39107,9.96101);
\draw[join=round] (6.36620,10)--(6.39107,9.96101)--(6.41593,9.92218)--(6.44080,9.88349)--(6.46567,9.84495)--(6.49054,9.80656)--(6.51541,9.76831)--(6.54027,9.73020)--(6.56514,9.69223)--(6.59001,9.65441)--(6.61488,9.61671)--(6.63975,9.57916)--(6.66461,9.54174)--(6.68948,9.50445)--(6.71435,9.46729)--(6.73922,9.43026)--(6.76409,9.39336)--(6.78895,9.35659)--(6.81382,9.31994)--(6.83869,9.28342)--(6.86356,9.24701)--(6.88842,9.21073)--(6.91329,9.17457)--(6.93816,9.13852)--(6.96303,9.10259)--(6.98790,9.06677)--(7.01276,9.03107)--(7.03763,8.99548)--(7.06250,8.96001)--(7.08737,8.92464)--(7.11224,8.88937)--(7.13710,8.85422)--(7.16197,8.81917)--(7.18684,8.78423)--(7.21171,8.74938)--(7.23658,8.71464)--(7.26144,8.68000)--(7.28631,8.64546)--(7.31118,8.61102)--(7.33605,8.57667)--(7.36092,8.54242)--(7.38578,8.50826)--(7.41065,8.47420)--(7.43552,8.44023)--(7.46039,8.40635)--(7.48526,8.37255)--(7.51012,8.33885)--(7.53499,8.30523)--(7.55986,8.27170)--(7.58473,8.23826)--(7.60960,8.20489)--(7.63446,8.17161)--(7.65933,8.13841)--(7.68420,8.10529)--(7.70907,8.07225)--(7.73394,8.03929)--(7.75880,8.00641)--(7.78367,7.97360)--(7.80854,7.94086)--(7.83341,7.90820)--(7.85828,7.87562)--(7.88314,7.84310)--(7.90801,7.81065)--(7.93288,7.77828)--(7.95775,7.74597)--(7.98262,7.71373)--(8.00748,7.68155)--(8.03235,7.64944)--(8.05722,7.61739)--(8.08209,7.58541)--(8.10695,7.55349)--(8.13182,7.52163)--(8.15669,7.48983)--(8.18156,7.45809)--(8.20643,7.42641)--(8.23129,7.39478)--(8.25616,7.36321)--(8.28103,7.33170)--(8.30590,7.30023)--(8.33077,7.26883)--(8.35563,7.23747)--(8.38050,7.20616)--(8.40537,7.17491)--(8.43024,7.14370)--(8.45511,7.11254)--(8.47997,7.08143)--(8.50484,7.05036)--(8.52971,7.01934)--(8.55458,6.98836)--(8.57945,6.95743)--(8.60431,6.92653)--(8.62918,6.89568)--(8.65405,6.86487)--(8.67892,6.83410)--(8.70379,6.80336)--(8.72865,6.77266)--(8.75352,6.74200)--(8.77839,6.71137)--(8.80326,6.68078)--(8.82813,6.65021)--(8.85299,6.61968)--(8.87786,6.58919)--(8.90273,6.55872)--(8.92760,6.52828)--(8.95247,6.49786)--(8.97733,6.46748)--(9.00220,6.43712)--(9.02707,6.40678)--(9.05194,6.37647)--(9.07681,6.34618)--(9.10167,6.31591)--(9.12654,6.28566)--(9.15141,6.25543)--(9.17628,6.22522)--(9.20115,6.19503)--(9.22601,6.16485)--(9.25088,6.13469)--(9.27575,6.10454)--(9.30062,6.07440)--(9.32548,6.04428)--(9.35035,6.01417)--(9.37522,5.98406)--(9.40009,5.95397)--(9.42496,5.92388)--(9.44982,5.89380)--(9.47469,5.86372)--(9.49956,5.83364)--(9.52443,5.80357)--(9.54930,5.77350)--(9.57416,5.74343)--(9.59903,5.71336)--(9.62390,5.68329)--(9.64877,5.65321)--(9.67364,5.62313)--(9.69850,5.59304)--(9.72337,5.56294)--(9.74824,5.53283)--(9.77311,5.50272)--(9.79798,5.47259)--(9.82284,5.44245)--(9.84771,5.41229)--(9.87258,5.38212)--(9.89745,5.35194)--(9.92232,5.32173)--(9.94718,5.29150)--(9.97205,5.26125)--(9.99692,5.23098)--(10.0218,5.20069)--(10.0467,5.17036)--(10.0715,5.14001)--(10.0964,5.10964)--(10.1213,5.07922)--(10.1461,5.04878)--(10.1710,5.01830)--(10.1959,4.98779)--(10.2207,4.95724)--(10.2456,4.92665)--(10.2705,4.89601)--(10.2953,4.86534)--(10.3202,4.83461)--(10.3451,4.80384)--(10.3699,4.77303)--(10.3948,4.74216)--(10.4197,4.71123)--(10.4445,4.68025)--(10.4694,4.64922)--(10.4943,4.61812)--(10.5191,4.58696)--(10.5440,4.55573)--(10.5689,4.52444)--(10.5938,4.49308)--(10.6186,4.46165)--(10.6435,4.43014)--(10.6684,4.39856)--(10.6932,4.36689)--(10.7181,4.33515)--(10.7430,4.30331)--(10.7678,4.27139)--(10.7927,4.23938)--(10.8176,4.20727)--(10.8424,4.17507)--(10.8673,4.14276)--(10.8922,4.11035)--(10.9170,4.07783)--(10.9419,4.04520)--(10.9668,4.01245)--(10.9916,3.97959)--(11.0165,3.94660)--(11.0414,3.91348)--(11.0662,3.88023)--(11.0911,3.84684)--(11.1160,3.81332)--(11.1408,3.77964)--(11.1657,3.74582)--(11.1906,3.71184)--(11.2154,3.67770)--(11.2403,3.64340)--(11.2652,3.60892)--(11.2901,3.57426)--(11.3149,3.53942)--(11.3398,3.50438)--(11.3647,3.46915)--(11.3895,3.43371)--(11.4144,3.39806)--(11.4393,3.36219)--(11.4641,3.32609)--(11.4890,3.28976)--(11.5139,3.25318)--(11.5387,3.21634)--(11.5636,3.17923)--(11.5885,3.14185)--(11.6133,3.10419)--(11.6382,3.06622)--(11.6631,3.02794)--(11.6879,2.98934)--(11.7128,2.95040)--(11.7377,2.91111)--(11.7625,2.87145)--(11.7874,2.83141)--(11.8123,2.79096)--(11.8371,2.75010)--(11.8620,2.70879)--(11.8869,2.66702)--(11.9118,2.62476)--(11.9366,2.58199)--(11.9615,2.53868)--(11.9864,2.49481)--(12.0112,2.45033)--(12.0361,2.40523)--(12.0610,2.35945)--(12.0858,2.31296)--(12.1107,2.26572)--(12.1356,2.21766)--(12.1604,2.16875)--(12.1853,2.11891)--(12.2102,2.06809)--(12.2350,2.01619)--(12.2599,1.96315)--(12.2848,1.90885)--(12.3096,1.85320)--(12.3345,1.79605)--(12.3594,1.73727)--(12.3842,1.67668)--(12.4091,1.61407)--(12.4340,1.54919)--(12.4588,1.48176)--(12.4837,1.41139)--(12.5086,1.33763)--(12.5335,1.25988)--(12.5583,1.17734)--(12.5832,1.08893)--(12.6081,0.993073)--(12.6329,0.887357)--(12.6578,0.767718)--(12.6827,0.626224)--(12.7075,0.442374)--(12.7324,0);
\end{tikzpicture}
\end{center}
\caption{The probability density function for the normalized spectrum is $1$ on $[0,2/\pi]$
and $x \mapsto x^{-1/2} (4/\pi-x)^{1/2}$ on $[2/\pi,4/\pi]$.}
\label{figure:pdf}
\end{figure}

\begin{conjecture} \label{conjecture:beta}
For each $c \ge 2$, the distribution of the normalized spectrum
$\Delta_n^{\textup{LP},N}(c)/N$ with $1 \le n \le N$ converges as $N \to
\infty$ to the probability distribution on the interval $[0,4/\pi]$ with
density function $1$ on $[0,2/\pi]$ and $x \mapsto x^{-1/2} (4/\pi-x)^{1/2}$
on $[2/\pi,4/\pi]$, shown in Figure~\ref{figure:pdf}.
\end{conjecture}

\begin{figure}
\begin{center}
\begin{tikzpicture}[x=0.525cm,y=0.525cm]
\draw[red] (0.0000,-10.0000)--(0.0781,-10.0000)--(0.0781,-9.9219)--(0.1563,-9.9219)--(0.1563,-9.8438)--(0.2344,-9.8438)--(0.2344,-9.7656)--(0.3125,-9.7656)--(0.3125,-9.6875)--(0.3906,-9.6875)--(0.3906,-9.6094)--(0.4688,-9.6094)--(0.4688,-9.5312)--(0.5469,-9.5312)--(0.5469,-9.4531)--(0.6250,-9.4531)--(0.6250,-9.3750)--(0.7031,-9.3750)--(0.7031,-9.2969)--(0.7813,-9.2969)--(0.7813,-9.2188)--(0.8594,-9.2188)--(0.8594,-9.1406)--(0.9375,-9.1406)--(0.9375,-9.0625)--(1.0156,-9.0625)--(1.0156,-8.9844)--(1.0938,-8.9844)--(1.0938,-8.9062)--(1.1719,-8.9062)--(1.1719,-8.8281)--(1.2500,-8.8281)--(1.2500,-8.7500)--(1.3281,-8.7500)--(1.3281,-8.6719)--(1.4063,-8.6719)--(1.4063,-8.5938)--(1.4844,-8.5938)--(1.4844,-8.5156)--(1.5625,-8.5156)--(1.5625,-8.4375)--(1.6406,-8.4375)--(1.6406,-8.3594)--(1.7188,-8.3594)--(1.7188,-8.2812)--(1.7969,-8.2812)--(1.7969,-8.2031)--(1.8750,-8.2031)--(1.8750,-8.1250)--(1.9531,-8.1250)--(1.9531,-8.0469)--(2.0313,-8.0469)--(2.0313,-7.9688)--(2.1094,-7.9688)--(2.1094,-7.8906)--(2.1875,-7.8906)--(2.1875,-7.8125)--(2.2656,-7.8125)--(2.2656,-7.7344)--(2.3438,-7.7344)--(2.3438,-7.6562)--(2.4219,-7.6562)--(2.4219,-7.5781)--(2.5000,-7.5781)--(2.5000,-7.5000)--(2.5781,-7.5000)--(2.5781,-7.4219)--(2.6563,-7.4219)--(2.6563,-7.3438)--(2.7344,-7.3438)--(2.7344,-7.2656)--(2.8125,-7.2656)--(2.8125,-7.1875)--(2.8906,-7.1875)--(2.8906,-7.1094)--(2.9688,-7.1094)--(2.9688,-7.0312)--(3.0469,-7.0312)--(3.0469,-6.9531)--(3.1250,-6.9531)--(3.1250,-6.8750)--(3.2031,-6.8750)--(3.2031,-6.7969)--(3.2813,-6.7969)--(3.2813,-6.7188)--(3.3594,-6.7188)--(3.3594,-6.6406)--(3.4375,-6.6406)--(3.4375,-6.5625)--(3.5156,-6.5625)--(3.5156,-6.4844)--(3.5938,-6.4844)--(3.5938,-6.4062)--(3.6719,-6.4062)--(3.6719,-6.3281)--(3.7500,-6.3281)--(3.7500,-6.2500)--(3.8281,-6.2500)--(3.8281,-6.1719)--(3.9063,-6.1719)--(3.9063,-6.0938)--(3.9844,-6.0938)--(3.9844,-6.0156)--(4.0625,-6.0156)--(4.0625,-5.9375)--(4.1406,-5.9375)--(4.1406,-5.8594)--(4.2188,-5.8594)--(4.2188,-5.7812)--(4.2969,-5.7812)--(4.2969,-5.7031)--(4.3750,-5.7031)--(4.3750,-5.6250)--(4.4531,-5.6250)--(4.4531,-5.5469)--(4.5313,-5.5469)--(4.5313,-5.4688)--(4.6094,-5.4688)--(4.6094,-5.3906)--(4.6875,-5.3906)--(4.6875,-5.3125)--(4.7656,-5.3125)--(4.7656,-5.2344)--(4.8438,-5.2344)--(4.8438,-5.1562)--(4.9219,-5.1562)--(4.9219,-5.0781)--(5.0000,-5.0781)--(5.0000,-5.0000)--(5.0781,-5.0000)--(5.0781,-4.9219)--(5.1563,-4.9219)--(5.1563,-4.8438)--(5.2344,-4.8438)--(5.2344,-4.7656)--(5.3125,-4.7656)--(5.3125,-4.6875)--(5.3906,-4.6875)--(5.3906,-4.6094)--(5.4688,-4.6094)--(5.4688,-4.5312)--(5.5469,-4.5312)--(5.5469,-4.4531)--(5.6251,-4.4531)--(5.6251,-4.3750)--(5.7032,-4.3750)--(5.7032,-4.2969)--(5.7814,-4.2969)--(5.7814,-4.2188)--(5.8597,-4.2188)--(5.8597,-4.1406)--(5.9379,-4.1406)--(5.9379,-4.0625)--(6.0164,-4.0625)--(6.0164,-3.9844)--(6.0948,-3.9844)--(6.0948,-3.9062)--(6.1737,-3.9062)--(6.1737,-3.8281)--(6.2523,-3.8281)--(6.2523,-3.7500)--(6.3320,-3.7500)--(6.3320,-3.6719)--(6.4113,-3.6719)--(6.4113,-3.5938)--(6.4925,-3.5938)--(6.4925,-3.5156)--(6.5719,-3.5156)--(6.5719,-3.4375)--(6.6554,-3.4375)--(6.6554,-3.3594)--(6.7363,-3.3594)--(6.7363,-3.2812)--(6.8214,-3.2812)--(6.8214,-3.2031)--(6.9032,-3.2031)--(6.9032,-3.1250)--(6.9926,-3.1250)--(6.9926,-3.0469)--(7.0747,-3.0469)--(7.0747,-2.9688)--(7.1659,-2.9688)--(7.1659,-2.8906)--(7.2528,-2.8906)--(7.2528,-2.8125)--(7.3435,-2.8125)--(7.3435,-2.7344)--(7.4360,-2.7344)--(7.4360,-2.6562)--(7.5252,-2.6562)--(7.5252,-2.5781)--(7.6235,-2.5781)--(7.6235,-2.5000)--(7.7166,-2.5000)--(7.7166,-2.4219)--(7.8142,-2.4219)--(7.8142,-2.3438)--(7.9149,-2.3438)--(7.9149,-2.2656)--(8.0107,-2.2656)--(8.0107,-2.1875)--(8.1158,-2.1875)--(8.1158,-2.1094)--(8.2193,-2.1094)--(8.2193,-2.0312)--(8.3222,-2.0312)--(8.3222,-1.9531)--(8.4324,-1.9531)--(8.4324,-1.8750)--(8.5432,-1.8750)--(8.5432,-1.7969)--(8.6517,-1.7969)--(8.6517,-1.7188)--(8.7651,-1.7188)--(8.7651,-1.6406)--(8.8851,-1.6406)--(8.8851,-1.5625)--(9.0031,-1.5625)--(9.0031,-1.4844)--(9.1217,-1.4844)--(9.1217,-1.4062)--(9.2451,-1.4062)--(9.2451,-1.3281)--(9.3747,-1.3281)--(9.3747,-1.2500)--(9.5076,-1.2500)--(9.5076,-1.1719)--(9.6418,-1.1719)--(9.6418,-1.0938)--(9.7794,-1.0938)--(9.7794,-1.0156)--(9.9193,-1.0156)--(9.9193,-0.9375)--(10.0655,-0.9375)--(10.0655,-0.8594)--(10.2174,-0.8594)--(10.2174,-0.7812)--(10.3737,-0.7812)--(10.3737,-0.7031)--(10.5393,-0.7031)--(10.5393,-0.6250)--(10.7112,-0.6250)--(10.7112,-0.5469)--(10.8914,-0.5469)--(10.8914,-0.4688)--(11.0815,-0.4688)--(11.0815,-0.3906)--(11.2853,-0.3906)--(11.2853,-0.3125)--(11.5065,-0.3125)--(11.5065,-0.2344)--(11.7496,-0.2344)--(11.7496,-0.1562)--(12.0271,-0.1562)--(12.0271,-0.0781)--(12.3694,-0.0781)--(12.3694,0.0000);
\draw[black!50] (6.3662,-10)--(6.3662,-3.634)--(0,-3.634);
\draw (0,-10)--(12.7324,-10)--(12.7324,0)--(0,0)--(0,-10)--(12.7324,-10);
\draw (-0.1,-10)--(0.1,-10); \draw (-0.1,-10) node[left] {\small $0$};
\draw (-0.1,0)--(0.1,0); \draw (-0.1,0) node[left] {\small $1$};
\draw (-0.1,-3.634)--(0.1,-3.634); \draw (-0.1,-3.634) node[left] {\small $2/\pi$};
\draw (12.7324,-9.9)--(12.7324,-10.1); \draw (12.7324,-10.1) node[below] {\small $4/\pi$};
\draw (6.3662,-9.9)--(6.3662,-10.1); \draw (6.3662,-10.1) node[below] {\small $2/\pi$};
\draw (0,-9.9)--(0,-10.1); \draw(0,-10.1) node[below] {\small $0$};
\draw[join=round] (0,-10)--(6.36620,-3.63380)--(6.39107,-3.60898)--(6.41593,-3.58426)--(6.44080,-3.55963)--(6.46567,-3.53510)--(6.49054,-3.51067)--(6.51541,-3.48633)--(6.54027,-3.46209)--(6.56514,-3.43794)--(6.59001,-3.41388)--(6.61488,-3.38992)--(6.63975,-3.36605)--(6.66461,-3.34228)--(6.68948,-3.31859)--(6.71435,-3.29500)--(6.73922,-3.27151)--(6.76409,-3.24810)--(6.78895,-3.22479)--(6.81382,-3.20157)--(6.83869,-3.17843)--(6.86356,-3.15539)--(6.88842,-3.13244)--(6.91329,-3.10958)--(6.93816,-3.08681)--(6.96303,-3.06413)--(6.98790,-3.04154)--(7.01276,-3.01904)--(7.03763,-2.99662)--(7.06250,-2.97430)--(7.08737,-2.95206)--(7.11224,-2.92991)--(7.13710,-2.90785)--(7.16197,-2.88587)--(7.18684,-2.86398)--(7.21171,-2.84218)--(7.23658,-2.82047)--(7.26144,-2.79884)--(7.28631,-2.77730)--(7.31118,-2.75584)--(7.33605,-2.73447)--(7.36092,-2.71318)--(7.38578,-2.69198)--(7.41065,-2.67087)--(7.43552,-2.64984)--(7.46039,-2.62889)--(7.48526,-2.60803)--(7.51012,-2.58725)--(7.53499,-2.56655)--(7.55986,-2.54594)--(7.58473,-2.52541)--(7.60960,-2.50497)--(7.63446,-2.48460)--(7.65933,-2.46432)--(7.68420,-2.44413)--(7.70907,-2.42401)--(7.73394,-2.40398)--(7.75880,-2.38403)--(7.78367,-2.36416)--(7.80854,-2.34437)--(7.83341,-2.32466)--(7.85828,-2.30504)--(7.88314,-2.28549)--(7.90801,-2.26603)--(7.93288,-2.24665)--(7.95775,-2.22734)--(7.98262,-2.20812)--(8.00748,-2.18898)--(8.03235,-2.16992)--(8.05722,-2.15093)--(8.08209,-2.13203)--(8.10695,-2.11321)--(8.13182,-2.09446)--(8.15669,-2.07580)--(8.18156,-2.05721)--(8.20643,-2.03870)--(8.23129,-2.02027)--(8.25616,-2.00192)--(8.28103,-1.98365)--(8.30590,-1.96546)--(8.33077,-1.94734)--(8.35563,-1.92931)--(8.38050,-1.91135)--(8.40537,-1.89347)--(8.43024,-1.87566)--(8.45511,-1.85794)--(8.47997,-1.84029)--(8.50484,-1.82272)--(8.52971,-1.80522)--(8.55458,-1.78781)--(8.57945,-1.77047)--(8.60431,-1.75320)--(8.62918,-1.73602)--(8.65405,-1.71891)--(8.67892,-1.70187)--(8.70379,-1.68492)--(8.72865,-1.66803)--(8.75352,-1.65123)--(8.77839,-1.63450)--(8.80326,-1.61785)--(8.82813,-1.60128)--(8.85299,-1.58478)--(8.87786,-1.56835)--(8.90273,-1.55200)--(8.92760,-1.53573)--(8.95247,-1.51953)--(8.97733,-1.50341)--(9.00220,-1.48737)--(9.02707,-1.47140)--(9.05194,-1.45550)--(9.07681,-1.43968)--(9.10167,-1.42394)--(9.12654,-1.40827)--(9.15141,-1.39268)--(9.17628,-1.37716)--(9.20115,-1.36172)--(9.22601,-1.34635)--(9.25088,-1.33105)--(9.27575,-1.31584)--(9.30062,-1.30069)--(9.32548,-1.28562)--(9.35035,-1.27063)--(9.37522,-1.25571)--(9.40009,-1.24087)--(9.42496,-1.22610)--(9.44982,-1.21141)--(9.47469,-1.19679)--(9.49956,-1.18224)--(9.52443,-1.16777)--(9.54930,-1.15338)--(9.57416,-1.13906)--(9.59903,-1.12481)--(9.62390,-1.11064)--(9.64877,-1.09655)--(9.67364,-1.08252)--(9.69850,-1.06858)--(9.72337,-1.05471)--(9.74824,-1.04091)--(9.77311,-1.02719)--(9.79798,-1.01354)--(9.82284,-0.999971)--(9.84771,-0.986474)--(9.87258,-0.973052)--(9.89745,-0.959706)--(9.92232,-0.946434)--(9.94718,-0.933238)--(9.97205,-0.920116)--(9.99692,-0.907070)--(10.0218,-0.894100)--(10.0467,-0.881204)--(10.0715,-0.868384)--(10.0964,-0.855640)--(10.1213,-0.842971)--(10.1461,-0.830378)--(10.1710,-0.817861)--(10.1959,-0.805419)--(10.2207,-0.793053)--(10.2456,-0.780764)--(10.2705,-0.768550)--(10.2953,-0.756413)--(10.3202,-0.744352)--(10.3451,-0.732368)--(10.3699,-0.720460)--(10.3948,-0.708629)--(10.4197,-0.696874)--(10.4445,-0.685197)--(10.4694,-0.673597)--(10.4943,-0.662074)--(10.5191,-0.650628)--(10.5440,-0.639260)--(10.5689,-0.627970)--(10.5938,-0.616757)--(10.6186,-0.605623)--(10.6435,-0.594567)--(10.6684,-0.583589)--(10.6932,-0.572690)--(10.7181,-0.561870)--(10.7430,-0.551129)--(10.7678,-0.540467)--(10.7927,-0.529885)--(10.8176,-0.519383)--(10.8424,-0.508960)--(10.8673,-0.498618)--(10.8922,-0.488356)--(10.9170,-0.478174)--(10.9419,-0.468074)--(10.9668,-0.458055)--(10.9916,-0.448118)--(11.0165,-0.438263)--(11.0414,-0.428489)--(11.0662,-0.418799)--(11.0911,-0.409191)--(11.1160,-0.399666)--(11.1408,-0.390225)--(11.1657,-0.380868)--(11.1906,-0.371595)--(11.2154,-0.362407)--(11.2403,-0.353304)--(11.2652,-0.344286)--(11.2901,-0.335355)--(11.3149,-0.326510)--(11.3398,-0.317751)--(11.3647,-0.309080)--(11.3895,-0.300497)--(11.4144,-0.292003)--(11.4393,-0.283597)--(11.4641,-0.275281)--(11.4890,-0.267054)--(11.5139,-0.258919)--(11.5387,-0.250875)--(11.5636,-0.242922)--(11.5885,-0.235063)--(11.6133,-0.227296)--(11.6382,-0.219624)--(11.6631,-0.212046)--(11.6879,-0.204564)--(11.7128,-0.197179)--(11.7377,-0.189891)--(11.7625,-0.182701)--(11.7874,-0.175610)--(11.8123,-0.168619)--(11.8371,-0.161729)--(11.8620,-0.154941)--(11.8869,-0.148257)--(11.9118,-0.141677)--(11.9366,-0.135203)--(11.9615,-0.128836)--(11.9864,-0.122577)--(12.0112,-0.116428)--(12.0361,-0.110390)--(12.0610,-0.104466)--(12.0858,-0.0986560)--(12.1107,-0.0929627)--(12.1356,-0.0873879)--(12.1604,-0.0819337)--(12.1853,-0.0766022)--(12.2102,-0.0713959)--(12.2350,-0.0663173)--(12.2599,-0.0613691)--(12.2848,-0.0565544)--(12.3096,-0.0518764)--(12.3345,-0.0473386)--(12.3594,-0.0429449)--(12.3842,-0.0386996)--(12.4091,-0.0346075)--(12.4340,-0.0306738)--(12.4588,-0.0269045)--(12.4837,-0.0233066)--(12.5086,-0.0198877)--(12.5335,-0.0166570)--(12.5583,-0.0136255)--(12.5832,-0.0108062)--(12.6081,-0.00821573)--(12.6329,-0.00587523)--(12.6578,-0.00381383)--(12.6827,-0.00207476)--(12.7075,-0.000733109)--(12.7324,0);
\end{tikzpicture}
\end{center}
\caption{The cumulative distribution functions for the normalized spectrum $\Delta_n^{\textup{LP},N}(4)/N$ with $1 \le n \le N$
when $N=128$ (red) and in the limit as $N \to \infty$ (black).}
\label{figure:cdf}
\end{figure}
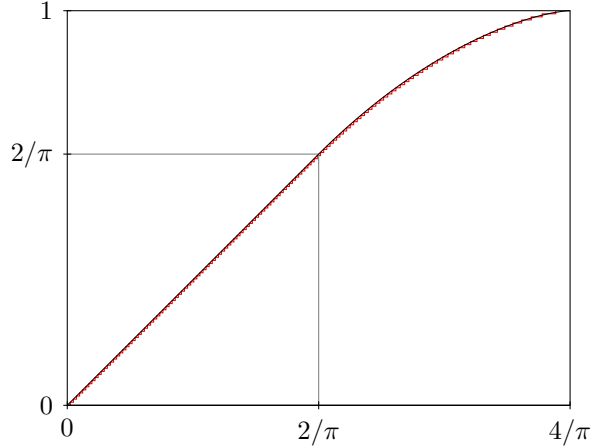

See Figure~\ref{figure:cdf} for a comparison with numerical data for $c=4$
and $N=128$. The behavior is similar for all $c \ge 2$, with slower
convergence as $c$ grows.

Conjecture~\ref{conjecture:beta} says that the distribution shifts from
uniform to a beta distribution around $n \sim (2/\pi)N$. The uniform
distribution corresponds to the 1d generalized free fermion spectrum, and the
beta distribution describes the root distribution of high-degree Laguerre
polynomials (see \cite[Theorem~1]{Gaw}). Specifically, if we normalize the
roots of the highest-degree polynomial $L^{(c-1)}_{4N-1}(4\pi \Delta)$ from
the truncated crossing equation \eqref{truncprim} by dividing by a factor of
$N$, then their distribution converges as $N \to \infty$ to the beta function
on $[0,4/\pi]$ with density $x \mapsto x^{-1/2} (4/\pi-x)^{1/2}/2$. From this
perspective, the transition in Conjecture~\ref{conjecture:beta} is between
the uniform limiting behavior as $N \to \infty$ and a generic root
distribution corresponding to high-degree Fourier eigenfunctions.

Conjecture~\ref{conjecture:beta} gives the following description of the
curves in Figure~\ref{figure:spectracomparison} via Theorem~4 in \cite{Gat}.
We wish to approximate $\Delta_n^{\textup{LP},N}(c)/N$ as $N \to \infty$ with
$n/N \to \alpha$ for some constant $\alpha \in [0,1]$. The conjecture says
that
\begin{equation}
\label{highspec}
\frac{\Delta_n^{\textup{LP},N}(c)}{N} \to \begin{cases} \alpha & \text{if $\alpha \le 2/\pi$, and}\\
(2/\pi)(1+\cos \beta) & \text{if $\alpha \ge 2/\pi$,}
\end{cases}
\end{equation}
where $\beta$ is the solution of $\beta-\sin \beta = (1-\alpha)\pi/2$ with $0
\le \beta \le \pi/2$. We will give some motivation for the high-energy
portion of this formula in Section~\ref{ss:heuristic}, after discussing
degeneracies.

\subsection{Deviations from the free fermion spectrum}

\begin{figure}
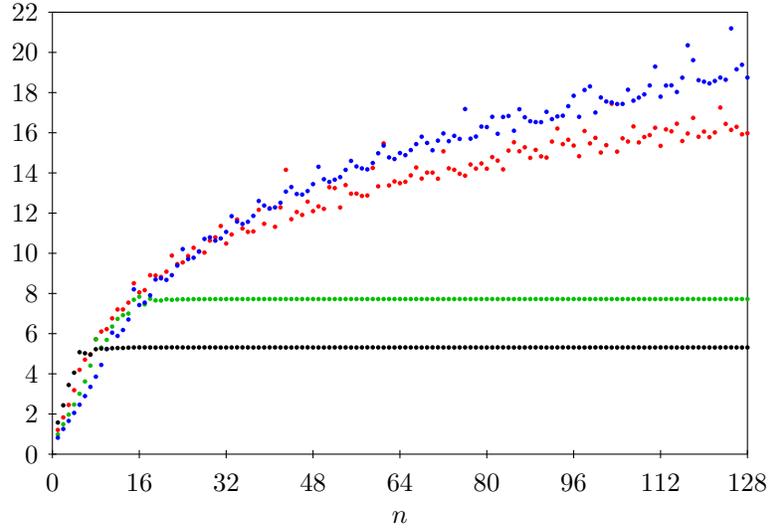

\begin{center}

\end{center}
\caption{The gaps $\log_{10}(1/|\Delta_n(c)-(n+(c-4)/8)|)$ for $c=16$
(black), $c=20$ (red), $c=24$ (green), and $c=28$ (blue).}
\label{figure:gapsDelta}
\end{figure}

Aside from $c=4$ or $12$, the spectrum $\Delta_n^\textup{LP}(c)$ is not
exactly equal to the 1d generalized free fermion spectrum $\Delta_n(c) = n +
(c-4)/8$. Instead, we always find some error in this approximation, and one
natural question is whether the error tends to $0$ as $n \to \infty$.
Figure~\ref{figure:gapsDelta} shows four test cases, namely $c \in \{16, 20,
24, 28\}$. In each of these cases, the error
$|\Delta^\textup{LP}_n(c)-(n+(c-4)/8)|$ becomes small, no more than $10^{-5}$
when $n$ is large. However, there is a striking difference between two
different scenarios: when $c=16$ or $24$, the error stabilizes above zero,
while it seems to converge to $0$ when $c=20$ or $28$. We have no conceptual
explanation of this behavior, which seems to be periodic modulo $8$ as $c$
varies, with multiples of $8$ being the worst case and $4$ modulo $8$ being
the best case (and the only case with convergence to zero). Note that $c=4$
and $12$ fall into the latter category.

\begin{figure}
\begin{center}
\begin{tikzpicture}[x=0.525cm,y=0.84cm]
\draw[red] (0,-10+10*5.651496129472318798043279/55)--(17.6,-10+10*37.651496129472318798043279/55);
\draw (0,-10)--(17.6,-10);
\draw (7.4,-0.5) node[right] {\small $\log_2(1/\varepsilon(c))$};
\draw (7.4,-1) node[right] {\small $c+\log_2(\pi)$};
\draw (7.03-0.28,-0.5)--(7.03+0.28,-0.5);
\draw[red] (7.03-0.28,-1)--(7.03+0.28,-1);
\foreach \x in {0,...,8}
\draw (17.6*\x/8,-10-0.065)--(17.6*\x/8,-10+0.065);
\foreach \x in {4, 8,...,36}
\draw ({17.6*\x/32-17.6*4/32},-10-0.15)  node[below] {\small \x};
\draw (0,0)--(17.6,0)--(17.6,-10);
\foreach \y in {-11,...,0} \draw (-0.104,{\y*10/11})--(0.104,{\y*10/11});
\draw (8.8,-10-0.65)
node[below] {\small central charge $c$}; \draw (-1.8,-5) node[rotate=90]
{\small $\phantom{\log_{10}(\text{kissing number})}$};
\draw (19.1,-5) node[rotate=-90]
{\small $\phantom{\log_{10}(\text{kissing number})}$};
\foreach \x in {0,5,...,55} \draw (-0.1,-10+\x*10/55) node[left] {\small \x};
\draw(0,-10)--(0,0); \draw
(0,0)--(17.6,0);
\draw[join=round] (0.00000,0)--(0.00275,-7.35468)--(0.00550,-7.53578)--(0.00825,-7.64186)--(0.01100,-7.71705)--(0.01375,-7.77531)--(0.01650,-7.82286)--(0.01925,-7.86301)--(0.02200,-7.89775)--(0.02475,-7.92835)--(0.02750,-7.95567)--(0.05500,-8.13409)--(0.08250,-8.23634)--(0.11000,-8.30701)--(0.13750,-8.36006)--(0.16500,-8.40176)--(0.19250,-8.43548)--(0.22000,-8.46325)--(0.24750,-8.48640)--(0.27500,-8.50587)--(0.30250,-8.52233)--(0.33000,-8.53629)--(0.35750,-8.54813)--(0.38500,-8.55817)--(0.41250,-8.56695)--(0.44000,-8.57443)--(0.46750,-8.58061)--(0.49500,-8.58562)--(0.52250,-8.58957)--(0.55000,-8.59258)--(0.57750,-8.59473)--(0.60500,-8.59689)--(0.63250,-8.59911)--(0.66000,-8.60073)--(0.68750,-8.60193)--(0.71500,-8.60262)--(0.74250,-8.60338)--(0.77000,-8.60392)--(0.79750,-8.60394)--(0.82500,-8.60348)--(0.85250,-8.60257)--(0.88000,-8.60123)--(0.90750,-8.59950)--(0.93500,-8.59752)--(0.96250,-8.59520)--(0.99000,-8.59255)--(1.01750,-8.58958)--(1.04500,-8.58631)--(1.07250,-8.58275)--(1.10000,-8.57900)--(1.12750,-8.57519)--(1.15500,-8.57110)--(1.18250,-8.56674)--(1.21000,-8.56212)--(1.23750,-8.55724)--(1.26500,-8.55211)--(1.29250,-8.54674)--(1.32000,-8.54144)--(1.34750,-8.53598)--(1.37500,-8.53031)--(1.40250,-8.52444)--(1.43000,-8.51838)--(1.45750,-8.51212)--(1.48500,-8.50567)--(1.51250,-8.49904)--(1.54000,-8.49223)--(1.56750,-8.48524)--(1.59500,-8.47809)--(1.62250,-8.47076)--(1.65000,-8.46327)--(1.67750,-8.45562)--(1.70500,-8.44782)--(1.73250,-8.43986)--(1.76000,-8.43176)--(1.78750,-8.42351)--(1.81500,-8.41512)--(1.84250,-8.40659)--(1.87000,-8.39792)--(1.89750,-8.38913)--(1.92500,-8.38020)--(1.95250,-8.37115)--(1.98000,-8.36197)--(2.00750,-8.35267)--(2.03500,-8.34326)--(2.06250,-8.33373)--(2.09000,-8.32415)--(2.11750,-8.31450)--(2.14500,-8.30473)--(2.17250,-8.29484)--(2.20000,-8.28483)--(2.22750,-8.27483)--(2.25500,-8.26480)--(2.28250,-8.25466)--(2.31000,-8.24441)--(2.33750,-8.23406)--(2.36500,-8.22360)--(2.39250,-8.21304)--(2.42000,-8.20239)--(2.44750,-8.19163)--(2.47500,-8.18077)--(2.50250,-8.16981)--(2.53000,-8.15876)--(2.55750,-8.14763)--(2.58500,-8.13642)--(2.61250,-8.12510)--(2.64000,-8.11367)--(2.66750,-8.10214)--(2.69500,-8.09050)--(2.72250,-8.07875)--(2.75000,-8.06689)--(2.77750,-8.05492)--(2.80500,-8.04283)--(2.83250,-8.03064)--(2.86000,-8.01832)--(2.88750,-8.00589)--(2.91500,-7.99337)--(2.94250,-7.98075)--(2.97000,-7.96799)--(2.99750,-7.95510)--(3.02500,-7.94208)--(3.05250,-7.92892)--(3.08000,-7.91562)--(3.10750,-7.90217)--(3.13500,-7.88856)--(3.16250,-7.87479)--(3.19000,-7.86085)--(3.21750,-7.84674)--(3.24500,-7.83246)--(3.27250,-7.81799)--(3.30000,-7.80334)--(3.32750,-7.78849)--(3.35500,-7.77343)--(3.38250,-7.75816)--(3.41000,-7.74269)--(3.43750,-7.72697)--(3.46500,-7.71102)--(3.49250,-7.69481)--(3.52000,-7.67833)--(3.54750,-7.66156)--(3.57500,-7.64450)--(3.60250,-7.62711)--(3.63000,-7.60939)--(3.65750,-7.59130)--(3.68500,-7.57286)--(3.71250,-7.55400)--(3.74000,-7.53470)--(3.76750,-7.51491)--(3.79500,-7.49460)--(3.82250,-7.47373)--(3.85000,-7.45223)--(3.87750,-7.43005)--(3.90500,-7.40711)--(3.93250,-7.38334)--(3.96000,-7.35863)--(3.98750,-7.33287)--(4.01500,-7.30592)--(4.04250,-7.27760)--(4.07000,-7.24770)--(4.09750,-7.21594)--(4.12500,-7.18198)--(4.15250,-7.14536)--(4.18000,-7.10545)--(4.20750,-7.06139)--(4.23500,-7.01190)--(4.26250,-6.95498)--(4.29000,-6.88734)--(4.31750,-6.80274)--(4.34500,-6.68721)--(4.37250,-6.49621)--(4.37525,-6.46765)--(4.37800,-6.43583)--(4.38075,-6.39989)--(4.38350,-6.35853)--(4.38625,-6.30979)--(4.38900,-6.25034)--(4.39175,-6.17396)--(4.39450,-6.06669)--(4.39725,-5.88407)--(4.40000,0)--(4.40275,-5.88223)--(4.40550,-6.06307)--(4.40825,-6.16851)--(4.41100,-6.24307)--(4.41375,-6.30070)--(4.41650,-6.34762)--(4.41925,-6.38716)--(4.42200,-6.42128)--(4.42475,-6.45128)--(4.42750,-6.47802)--(4.45500,-6.65082)--(4.48250,-6.74813)--(4.51000,-6.81450)--(4.53750,-6.86391)--(4.56500,-6.90257)--(4.59250,-6.93380)--(4.62000,-6.95959)--(4.64750,-6.98121)--(4.67500,-6.99953)--(4.70250,-7.01518)--(4.73000,-7.02862)--(4.75750,-7.04018)--(4.78500,-7.05016)--(4.81250,-7.05875)--(4.84000,-7.06614)--(4.86750,-7.07246)--(4.89500,-7.07783)--(4.92250,-7.08236)--(4.95000,-7.08612)--(4.97750,-7.08918)--(5.00500,-7.09161)--(5.03250,-7.09346)--(5.06000,-7.09478)--(5.08750,-7.09560)--(5.11500,-7.09596)--(5.14250,-7.09589)--(5.17000,-7.09543)--(5.19750,-7.09460)--(5.22500,-7.09341)--(5.25250,-7.09190)--(5.28000,-7.09007)--(5.30750,-7.08794)--(5.33500,-7.08554)--(5.36250,-7.08286)--(5.39000,-7.07993)--(5.41750,-7.07676)--(5.44500,-7.07336)--(5.47250,-7.06974)--(5.50000,-7.06590)--(5.52750,-7.06186)--(5.55500,-7.05762)--(5.58250,-7.05319)--(5.61000,-7.04857)--(5.63750,-7.04378)--(5.66500,-7.03882)--(5.69250,-7.03369)--(5.72000,-7.02841)--(5.74750,-7.02296)--(5.77500,-7.01737)--(5.80250,-7.01162)--(5.83000,-7.00574)--(5.85750,-6.99971)--(5.88500,-6.99355)--(5.91250,-6.98725)--(5.94000,-6.98083)--(5.96750,-6.97427)--(5.99500,-6.96759)--(6.02250,-6.96078)--(6.05000,-6.95385)--(6.07750,-6.94681)--(6.10500,-6.93964)--(6.13250,-6.93236)--(6.16000,-6.92497)--(6.18750,-6.91746)--(6.21500,-6.90984)--(6.24250,-6.90212)--(6.27000,-6.89428)--(6.29750,-6.88634)--(6.32500,-6.87829)--(6.35250,-6.87008)--(6.38000,-6.86182)--(6.40750,-6.85346)--(6.43500,-6.84499)--(6.46250,-6.83642)--(6.49000,-6.82774)--(6.51750,-6.81897)--(6.54500,-6.81009)--(6.57250,-6.80111)--(6.60000,-6.79205)--(6.62750,-6.78289)--(6.65500,-6.77363)--(6.68250,-6.76427)--(6.71000,-6.75480)--(6.73750,-6.74523)--(6.76500,-6.73556)--(6.79250,-6.72579)--(6.82000,-6.71591)--(6.84750,-6.70593)--(6.87500,-6.69585)--(6.90250,-6.68566)--(6.93000,-6.67536)--(6.95750,-6.66495)--(6.98500,-6.65444)--(7.01250,-6.64382)--(7.04000,-6.63309)--(7.06750,-6.62224)--(7.09500,-6.61129)--(7.12250,-6.60021)--(7.15000,-6.58902)--(7.17750,-6.57772)--(7.20500,-6.56629)--(7.23250,-6.55474)--(7.26000,-6.54306)--(7.28750,-6.53126)--(7.31500,-6.51933)--(7.34250,-6.50726)--(7.37000,-6.49506)--(7.39750,-6.48272)--(7.42500,-6.47024)--(7.45250,-6.45761)--(7.48000,-6.44483)--(7.50750,-6.43189)--(7.53500,-6.41880)--(7.56250,-6.40555)--(7.59000,-6.39212)--(7.61750,-6.37852)--(7.64500,-6.36474)--(7.67250,-6.35077)--(7.70000,-6.33661)--(7.72750,-6.32225)--(7.75500,-6.30767)--(7.78250,-6.29288)--(7.81000,-6.27786)--(7.83750,-6.26260)--(7.86500,-6.24708)--(7.89250,-6.23131)--(7.92000,-6.21525)--(7.94750,-6.19891)--(7.97500,-6.18225)--(8.00250,-6.16527)--(8.03000,-6.14794)--(8.05750,-6.13024)--(8.08500,-6.11214)--(8.11250,-6.09362)--(8.14000,-6.07464)--(8.16750,-6.05518)--(8.19500,-6.03518)--(8.22250,-6.01461)--(8.25000,-5.99340)--(8.27750,-5.97151)--(8.30500,-5.94885)--(8.33250,-5.92535)--(8.36000,-5.90091)--(8.38750,-5.87540)--(8.41500,-5.84869)--(8.44250,-5.82061)--(8.47000,-5.79094)--(8.49750,-5.75940)--(8.52500,-5.72566)--(8.55250,-5.68924)--(8.58000,-5.64953)--(8.60750,-5.60566)--(8.63500,-5.55635)--(8.66250,-5.49961)--(8.69000,-5.43214)--(8.71750,-5.34773)--(8.74500,-5.23241)--(8.77250,-5.04166)--(8.77525,-5.01314)--(8.77800,-4.98137)--(8.78075,-4.94548)--(8.78350,-4.90418)--(8.78625,-4.85552)--(8.78900,-4.79619)--(8.79175,-4.72000)--(8.79450,-4.61309)--(8.79725,-4.43141)--(8.80000,-3.16457)--(8.80275,-4.42837)--(8.80550,-4.60884)--(8.80825,-4.71413)--(8.81100,-4.78861)--(8.81375,-4.84619)--(8.81650,-4.89307)--(8.81925,-4.93257)--(8.82200,-4.96667)--(8.82475,-4.99664)--(8.82750,-5.02336)--(8.85500,-5.19598)--(8.88250,-5.29315)--(8.91000,-5.35941)--(8.93750,-5.40870)--(8.96500,-5.44725)--(8.99250,-5.47838)--(9.02000,-5.50406)--(9.04750,-5.52559)--(9.07500,-5.54381)--(9.10250,-5.55937)--(9.13000,-5.57272)--(9.15750,-5.58421)--(9.18500,-5.59410)--(9.21250,-5.60262)--(9.24000,-5.60993)--(9.26750,-5.61619)--(9.29500,-5.62150)--(9.32250,-5.62596)--(9.35000,-5.62967)--(9.37750,-5.63268)--(9.40500,-5.63506)--(9.43250,-5.63686)--(9.46000,-5.63813)--(9.48750,-5.63891)--(9.51500,-5.63923)--(9.54250,-5.63914)--(9.57000,-5.63864)--(9.59750,-5.63777)--(9.62500,-5.63656)--(9.65250,-5.63502)--(9.68000,-5.63317)--(9.70750,-5.63102)--(9.73500,-5.62860)--(9.76250,-5.62591)--(9.79000,-5.62297)--(9.81750,-5.61980)--(9.84500,-5.61639)--(9.87250,-5.61276)--(9.90000,-5.60892)--(9.92750,-5.60488)--(9.95500,-5.60065)--(9.98250,-5.59623)--(10.01000,-5.59162)--(10.03750,-5.58684)--(10.06500,-5.58189)--(10.09250,-5.57678)--(10.12000,-5.57151)--(10.14750,-5.56608)--(10.17500,-5.56051)--(10.20250,-5.55479)--(10.23000,-5.54892)--(10.25750,-5.54292)--(10.28500,-5.53677)--(10.31250,-5.53050)--(10.34000,-5.52410)--(10.36750,-5.51757)--(10.39500,-5.51091)--(10.42250,-5.50413)--(10.45000,-5.49723)--(10.47750,-5.49021)--(10.50500,-5.48308)--(10.53250,-5.47583)--(10.56000,-5.46847)--(10.58750,-5.46099)--(10.61500,-5.45341)--(10.64250,-5.44571)--(10.67000,-5.43791)--(10.69750,-5.43000)--(10.72500,-5.42199)--(10.75250,-5.41387)--(10.78000,-5.40564)--(10.80750,-5.39731)--(10.83500,-5.38888)--(10.86250,-5.38035)--(10.89000,-5.37171)--(10.91750,-5.36297)--(10.94500,-5.35413)--(10.97250,-5.34519)--(11.00000,-5.33615)--(11.02750,-5.32701)--(11.05500,-5.31776)--(11.08250,-5.30842)--(11.11000,-5.29897)--(11.13750,-5.28942)--(11.16500,-5.27977)--(11.19250,-5.27002)--(11.22000,-5.26016)--(11.24750,-5.25021)--(11.27500,-5.24014)--(11.30250,-5.22997)--(11.33000,-5.21970)--(11.35750,-5.20932)--(11.38500,-5.19883)--(11.41250,-5.18823)--(11.44000,-5.17752)--(11.46750,-5.16670)--(11.49500,-5.15576)--(11.52250,-5.14471)--(11.55000,-5.13354)--(11.57750,-5.12226)--(11.60500,-5.11085)--(11.63250,-5.09933)--(11.66000,-5.08767)--(11.68750,-5.07589)--(11.71500,-5.06398)--(11.74250,-5.05194)--(11.77000,-5.03976)--(11.79750,-5.02744)--(11.82500,-5.01498)--(11.85250,-5.00238)--(11.88000,-4.98962)--(11.90750,-4.97671)--(11.93500,-4.96364)--(11.96250,-4.95040)--(11.99000,-4.93700)--(12.01750,-4.92342)--(12.04500,-4.90966)--(12.07250,-4.89571)--(12.10000,-4.88157)--(12.12750,-4.86722)--(12.15500,-4.85267)--(12.18250,-4.83789)--(12.21000,-4.82289)--(12.23750,-4.80764)--(12.26500,-4.79215)--(12.29250,-4.77639)--(12.32000,-4.76035)--(12.34750,-4.74402)--(12.37500,-4.72739)--(12.40250,-4.71042)--(12.43000,-4.69310)--(12.45750,-4.67542)--(12.48500,-4.65734)--(12.51250,-4.63883)--(12.54000,-4.61987)--(12.56750,-4.60042)--(12.59500,-4.58044)--(12.62250,-4.55988)--(12.65000,-4.53869)--(12.67750,-4.51680)--(12.70500,-4.49416)--(12.73250,-4.47067)--(12.76000,-4.44624)--(12.78750,-4.42074)--(12.81500,-4.39404)--(12.84250,-4.36597)--(12.87000,-4.33631)--(12.89750,-4.30478)--(12.92500,-4.27104)--(12.95250,-4.23464)--(12.98000,-4.19494)--(13.00750,-4.15108)--(13.03500,-4.10177)--(13.06250,-4.04504)--(13.09000,-3.97757)--(13.11750,-3.89315)--(13.14500,-3.77779)--(13.17250,-3.58697)--(13.17525,-3.55844)--(13.17525,-3.55844)--(13.17800,-3.52665)--(13.17800,-3.52665)--(13.18075,-3.49073)--(13.18075,-3.49073)--(13.18350,-3.44940)--(13.18350,-3.44940)--(13.18625,-3.40070)--(13.18625,-3.40070)--(13.18900,-3.34130)--(13.18900,-3.34130)--(13.19175,-3.26499)--(13.19175,-3.26499)--(13.19450,-3.15785)--(13.19450,-3.15785)--(13.19725,-2.97549)--(13.19725,-2.97549)--(13.20000,-1.43661)--(13.22750,-3.56879)--(13.25500,-3.74143)--(13.28250,-3.83860)--(13.31000,-3.90485)--(13.33750,-3.95413)--(13.36500,-3.99268)--(13.39250,-4.02380)--(13.42000,-4.04948)--(13.44750,-4.07100)--(13.47500,-4.08923)--(13.50250,-4.10478)--(13.53000,-4.11812)--(13.55750,-4.12961)--(13.58500,-4.13950)--(13.61250,-4.14801)--(13.64000,-4.15532)--(13.66750,-4.16158)--(13.69500,-4.16688)--(13.72250,-4.17134)--(13.75000,-4.17504)--(13.77750,-4.17805)--(13.80500,-4.18043)--(13.83250,-4.18223)--(13.86000,-4.18350)--(13.88750,-4.18428)--(13.91500,-4.18460)--(13.94250,-4.18450)--(13.97000,-4.18400)--(13.99750,-4.18314)--(14.02500,-4.18192)--(14.05250,-4.18038)--(14.08000,-4.17852)--(14.10750,-4.17638)--(14.13500,-4.17396)--(14.16250,-4.17127)--(14.19000,-4.16833)--(14.21750,-4.16515)--(14.24500,-4.16174)--(14.27250,-4.15812)--(14.30000,-4.15428)--(14.32750,-4.15024)--(14.35500,-4.14600)--(14.38250,-4.14158)--(14.41000,-4.13698)--(14.43750,-4.13220)--(14.46500,-4.12725)--(14.49250,-4.12214)--(14.52000,-4.11687)--(14.54750,-4.11144)--(14.57500,-4.10587)--(14.60250,-4.10014)--(14.63000,-4.09428)--(14.65750,-4.08827)--(14.68500,-4.08214)--(14.71250,-4.07586)--(14.74000,-4.06946)--(14.76750,-4.06293)--(14.79500,-4.05628)--(14.82250,-4.04950)--(14.85000,-4.04260)--(14.87750,-4.03558)--(14.90500,-4.02845)--(14.93250,-4.02120)--(14.96000,-4.01384)--(14.98750,-4.00637)--(15.01500,-3.99878)--(15.04250,-3.99109)--(15.07000,-3.98329)--(15.09750,-3.97538)--(15.12500,-3.96737)--(15.15250,-3.95925)--(15.18000,-3.95103)--(15.20750,-3.94270)--(15.23500,-3.93427)--(15.26250,-3.92574)--(15.29000,-3.91711)--(15.31750,-3.90837)--(15.34500,-3.89953)--(15.37250,-3.89059)--(15.40000,-3.88155)--(15.42750,-3.87241)--(15.45500,-3.86317)--(15.48250,-3.85382)--(15.51000,-3.84438)--(15.53750,-3.83483)--(15.56500,-3.82518)--(15.59250,-3.81543)--(15.62000,-3.80558)--(15.64750,-3.79562)--(15.67500,-3.78556)--(15.70250,-3.77538)--(15.73000,-3.76511)--(15.75750,-3.75473)--(15.78500,-3.74424)--(15.81250,-3.73364)--(15.84000,-3.72293)--(15.86750,-3.71211)--(15.89500,-3.70118)--(15.92250,-3.69013)--(15.95000,-3.67897)--(15.97750,-3.66768)--(16.00500,-3.65628)--(16.03250,-3.64475)--(16.06000,-3.63310)--(16.08750,-3.62132)--(16.11500,-3.60941)--(16.14250,-3.59737)--(16.17000,-3.58519)--(16.19750,-3.57288)--(16.22500,-3.56042)--(16.25250,-3.54781)--(16.28000,-3.53505)--(16.30750,-3.52214)--(16.33500,-3.50907)--(16.36250,-3.49584)--(16.39000,-3.48243)--(16.41750,-3.46885)--(16.44500,-3.45510)--(16.47250,-3.44115)--(16.50000,-3.42701)--(16.52750,-3.41266)--(16.55500,-3.39811)--(16.58250,-3.38334)--(16.61000,-3.36833)--(16.63750,-3.35309)--(16.66500,-3.33760)--(16.69250,-3.32184)--(16.72000,-3.30580)--(16.74750,-3.28948)--(16.77500,-3.27284)--(16.80250,-3.25587)--(16.83000,-3.23856)--(16.85750,-3.22087)--(16.88500,-3.20279)--(16.91250,-3.18429)--(16.94000,-3.16533)--(16.96750,-3.14588)--(16.99500,-3.12589)--(17.02250,-3.10534)--(17.05000,-3.08415)--(17.07750,-3.06226)--(17.10500,-3.03962)--(17.13250,-3.01613)--(17.16000,-2.99170)--(17.18750,-2.96620)--(17.21500,-2.93951)--(17.24250,-2.91143)--(17.27000,-2.88177)--(17.29750,-2.85025)--(17.32500,-2.81651)--(17.35250,-2.78011)--(17.38000,-2.74041)--(17.40750,-2.69655)--(17.43500,-2.64724)--(17.46250,-2.59051)--(17.49000,-2.52305)--(17.51750,-2.43862)--(17.54500,-2.32327)--(17.57250,-2.13247)--(17.57525,-2.10394)--(17.57800,-2.07215)--(17.58075,-2.03624)--(17.58350,-1.99492)--(17.58625,-1.94622)--(17.58900,-1.88683)--(17.59175,-1.81055)--(17.59450,-1.70346)--(17.59725,-1.52125)--(17.60000,-0.06865);
\end{tikzpicture}
\end{center}
\caption{The maximum $\varepsilon(c)$ of $|\Delta^\textup{LP}_n(c)-(n+(c-4)/8)|$
over $2c \le n \le 128$.}
\label{figure:maximumDelta}
\end{figure}
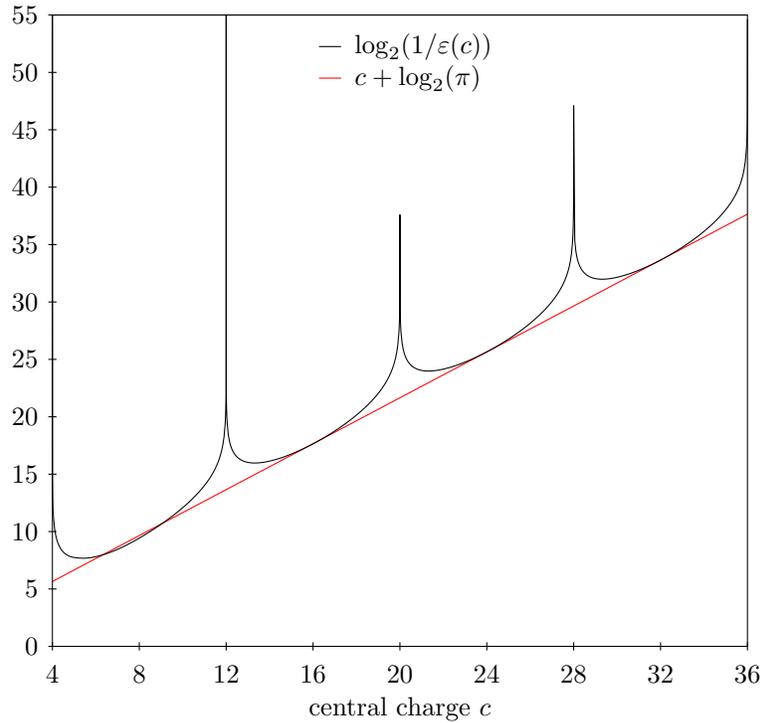

Figure~\ref{figure:maximumDelta} shows another aspect of this periodicity. In
this figure, we plot
\begin{equation}
\varepsilon(c) = \max_{2c \le n \le 128}|\Delta^\textup{LP}_n(c)-(n+(c-4)/8)|.
\end{equation}
Here the upper bound of $128$ for $n$ should be viewed as a stand-in for
infinity, and the lower bound $2c$ is intended to exclude sporadic behavior
for small $n$ before convergence; in other words, we view $\varepsilon(c)$ as
an approximation to
\begin{equation}
\limsup_{n \to \infty} |\Delta^\textup{LP}_n(c)-(n+(c-4)/8)|.
\end{equation}
In the plot, there are singularities at $c=4$ and $12$ that correspond to
$\varepsilon(c)=0$, with similar cusps at $c=20$, $28$, and $36$, and again
we see the largest error terms when $c$ is a multiple of $8$. This unexpected
periodicity shows that the spinless modular bootstrap has a much richer
number-theoretic structure than is apparent from the smooth plot of
$\Delta_1^\textup{LP}(c)$ in Figure~\ref{figure:ratioDeltac}.

We can summarize Figures~\ref{figure:gapsDelta} and~\ref{figure:maximumDelta}
with the following conjecture:

\begin{conjecture} \label{conjecture:Deltagaps}
For $c \ge 4$, the quantity
\begin{equation}
\limsup_{n \to \infty} |\Delta^\textup{LP}_n(c)-(n+(c-4)/8)|
\end{equation}
vanishes if and only if $c$ is an integer congruent to $4$ modulo $8$, and
there is a constant $\alpha$ such that
\begin{equation}
\limsup_{n \to \infty} |\Delta^\textup{LP}_n(c)-(n+(c-4)/8)| \le \alpha 2^{-c}
\end{equation}
for all $c \ge 4$. Furthermore, we can take $\alpha = 1/\pi + o(1)$ as $c \to \infty$.
\end{conjecture}

\subsection{Growth rate of the degeneracies}

Solving the truncated crossing equation \eqref{truncprim} yields not just
scaling dimensions $\Delta_1^{\textup{LP},N}(c)$ but also corresponding
degeneracies $d_n^{\textup{LP},N}(c)$, which converge as $N \to \infty$ to
the degeneracies $d_n^\textup{LP}(c)$ of a hypothetical CFT that attains the
spinless modular bootstrap bound. For $c \le 50$ and $c \not\in
\{1/2,1,4,12\}$, numerical calculations show that these degeneracies are not
integers, and thus they cannot come from an actual CFT.\footnote{They could
still come from a sphere packing. In that case $d_n$ would be the average
number of sphere centers at distance $\sqrt{2\Delta_n}$ from a given sphere
center, which need not be an integer.} For larger $c$, it is difficult to
assess integrality, because the degeneracies grow exponentially as $c \to
\infty$ and must therefore be computed to high precision.

The cumulative growth rate of the degeneracies is determined by modularity as
follows. Because $\eta(-1/\tau) = (\tau/i)^{1/2} \eta(\tau)$, the modular
invariance of the partition function $\cZ(\tau)=\eta(\tau)^{-2c} \sum_{\Delta} d_{\Delta}
e^{2\pi i \tau \Delta}$ implies the identity
\begin{equation}
\sum_{\Delta} d_{\Delta} e^{2\pi i \tau \Delta} = (i/\tau)^{c} \sum_{\Delta} e^{-2\pi i \Delta/\tau}.
\end{equation}
If we set $\tau = i \beta/(2\pi)$ and let $\beta \to 0$, we find that
\begin{equation}
\sum_{\Delta} d_{\Delta} e^{-\beta \Delta} \sim (2\pi/\beta)^c.
\end{equation}
Now the Karamata Tauberian theorem \cite[Theorem~4.3 of Chapter~V]{W} implies
that
\begin{equation} \label{eq:Karamata}
\sum_{\Delta \le A} d_{\Delta} \sim \frac{(2\pi A)^c}{\Gamma(c+1)}
\end{equation}
as $A \to \infty$.

In other words, the function
\begin{equation}
 \rho_c(\Delta) = \frac{(2\pi)^c \Delta^{c-1}}{\Gamma(c)}
\end{equation}
is the $\textup{U}(1)^c$ analogue of the Cardy formula \cite{Cardy} for degeneracies,
because
\begin{equation}
\int_0^A d\Delta\, \rho_c(\Delta) = \frac{(2\pi A)^c}{\Gamma(c+1)}.
\end{equation}
A similar formula applies to operator-product coefficients that appear in the
bootstrap equations for conformal correlators \cite{1709.00008}.

Because the scaling dimensions $\Delta_n^\textup{LP}(c)$ are uniformly spaced
with distance $1$
asymptotically, we expect that $d_n^\textup{LP}(c)$ will be roughly
$\rho_c(\Delta_n^{\textup{LP}}(c))$ as $n \to \infty$. The asymptotic formula
\eqref{eq:Karamata} gives a sense in which this approximation is true on
average, but we will see that the precise behavior is far more delicate. In
the Virasoro case, a more fine-grained understanding of the asymptotic
spectrum has been obtained recently using complex Tauberian theorems
\cite{1904.06359, 1905.12636, 1910.07727, 2003.14316, 2004.12557}. It would
be interesting to do the same for sphere packing. This could perhaps explain
the linear portion of the large-$c$ spectrum, where the level spacing is very
close to $1$.

\subsection{Degeneracies for $c=4$ and $12$}

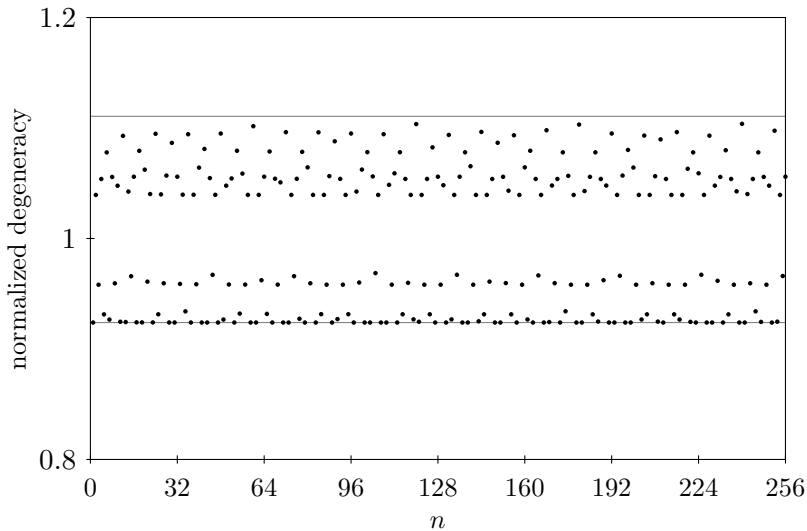
\begin{figure}
\begin{center}
\begin{tikzpicture}[x=0.525cm,y=0.84cm]
\draw[black!50] (0,-7/2+35/2*1.1106-35/2)--(17.6,-7/2+35/2*1.1106-35/2);
\draw[black!50] (0,-7/2+35/2*0.9239-35/2)--(17.6,-7/2+35/2*0.9239-35/2);
\draw (0,0)--(17.6,0)--(17.6,-7)--(0,-7)--(0,0);
\foreach \x in {0,...,8}
\draw (17.6*\x/8,-7-0.065)--(17.6*\x/8,-7+0.065);
\foreach \x in {0,32,...,256}
\draw ({17.6*\x/256},{-7-0.15})  node[below] {\small \x};
\draw (-0.104,-7)--(0.104,-7); \draw (-0.1,-7) node[left] {$0.8$};
\draw (-0.104,-3.5)--(0.104,-3.5); \draw (-0.1,-3.5) node[left] {$1$};
\draw (-0.104,0)--(0.104,0); \draw (-0.1,0) node[left] {$1.2$};
\draw (8.8,-7-0.75)
node[below] {\small $n$}; \draw (-1.8,-3.5) node[rotate=90]
{\small normalized degeneracy};
\draw (19.1,-3.5) node[rotate=-90]
{\small $\phantom{\text{normalized degeneracy}}$};
\fill (0.0687500,-4.83108) circle (0.03cm);
\fill (0.137500,-2.80996) circle (0.03cm);
\fill (0.206250,-4.23223) circle (0.03cm);
\fill (0.275000,-2.55732) circle (0.03cm);
\fill (0.343750,-4.70173) circle (0.03cm);
\fill (0.412500,-2.13626) circle (0.03cm);
\fill (0.481250,-4.78394) circle (0.03cm);
\fill (0.550000,-2.52574) circle (0.03cm);
\fill (0.618750,-4.21005) circle (0.03cm);
\fill (0.687500,-2.66444) circle (0.03cm);
\fill (0.756250,-4.81893) circle (0.03cm);
\fill (0.825000,-1.87426) circle (0.03cm);
\fill (0.893750,-4.82372) circle (0.03cm);
\fill (0.962500,-2.75693) circle (0.03cm);
\fill (1.03125,-4.09809) circle (0.03cm);
\fill (1.10000,-2.52180) circle (0.03cm);
\fill (1.16875,-4.82779) circle (0.03cm);
\fill (1.23750,-2.11131) circle (0.03cm);
\fill (1.30625,-4.82872) circle (0.03cm);
\fill (1.37500,-2.40978) circle (0.03cm);
\fill (1.44375,-4.18334) circle (0.03cm);
\fill (1.51250,-2.79630) circle (0.03cm);
\fill (1.58125,-4.82975) circle (0.03cm);
\fill (1.65000,-1.84151) circle (0.03cm);
\fill (1.71875,-4.70069) circle (0.03cm);
\fill (1.78750,-2.80168) circle (0.03cm);
\fill (1.85625,-4.20923) circle (0.03cm);
\fill (1.92500,-2.50355) circle (0.03cm);
\fill (1.99375,-4.83041) circle (0.03cm);
\fill (2.06250,-1.98535) circle (0.03cm);
\fill (2.13125,-4.83054) circle (0.03cm);
\fill (2.20000,-2.52130) circle (0.03cm);
\fill (2.26875,-4.21963) circle (0.03cm);
\fill (2.33750,-2.80626) circle (0.03cm);
\fill (2.40625,-4.65421) circle (0.03cm);
\fill (2.47500,-1.84896) circle (0.03cm);
\fill (2.54375,-4.83076) circle (0.03cm);
\fill (2.61250,-2.80731) circle (0.03cm);
\fill (2.68125,-4.22460) circle (0.03cm);
\fill (2.75000,-2.37795) circle (0.03cm);
\fill (2.81875,-4.83084) circle (0.03cm);
\fill (2.88750,-2.08126) circle (0.03cm);
\fill (2.95625,-4.83087) circle (0.03cm);
\fill (3.02500,-2.54347) circle (0.03cm);
\fill (3.09375,-4.07573) circle (0.03cm);
\fill (3.16250,-2.80847) circle (0.03cm);
\fill (3.23125,-4.83092) circle (0.03cm);
\fill (3.30000,-1.83742) circle (0.03cm);
\fill (3.36875,-4.78380) circle (0.03cm);
\fill (3.43750,-2.66328) circle (0.03cm);
\fill (3.50625,-4.22882) circle (0.03cm);
\fill (3.57500,-2.54893) circle (0.03cm);
\fill (3.64375,-4.83097) circle (0.03cm);
\fill (3.71250,-2.11038) circle (0.03cm);
\fill (3.78125,-4.68948) circle (0.03cm);
\fill (3.85000,-2.47188) circle (0.03cm);
\fill (3.91875,-4.22978) circle (0.03cm);
\fill (3.98750,-2.80922) circle (0.03cm);
\fill (4.05625,-4.83100) circle (0.03cm);
\fill (4.12500,-1.72126) circle (0.03cm);
\fill (4.19375,-4.83101) circle (0.03cm);
\fill (4.26250,-2.80935) circle (0.03cm);
\fill (4.33125,-4.16110) circle (0.03cm);
\fill (4.40000,-2.52124) circle (0.03cm);
\fill (4.46875,-4.69431) circle (0.03cm);
\fill (4.53750,-2.12208) circle (0.03cm);
\fill (4.60625,-4.83102) circle (0.03cm);
\fill (4.67500,-2.55357) circle (0.03cm);
\fill (4.74375,-4.23085) circle (0.03cm);
\fill (4.81250,-2.61099) circle (0.03cm);
\fill (4.88125,-4.83103) circle (0.03cm);
\fill (4.95000,-1.81617) circle (0.03cm);
\fill (5.01875,-4.83104) circle (0.03cm);
\fill (5.08750,-2.80960) circle (0.03cm);
\fill (5.15625,-4.09701) circle (0.03cm);
\fill (5.22500,-2.55463) circle (0.03cm);
\fill (5.29375,-4.77175) circle (0.03cm);
\fill (5.36250,-2.12767) circle (0.03cm);
\fill (5.43125,-4.83105) circle (0.03cm);
\fill (5.50000,-2.37397) circle (0.03cm);
\fill (5.56875,-4.20920) circle (0.03cm);
\fill (5.63750,-2.80970) circle (0.03cm);
\fill (5.70625,-4.83105) circle (0.03cm);
\fill (5.77500,-1.81850) circle (0.03cm);
\fill (5.84375,-4.69841) circle (0.03cm);
\fill (5.91250,-2.80973) circle (0.03cm);
\fill (5.98125,-4.23154) circle (0.03cm);
\fill (6.05000,-2.51186) circle (0.03cm);
\fill (6.11875,-4.83106) circle (0.03cm);
\fill (6.18750,-1.96020) circle (0.03cm);
\fill (6.25625,-4.77656) circle (0.03cm);
\fill (6.32500,-2.55581) circle (0.03cm);
\fill (6.39375,-4.23167) circle (0.03cm);
\fill (6.46250,-2.80979) circle (0.03cm);
\fill (6.53125,-4.69935) circle (0.03cm);
\fill (6.60000,-1.83691) circle (0.03cm);
\fill (6.66875,-4.83106) circle (0.03cm);
\fill (6.73750,-2.75678) circle (0.03cm);
\fill (6.80625,-4.19743) circle (0.03cm);
\fill (6.87500,-2.40860) circle (0.03cm);
\fill (6.94375,-4.83106) circle (0.03cm);
\fill (7.01250,-2.13242) circle (0.03cm);
\fill (7.08125,-4.83106) circle (0.03cm);
\fill (7.15000,-2.51733) circle (0.03cm);
\fill (7.21875,-4.04881) circle (0.03cm);
\fill (7.28750,-2.80984) circle (0.03cm);
\fill (7.35625,-4.83106) circle (0.03cm);
\fill (7.42500,-1.84803) circle (0.03cm);
\fill (7.49375,-4.83107) circle (0.03cm);
\fill (7.56250,-2.65067) circle (0.03cm);
\fill (7.63125,-4.23190) circle (0.03cm);
\fill (7.70000,-2.46792) circle (0.03cm);
\fill (7.76875,-4.83107) circle (0.03cm);
\fill (7.83750,-2.13351) circle (0.03cm);
\fill (7.90625,-4.70039) circle (0.03cm);
\fill (7.97500,-2.55657) circle (0.03cm);
\fill (8.04375,-4.20241) circle (0.03cm);
\fill (8.11250,-2.80987) circle (0.03cm);
\fill (8.18125,-4.78064) circle (0.03cm);
\fill (8.25000,-1.68824) circle (0.03cm);
\fill (8.31875,-4.81892) circle (0.03cm);
\fill (8.38750,-2.80988) circle (0.03cm);
\fill (8.45625,-4.23199) circle (0.03cm);
\fill (8.52500,-2.55670) circle (0.03cm);
\fill (8.59375,-4.70068) circle (0.03cm);
\fill (8.66250,-2.05624) circle (0.03cm);
\fill (8.73125,-4.83107) circle (0.03cm);
\fill (8.80000,-2.52123) circle (0.03cm);
\fill (8.86875,-4.23202) circle (0.03cm);
\fill (8.93750,-2.65610) circle (0.03cm);
\fill (9.00625,-4.83107) circle (0.03cm);
\fill (9.07500,-1.85989) circle (0.03cm);
\fill (9.14375,-4.78157) circle (0.03cm);
\fill (9.21250,-2.80990) circle (0.03cm);
\fill (9.28125,-4.07490) circle (0.03cm);
\fill (9.35000,-2.52198) circle (0.03cm);
\fill (9.41875,-4.83107) circle (0.03cm);
\fill (9.48750,-2.13471) circle (0.03cm);
\fill (9.55625,-4.83107) circle (0.03cm);
\fill (9.62500,-2.35558) circle (0.03cm);
\fill (9.69375,-4.23207) circle (0.03cm);
\fill (9.76250,-2.80991) circle (0.03cm);
\fill (9.83125,-4.81156) circle (0.03cm);
\fill (9.90000,-1.81207) circle (0.03cm);
\fill (9.96875,-4.70106) circle (0.03cm);
\fill (10.0375,-2.80992) circle (0.03cm);
\fill (10.1063,-4.18320) circle (0.03cm);
\fill (10.1750,-2.55696) circle (0.03cm);
\fill (10.2438,-4.83107) circle (0.03cm);
\fill (10.3125,-1.98414) circle (0.03cm);
\fill (10.3813,-4.83107) circle (0.03cm);
\fill (10.4500,-2.52305) circle (0.03cm);
\fill (10.5188,-4.20663) circle (0.03cm);
\fill (10.5875,-2.74322) circle (0.03cm);
\fill (10.6563,-4.70118) circle (0.03cm);
\fill (10.7250,-1.86556) circle (0.03cm);
\fill (10.7938,-4.83107) circle (0.03cm);
\fill (10.8625,-2.80993) circle (0.03cm);
\fill (10.9313,-4.23212) circle (0.03cm);
\fill (11.0000,-2.37347) circle (0.03cm);
\fill (11.0688,-4.78261) circle (0.03cm);
\fill (11.1375,-2.11035) circle (0.03cm);
\fill (11.2063,-4.83107) circle (0.03cm);
\fill (11.2750,-2.55706) circle (0.03cm);
\fill (11.3438,-4.08539) circle (0.03cm);
\fill (11.4125,-2.80993) circle (0.03cm);
\fill (11.4813,-4.83107) circle (0.03cm);
\fill (11.5500,-1.78566) circle (0.03cm);
\fill (11.6188,-4.82372) circle (0.03cm);
\fill (11.6875,-2.66071) circle (0.03cm);
\fill (11.7563,-4.20760) circle (0.03cm);
\fill (11.8250,-2.55709) circle (0.03cm);
\fill (11.8938,-4.83107) circle (0.03cm);
\fill (11.9625,-2.13548) circle (0.03cm);
\fill (12.0313,-4.65317) circle (0.03cm);
\fill (12.1000,-2.50791) circle (0.03cm);
\fill (12.1688,-4.23215) circle (0.03cm);
\fill (12.2375,-2.80994) circle (0.03cm);
\fill (12.3063,-4.83108) circle (0.03cm);
\fill (12.3750,-1.69575) circle (0.03cm);
\fill (12.4438,-4.83108) circle (0.03cm);
\fill (12.5125,-2.74863) circle (0.03cm);
\fill (12.5813,-4.23216) circle (0.03cm);
\fill (12.6500,-2.52422) circle (0.03cm);
\fill (12.7188,-4.70140) circle (0.03cm);
\fill (12.7875,-2.13562) circle (0.03cm);
\fill (12.8563,-4.81564) circle (0.03cm);
\fill (12.9250,-2.55715) circle (0.03cm);
\fill (12.9938,-4.16028) circle (0.03cm);
\fill (13.0625,-2.66177) circle (0.03cm);
\fill (13.1313,-4.83108) circle (0.03cm);
\fill (13.2000,-1.83684) circle (0.03cm);
\fill (13.2688,-4.83108) circle (0.03cm);
\fill (13.3375,-2.80994) circle (0.03cm);
\fill (13.4063,-4.09039) circle (0.03cm);
\fill (13.4750,-2.50340) circle (0.03cm);
\fill (13.5438,-4.83108) circle (0.03cm);
\fill (13.6125,-2.09711) circle (0.03cm);
\fill (13.6813,-4.83108) circle (0.03cm);
\fill (13.7500,-2.37677) circle (0.03cm);
\fill (13.8188,-4.23217) circle (0.03cm);
\fill (13.8875,-2.80995) circle (0.03cm);
\fill (13.9563,-4.78327) circle (0.03cm);
\fill (14.0250,-1.87037) circle (0.03cm);
\fill (14.0938,-4.70149) circle (0.03cm);
\fill (14.1625,-2.80995) circle (0.03cm);
\fill (14.2313,-4.20867) circle (0.03cm);
\fill (14.3000,-2.51339) circle (0.03cm);
\fill (14.3688,-4.81657) circle (0.03cm);
\fill (14.4375,-1.92991) circle (0.03cm);
\fill (14.5063,-4.83108) circle (0.03cm);
\fill (14.5750,-2.55720) circle (0.03cm);
\fill (14.6438,-4.23218) circle (0.03cm);
\fill (14.7125,-2.80995) circle (0.03cm);
\fill (14.7813,-4.70152) circle (0.03cm);
\fill (14.8500,-1.81523) circle (0.03cm);
\fill (14.9188,-4.78339) circle (0.03cm);
\fill (14.9875,-2.80995) circle (0.03cm);
\fill (15.0563,-4.23219) circle (0.03cm);
\fill (15.1250,-2.39581) circle (0.03cm);
\fill (15.1938,-4.82043) circle (0.03cm);
\fill (15.2625,-2.13589) circle (0.03cm);
\fill (15.3313,-4.83108) circle (0.03cm);
\fill (15.4000,-2.46743) circle (0.03cm);
\fill (15.4688,-4.07466) circle (0.03cm);
\fill (15.5375,-2.80995) circle (0.03cm);
\fill (15.6063,-4.83108) circle (0.03cm);
\fill (15.6750,-1.87147) circle (0.03cm);
\fill (15.7438,-4.83108) circle (0.03cm);
\fill (15.8125,-2.66294) circle (0.03cm);
\fill (15.8813,-4.17071) circle (0.03cm);
\fill (15.9500,-2.52499) circle (0.03cm);
\fill (16.0188,-4.83108) circle (0.03cm);
\fill (16.0875,-2.10271) circle (0.03cm);
\fill (16.1563,-4.70157) circle (0.03cm);
\fill (16.2250,-2.55723) circle (0.03cm);
\fill (16.2938,-4.23219) circle (0.03cm);
\fill (16.3625,-2.75322) circle (0.03cm);
\fill (16.4313,-4.83108) circle (0.03cm);
\fill (16.5000,-1.68412) circle (0.03cm);
\fill (16.5688,-4.83108) circle (0.03cm);
\fill (16.6375,-2.79629) circle (0.03cm);
\fill (16.7063,-4.20920) circle (0.03cm);
\fill (16.7750,-2.55724) circle (0.03cm);
\fill (16.8438,-4.65407) circle (0.03cm);
\fill (16.9125,-2.13598) circle (0.03cm);
\fill (16.9813,-4.82136) circle (0.03cm);
\fill (17.0500,-2.52512) circle (0.03cm);
\fill (17.1188,-4.23220) circle (0.03cm);
\fill (17.1875,-2.66327) circle (0.03cm);
\fill (17.2563,-4.83108) circle (0.03cm);
\fill (17.3250,-1.79313) circle (0.03cm);
\fill (17.3938,-4.81760) circle (0.03cm);
\fill (17.4625,-2.80995) circle (0.03cm);
\fill (17.5313,-4.09465) circle (0.03cm);
\fill (17.6000,-2.52123) circle (0.03cm);
\end{tikzpicture}
\end{center}
\caption{The normalized degeneracies $d_n^\textup{LP}(4)/\rho_4(\Delta_n^\textup{LP}(4))$ for $1 \le n \le 256$.
As $n \to \infty$, they fluctuate between $1/\zeta(4)=0.9239\dots$ and $\zeta(3)/\zeta(4)=1.1106\dots$,
shown with gray lines, and they are bounded away from $1$.}
\label{figure:E8normalizeddegeneracy}
\end{figure}

The degeneracies $d_n^\textup{LP}(4)$ and $d_n^\textup{LP}(12)$ are the
coefficients of the theta series of the $E_8$ and Leech lattices,
respectively. Much is known about these modular forms, including precise
descriptions of their coefficients (see, for example, \cite[p.~122 and
p.~134]{SPLAG}). The degeneracies are well understood, but far more subtle
than the scaling dimensions $\Delta_n^\textup{LP}(4) = n$ and
$\Delta_n^\textup{LP}(12) = n+1$.

Figure~\ref{figure:E8normalizeddegeneracy} shows the normalized degeneracies
$d_n^\textup{LP}(4)/\rho_4(\Delta_n^\textup{LP}(4))$ for $c=4$. They are
bounded above and below, but do not converge to $1$; instead, they are
strictly bounded away from $1$. The most noteworthy aspect of
Figure~\ref{figure:E8normalizeddegeneracy} is that the normalized
degeneracies are almost, but not quite, periodic.\footnote{See
\cite[Theorem~5.13A.1]{Simon} for the theory of almost periodic functions.}
This near periodicity is explained by a classical formula for coefficients of
Eisenstein series: $d_n^\textup{LP}(4)/\rho_4(\Delta_n^\textup{LP}(4)) =
\sigma_3(n)/(\zeta(4) n^3)$, where $\sigma_k(n)$ denotes the sum of the
$k$-th powers of the divisors of $n$, and $\zeta$ is the Riemann zeta
function. This function is not periodic, but for each $\varepsilon>0$, there
exists a natural number $m$ such that if $n_1 \equiv n_2 \pmod{m}$, then
$|\sigma_3(n_1)/n_1^3 - \sigma_3(n_2)/n_2^3| < \varepsilon$. In other words,
for each $\varepsilon>0$ it is approximately periodic to within
$\varepsilon$, with the period length growing as $\varepsilon \to 0$.
Similarly, the function $n \mapsto
d_n^\textup{LP}(12)/\rho_{12}(\Delta_n^\textup{LP}(12))$ is the sum of the
almost periodic function $n \mapsto \sigma_{11}(n)/(\zeta(12) n^3)$ and a
term converging to zero as $n \to \infty$.

\subsection{Degeneracies for arbitrary $c$}

\begin{figure}
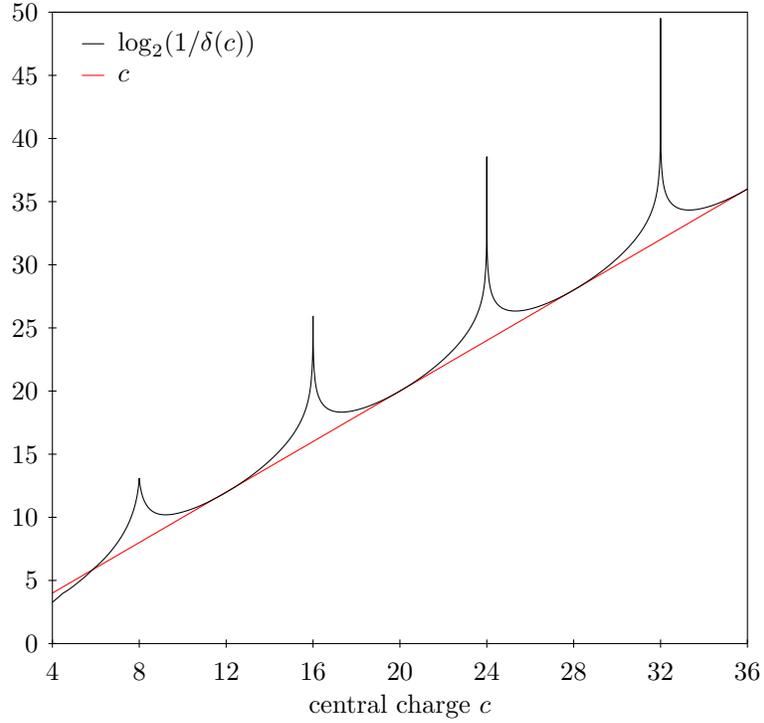

\begin{center}

\end{center}
\caption{The maximum $\delta(c)$ of $|d_n^\textup{LP}(c)/\rho_c(\Delta_n^\textup{LP}(c))-1|$
over $2c \le n \le 128$.}
\label{figure:maximumd}
\end{figure}

\begin{figure}
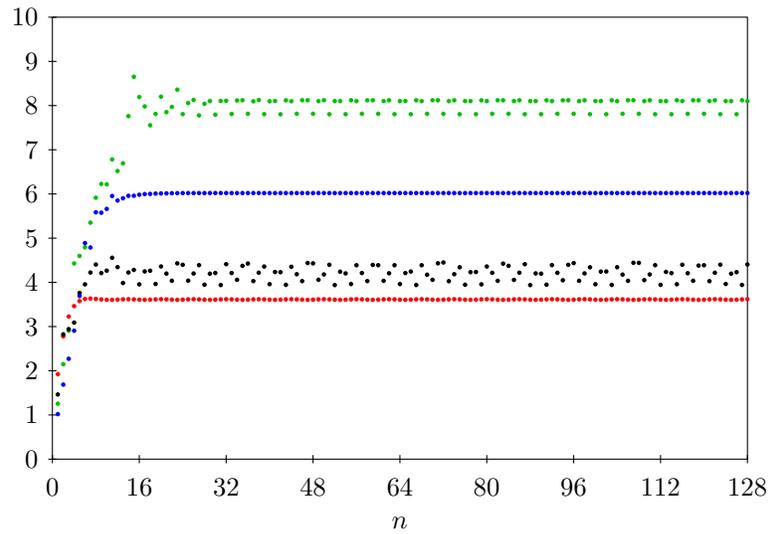

\begin{center}
%
\end{center}
\caption{The gaps $\log_{10}(1/|d_n^\textup{LP}(c)/\rho_c(\Delta_n^\textup{LP}(c))-1|)$ for $c=8$ (black), $c=12$ (red), $c=16$ (green), and $c=20$ (blue).}
\label{figure:gapsd}
\end{figure}

The unexpected periodicity modulo $8$ for scaling dimensions has a
counterpart for degeneracies, as shown in Figure~\ref{figure:maximumd}. Here
the multiples of $8$ are the best case for the accuracy of the $\textup{U}(1)^c$ Cardy
formula, while integers that are $4$ modulo $8$ are the worst case. Unlike
the case of scaling dimensions, the cusps in Figure~\ref{figure:maximumd} do
not seem to correspond to zero error in the limit as $n \to \infty$. Instead,
see Figure~\ref{figure:gapsd}. Perhaps this discrepancy indicates that
$\rho_c(\Delta)$ should be replaced with some better approximation.

Based on this data and the cases $c=4$ and $12$, we make the following
conjecture, which is a more precise analogue of
Conjecture~\ref{conjecture:Deltagaps}:

\begin{conjecture}
For $c > 2$,
\begin{equation}
\limsup_{n \to \infty} \frac{d_n^\textup{LP}(c)}{\rho_c(\Delta^\textup{LP}_n(c))} \le \frac{\zeta(c-1)}{\zeta(c)} = 1 + 2^{-c} + O(3^{-c})
\end{equation}
and
\begin{equation}
\liminf_{n \to \infty} \frac{d_n^\textup{LP}(c)}{\rho_c(\Delta_n^\textup{LP}(c))} \ge \frac{1}{\zeta(c)} = 1 - 2^{-c} + O(3^{-c}),
\end{equation}
with equality whenever $c$ is an integer that is congruent to $4$ modulo $8$.
\end{conjecture}

Equality provably holds for $c=4$ or $12$.

\subsection{Possible explanation of the high-energy spectrum}\label{ss:heuristic}

The formula for the upper portion of the high-energy spectrum, given by the
second line in \eqref{highspec}, was motivated by the following calculation.
It does not fully explain the formula, but it indicates why it is a
reasonable guess.

The counterpart of the Cardy formula in terms of functionals acting on the
crossing equation is the integral identity
\begin{equation}
\int_0^{\infty} d\Delta \frac{(2\pi)^c \Delta^{(c-1)}}{\Gamma(c)}e^{-2\pi \Delta}L_k^{(c-1)}(4\pi \Delta) = (-1)^{k}L_k^{(c-1)}(0),
\end{equation}
which is the evaluation at $0$ of the condition of being a radial Fourier
eigenfunction. We will use this identity to generate an exact solution of the
truncated crossing equations, which has too many states but is otherwise
suggestive of the optimal solution.

For small enough $k$, we can evaluate the integral exactly using the
Gauss-Laguerre quadrature formula
\begin{equation}
\int_0^\infty dx \, x^{c-1} e^{-x}  p(x) = \sum_{m=1}^n w_m p(x_m),
\end{equation}
which holds for all polynomials $p$ of degree at most $2n-1$ if we use the
roots of $L_n^{(c-1)}(x)$ as the quadrature nodes $x_m$ with weights
\begin{equation}
w_m = \frac{\Gamma(n+c)x_m}{n!(n+1)^2(L_{n+1}^{(c-1)}(x_m))^2}.
\end{equation}
If we take $n=2N$ and $p(x) = L_k^{(c-1)}(2x)$, we find that for $k \le 4N-1$,
\begin{equation}
\int_0^{\infty} d\Delta \frac{(2\pi)^c \Delta^{(c-1)}}{\Gamma(c)}e^{-2\pi \Delta}L_k^{(c-1)}(4\pi \Delta)
=
\sum_{m=1}^{2N} d_m L_k^{(c-1)}(4\pi \Delta_m),
\end{equation}
where $d_m = w_m/\Gamma(c)$ and the dimensions $\Delta_m$ satisfy $L_{2N}^{(c-1)}(2\pi \Delta_m) = 0$.
In other words, we have generated a solution to the equations
\begin{equation}
L_k^{(c-1)}(0) +  \sum_{m=1}^{2N} d_m L_k^{(c-1)}(4\pi \Delta_m) = 0
\end{equation}
for odd $k \leq 4N-1$. This solution has $2N+1$ states, while our numerical
method involves finding a solution to these truncated crossing equations with
only $N+1$ states, so this solution has no direct bearing on the bound.
However we observe numerically that the high energy spectrum approximately
agrees. The roots of the Laguerre polynomial $L_{2N}^{(c-1)}(2\pi \Delta)$
for large $N$ and $\Delta$ are given by the beta distribution described in
Section~\ref{ss:betacon}, which motivates its appearance.

\section{Spherical codes and implied kissing numbers}
\label{sec:kissing}

In sphere packing terms, the scaling dimensions $\Delta_n$ measure the
distances $\sqrt{2\Delta_n}$ between distinct sphere centers in the packing,
and the corresponding degeneracy is the average number of centers at that
distance from a given center. In particular, the first degeneracy $d_1$ is
the average number of tangencies for spheres in the packing, i.e., the
\emph{average kissing number} of the packing. De Laat, Oliveira, and
Vallentin \cite{LOV12} showed how to strengthen the linear programming bound
for sphere packing by incorporating geometric bounds for the degeneracies,
which go beyond the modular invariance of the partition function. In this
section, we will use this idea systematically to explore when the linear
programming bound can be sharp and how to improve on it. Along the way, we
will review the Kabatyanskii-Levenshtein bound.

\begin{figure}
\begin{center}
\begin{tikzpicture}[x=0.525cm,y=0.84cm]
\draw (0,-10)--(17.6,-10);
\draw (1.4,-0.5) node[right] {\small Implied kissing number};
\draw (1.4,-1) node[right] {\small Upper bound};
\draw (1.4,-1.5) node[right] {\small Record kissing number};
\fill (1.03,-0.5) circle (0.03cm);
\fill[red] (1.03,-1) circle (0.03cm);
\fill[green!75!black] (1.03,-1.5) circle (0.03cm);
\foreach \x in {0,...,6}
\draw (1.46666*\x*2,-10-0.065)--(1.466666*\x*2,-10+0.065);
\foreach \x in {0,4,...,24}
\draw ({1.46666*\x/2},-10-0.15)  node[below] {\small \x};
\draw (0,0)--(17.6,0)--(17.6,-10);
\foreach \y in {-6,...,0} \draw (-0.104,{\y*10/6})--(0.104,{\y*10/6});
\draw (8.8,-10-0.65)
node[below] {\small dimension}; \draw (-1.8,-5) node[rotate=90]
{\small $\log_{10}(\text{kissing number})$};
\draw (19.1,-5) node[rotate=-90]
{\small $\phantom{\log_{10}(\text{kissing number})}$};
\foreach \x in {0,1,...,6} \draw (-0.1,-10+\x*10/6) node[left] {\small \x};
\draw(0,-10)--(0,0); \draw
(0,0)--(17.6,0);
\fill[green!75!black] (0.733333,-9.49828) circle (0.03cm);
\fill[green!75!black] (1.46667,-8.70308) circle (0.03cm);
\fill[green!75!black] (2.20000,-8.20136) circle (0.03cm);
\fill[green!75!black] (2.93333,-7.69965) circle (0.03cm);
\fill[green!75!black] (3.66667,-7.32990) circle (0.03cm);
\fill[green!75!black] (4.40000,-6.90445) circle (0.03cm);
\fill[green!75!black] (5.13333,-6.49938) circle (0.03cm);
\fill[green!75!black] (5.86667,-6.03298) circle (0.03cm);
\fill[green!75!black] (6.60000,-5.85713) circle (0.03cm);
\fill[green!75!black] (7.33333,-5.50172) circle (0.03cm);
\fill[green!75!black] (8.06667,-5.39180) circle (0.03cm);
\fill[green!75!black] (8.80000,-5.12620) circle (0.03cm);
\fill[green!75!black] (9.53333,-4.89632) circle (0.03cm);
\fill[green!75!black] (10.2667,-4.65709) circle (0.03cm);
\fill[green!75!black] (11.0000,-4.31847) circle (0.03cm);
\fill[green!75!black] (11.7333,-3.94086) circle (0.03cm);
\fill[green!75!black] (12.4667,-3.78662) circle (0.03cm);
\fill[green!75!black] (13.2000,-3.55148) circle (0.03cm);
\fill[green!75!black] (13.9333,-3.28653) circle (0.03cm);
\fill[green!75!black] (14.6667,-2.93242) circle (0.03cm);
\fill[green!75!black] (15.4000,-2.59534) circle (0.03cm);
\fill[green!75!black] (16.1333,-2.16989) circle (0.03cm);
\fill[green!75!black] (16.8667,-1.71803) circle (0.03cm);
\fill[green!75!black] (17.6000,-1.17751) circle (0.03cm);
\fill[red] (0.733333,-9.49828) circle (0.03cm);
\fill[red] (1.46667,-8.70308) circle (0.03cm);
\fill[red] (2.20000,-8.20136) circle (0.03cm);
\fill[red] (2.93333,-7.69965) circle (0.03cm);
\fill[red] (3.66667,-7.26091) circle (0.03cm);
\fill[red] (4.40000,-6.84651) circle (0.03cm);
\fill[red] (5.13333,-6.45483) circle (0.03cm);
\fill[red] (5.86667,-6.03298) circle (0.03cm);
\fill[red] (6.60000,-5.73150) circle (0.03cm);
\fill[red] (7.33333,-5.42879) circle (0.03cm);
\fill[red] (8.06667,-5.10163) circle (0.03cm);
\fill[red] (8.80000,-4.77957) circle (0.03cm);
\fill[red] (9.53333,-4.47478) circle (0.03cm);
\fill[red] (10.2667,-4.16330) circle (0.03cm);
\fill[red] (11.0000,-3.85590) circle (0.03cm);
\fill[red] (11.7333,-3.55796) circle (0.03cm);
\fill[red] (12.4667,-3.26342) circle (0.03cm);
\fill[red] (13.2000,-2.97222) circle (0.03cm);
\fill[red] (13.9333,-2.68251) circle (0.03cm);
\fill[red] (14.6667,-2.39812) circle (0.03cm);
\fill[red] (15.4000,-2.11431) circle (0.03cm);
\fill[red] (16.1333,-1.81584) circle (0.03cm);
\fill[red] (16.8667,-1.51490) circle (0.03cm);
\fill (0.7333,-9.498) circle (0.03cm);
\fill (1.4667,-8.703) circle (0.03cm);
\fill (2.2000,-8.121) circle (0.03cm);
\fill (2.9333,-7.630) circle (0.03cm);
\fill (3.6667,-7.189) circle (0.03cm);
\fill (4.4000,-6.782) circle (0.03cm);
\fill (5.1333,-6.399) circle (0.03cm);
\fill (5.8667,-6.033) circle (0.03cm);
\fill (6.6000,-5.681) circle (0.03cm);
\fill (7.3333,-5.341) circle (0.03cm);
\fill (8.0667,-5.010) circle (0.03cm);
\fill (8.8000,-4.688) circle (0.03cm);
\fill (9.5333,-4.372) circle (0.03cm);
\fill (10.2667,-4.062) circle (0.03cm);
\fill (11.0000,-3.757) circle (0.03cm);
\fill (11.7333,-3.457) circle (0.03cm);
\fill (12.4667,-3.161) circle (0.03cm);
\fill (13.2000,-2.869) circle (0.03cm);
\fill (13.9333,-2.580) circle (0.03cm);
\fill (14.6667,-2.294) circle (0.03cm);
\fill (15.4000,-2.011) circle (0.03cm);
\fill (16.1333,-1.731) circle (0.03cm);
\fill (16.8667,-1.453) circle (0.03cm);
\fill (17.6000,-1.178) circle (0.03cm);
\end{tikzpicture}
\end{center}
\caption{The implied kissing number from the linear programming bound, compared with
the best upper bound known for the actual kissing number \cite{1609.05167} and the current record \cite{SPLAG}.}
\label{figure:impliedkissinglow}
\end{figure}
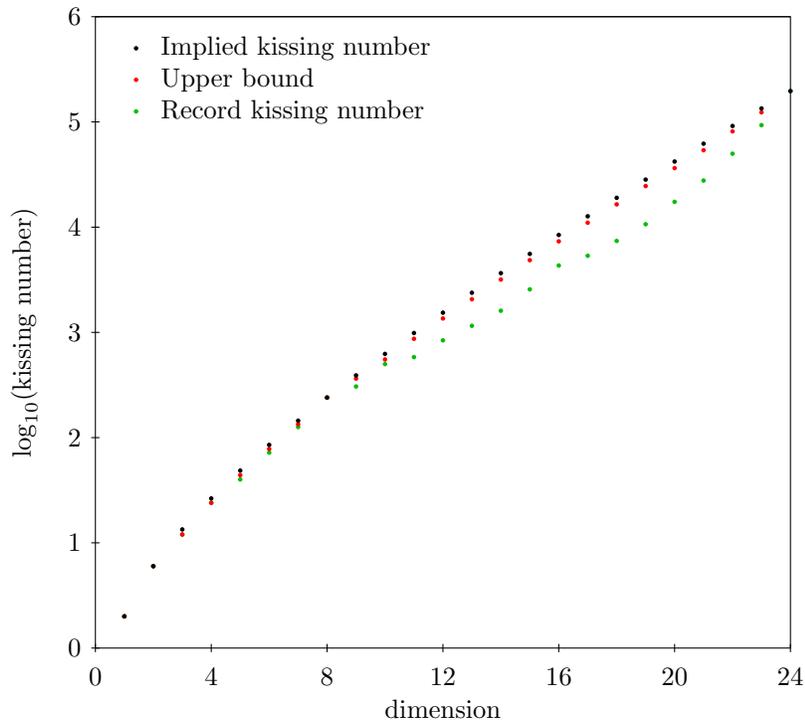

As motivation for this line of work, consider the \emph{implied kissing
number} $d_1^\textup{LP}(d/2)$, which is the average kissing number in a
hypothetical $d$-dimensional packing achieving the linear programming bound.
When $d = 1$, $2$, $8$, or $24$, we of course obtain the kissing number of
the optimal packing, but in general we obtain unrealistic numbers. For
example, when $d=4$ the implied kissing number is $26.43\dots$, which exceeds
Musin's optimal bound of $24$ for the four-dimensional kissing number
\cite{Musin}; it is therefore impossible for any packing to achieve the exact
linear programming bound in $\R^4$. As Figure~\ref{figure:impliedkissinglow}
shows, the implied kissing number is impossibly high for every $d \le 24$
except the known sharp cases. Within this range of dimensions, it perfectly
delineates which cases are sharp.

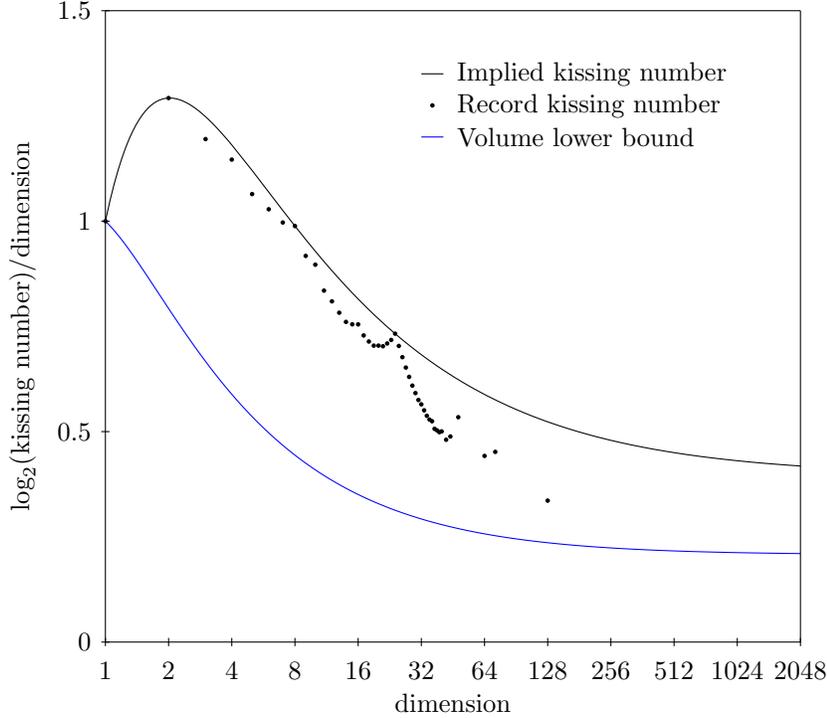
\begin{figure}
\begin{center}
\begin{tikzpicture}[x=0.525cm,y=0.84cm]
\draw[blue,join=round] (0.0000,-3.3333)--(0.1222,-3.4113)--(0.2444,-3.4978)--(0.3667,-3.5913)--(0.4889,-3.6908)--(0.6111,-3.7951)--(0.7333,-3.9032)--(0.8556,-4.0144)--(0.9778,-4.1277)--(1.1000,-4.2426)--(1.2222,-4.3584)--(1.3444,-4.4745)--(1.4667,-4.5907)--(1.5889,-4.7063)--(1.7111,-4.8211)--(1.8333,-4.9348)--(1.9556,-5.0471)--(2.0778,-5.1578)--(2.2000,-5.2667)--(2.3222,-5.3737)--(2.4444,-5.4785)--(2.5667,-5.5812)--(2.6889,-5.6815)--(2.8111,-5.7795)--(2.9333,-5.8752)--(3.0556,-5.9684)--(3.1778,-6.0592)--(3.3000,-6.1475)--(3.4222,-6.2334)--(3.5444,-6.3168)--(3.6667,-6.3978)--(3.7889,-6.4764)--(3.9111,-6.5527)--(4.0333,-6.6266)--(4.1556,-6.6982)--(4.2778,-6.7676)--(4.4000,-6.8347)--(4.5222,-6.8997)--(4.6444,-6.9626)--(4.7667,-7.0234)--(4.8889,-7.0822)--(5.0111,-7.1391)--(5.1333,-7.1940)--(5.2556,-7.2471)--(5.3778,-7.2983)--(5.5000,-7.3478)--(5.6222,-7.3956)--(5.7444,-7.4417)--(5.8667,-7.4862)--(5.9889,-7.5292)--(6.1111,-7.5706)--(6.2333,-7.6106)--(6.3556,-7.6491)--(6.4778,-7.6863)--(6.6000,-7.7221)--(6.7222,-7.7567)--(6.8444,-7.7900)--(6.9667,-7.8220)--(7.0889,-7.8530)--(7.2111,-7.8828)--(7.3333,-7.9115)--(7.4556,-7.9391)--(7.5778,-7.9657)--(7.7000,-7.9914)--(7.8222,-8.0161)--(7.9444,-8.0399)--(8.0667,-8.0627)--(8.1889,-8.0848)--(8.3111,-8.1060)--(8.4333,-8.1264)--(8.5556,-8.1460)--(8.6778,-8.1649)--(8.8000,-8.1831)--(8.9222,-8.2006)--(9.0444,-8.2174)--(9.1667,-8.2335)--(9.2889,-8.2491)--(9.4111,-8.2640)--(9.5333,-8.2784)--(9.6556,-8.2922)--(9.7778,-8.3055)--(9.9000,-8.3183)--(10.0222,-8.3305)--(10.1444,-8.3423)--(10.2667,-8.3537)--(10.3889,-8.3645)--(10.5111,-8.3750)--(10.6333,-8.3850)--(10.7556,-8.3947)--(10.8778,-8.4039)--(11.0000,-8.4128)--(11.1222,-8.4213)--(11.2444,-8.4295)--(11.3667,-8.4374)--(11.4889,-8.4449)--(11.6111,-8.4522)--(11.7333,-8.4591)--(11.8556,-8.4658)--(11.9778,-8.4722)--(12.1000,-8.4783)--(12.2222,-8.4842)--(12.3444,-8.4898)--(12.4667,-8.4953)--(12.5889,-8.5004)--(12.7111,-8.5054)--(12.8333,-8.5102)--(12.9556,-8.5148)--(13.0778,-8.5191)--(13.2000,-8.5233)--(13.3222,-8.5274)--(13.4444,-8.5312)--(13.5667,-8.5349)--(13.6889,-8.5385)--(13.8111,-8.5419)--(13.9333,-8.5451)--(14.0556,-8.5482)--(14.1778,-8.5512)--(14.3000,-8.5541)--(14.4222,-8.5568)--(14.5444,-8.5594)--(14.6667,-8.5619)--(14.7889,-8.5643)--(14.9111,-8.5666)--(15.0333,-8.5688)--(15.1556,-8.5709)--(15.2778,-8.5729)--(15.4000,-8.5749)--(15.5222,-8.5767)--(15.6444,-8.5785)--(15.7667,-8.5802)--(15.8889,-8.5818)--(16.0111,-8.5833)--(16.1333,-8.5848)--(16.2556,-8.5862)--(16.3778,-8.5876)--(16.5000,-8.5889)--(16.6222,-8.5901)--(16.7444,-8.5913)--(16.8667,-8.5924)--(16.9889,-8.5935)--(17.1111,-8.5945)--(17.2333,-8.5955)--(17.3556,-8.5965)--(17.4778,-8.5974)--(17.6000,-8.5982);
\draw (8.65,-1) node[right] {\small Implied kissing number};
\draw (8.65,-1.5) node[right] {\small Record kissing number};
\draw (8.65,-2) node[right] {\small Volume lower bound};
\draw (8,-1)--(8.56,-1);
\fill (8.28,-1.5) circle (0.03cm);
\draw[blue] (8,-2)--(8.56,-2);
\draw (0,-10)--(17.6,-10)--(17.6,0);
\foreach \x in {0,...,11}
\draw (1.6*\x,-10-0.065)--(1.6*\x,-10+0.065); \draw (0,-10-0.15) node[below] {\small $1$}; \draw
(1.6,-10-0.15) node[below] {\small $2$}; \draw (3.2,-10-0.15) node[below] {\small $4$}; \draw
(4.8,-10-0.15) node[below] {\small $8$}; \draw (6.4,-10-0.15) node[below] {\small $16$}; \draw
(8.0,-10-0.15) node[below] {\small $32$}; \draw (9.6,-10-0.15) node[below] {\small $64$}; \draw
(11.2,-10-0.15) node[below] {\small $128$}; \draw (12.8,-10-0.15) node[below] {\small $256$}; \draw
(14.4,-10-0.15) node[below] {\small $512$}; \draw (16,-10-0.15) node[below] {\small $1024$};  \draw (17.6,-10-0.15) node[below] {\small $2048$};
\draw (8.8,-10-0.65)
node[below] {\small dimension}; \draw (-2.1,-5) node[rotate=90]
{\small $\log_2(\text{kissing number})/\text{dimension}$};
\draw (19.4,-5) node[rotate=-90]
{\small $\phantom{\log_2(\text{kissing number})/\text{dimension}}$};
\draw (-0.104,-10)--(0.104,-10);
\draw (-0.104,-6.6667)--(0.104,-6.6667);
\draw (-0.104,-3.3333)--(0.104,-3.3333);
\draw (-0.104,-0)--(0.104,0);
\draw (-0.1,-10) node[left] {\small $0$};
\draw (-0.1,-6.6667) node[left] {\small $0.5$};
\draw (-0.1,-3.3333) node[left] {\small $1$};
\draw (-0.1,0) node[left] {\small $1.5$};
\draw(0,-10)--(0,0); \draw
(0,0)--(17.6,0);
\draw[join=round] (0.0000,-3.333)--(0.1126,-3.017)--(0.2200,-2.749)--(0.3226,-2.522)--(0.4209,-2.329)--(0.5151,-2.165)--(0.6056,-2.025)--(0.6927,-1.907)--(0.7767,-1.806)--(0.8577,-1.722)--(0.9359,-1.650)--(1.0116,-1.590)--(1.0849,-1.540)--(1.1559,-1.499)--(1.2249,-1.466)--(1.2918,-1.440)--(1.3568,-1.419)--(1.4200,-1.404)--(1.4816,-1.393)--(1.5416,-1.387)--(1.6000,-1.383)--(1.7126,-1.386)--(1.8200,-1.399)--(1.9226,-1.420)--(2.0209,-1.447)--(2.1151,-1.479)--(2.2056,-1.514)--(2.2927,-1.553)--(2.3767,-1.593)--(2.4577,-1.636)--(2.5359,-1.679)--(2.6116,-1.723)--(2.6849,-1.768)--(2.7559,-1.813)--(2.8249,-1.859)--(2.8918,-1.904)--(2.9568,-1.949)--(3.0200,-1.994)--(3.0816,-2.038)--(3.1416,-2.082)--(3.2000,-2.126)--(3.7151,-2.530)--(4.1359,-2.874)--(4.4918,-3.164)--(4.8000,-3.411)--(5.0719,-3.624)--(5.3151,-3.810)--(5.5351,-3.973)--(5.7359,-4.118)--(5.9207,-4.247)--(6.0918,-4.364)--(6.2510,-4.470)--(6.4000,-4.566)--(6.5399,-4.654)--(6.6719,-4.736)--(6.7967,-4.811)--(6.9151,-4.880)--(7.0277,-4.945)--(7.1351,-5.006)--(7.2377,-5.062)--(7.3359,-5.115)--(7.4302,-5.165)--(7.5207,-5.213)--(7.6078,-5.257)--(7.6918,-5.299)--(7.7728,-5.339)--(7.8510,-5.377)--(7.9267,-5.414)--(8.0000,-5.448)--(8.0710,-5.481)--(8.1399,-5.512)--(8.2069,-5.543)--(8.2719,-5.571)--(8.3351,-5.599)--(8.3967,-5.626)--(8.4566,-5.651)--(8.5151,-5.676)--(8.5721,-5.699)--(8.6277,-5.722)--(8.6820,-5.744)--(8.7351,-5.765)--(8.7870,-5.786)--(8.8377,-5.806)--(8.8873,-5.825)--(8.9359,-5.843)--(8.9835,-5.861)--(9.0302,-5.879)--(9.0759,-5.896)--(9.1207,-5.912)--(9.1647,-5.928)--(9.2078,-5.944)--(9.2502,-5.959)--(9.2918,-5.973)--(9.3326,-5.988)--(9.3728,-6.001)--(9.4122,-6.015)--(9.4510,-6.028)--(9.4892,-6.041)--(9.5267,-6.053)--(9.5636,-6.066)--(9.6000,-6.078)--(9.6358,-6.089)--(9.6710,-6.101)--(9.7057,-6.112)--(9.7399,-6.122)--(9.7736,-6.133)--(9.8069,-6.143)--(9.8396,-6.154)--(9.8719,-6.163)--(9.9037,-6.173)--(9.9351,-6.183)--(9.9661,-6.192)--(9.9967,-6.201)--(10.0269,-6.210)--(10.0566,-6.219)--(10.0860,-6.227)--(10.1151,-6.236)--(10.1438,-6.244)--(10.1721,-6.252)--(10.2001,-6.260)--(10.2277,-6.268)--(10.2550,-6.276)--(10.2820,-6.283)--(10.3087,-6.290)--(10.3351,-6.298)--(10.3612,-6.305)--(10.3870,-6.312)--(10.4125,-6.319)--(10.4377,-6.326)--(10.4627,-6.332)--(10.4873,-6.339)--(10.5118,-6.345)--(10.5359,-6.351)--(10.5599,-6.358)--(10.5835,-6.364)--(10.6070,-6.370)--(10.6302,-6.376)--(10.6531,-6.382)--(10.6759,-6.387)--(10.6984,-6.393)--(10.7207,-6.399)--(10.7428,-6.404)--(10.7647,-6.410)--(10.7863,-6.415)--(10.8078,-6.420)--(10.8291,-6.425)--(10.8502,-6.431)--(10.8711,-6.436)--(10.8918,-6.441)--(10.9123,-6.445)--(10.9326,-6.450)--(10.9528,-6.455)--(10.9728,-6.460)--(10.9926,-6.464)--(11.0122,-6.469)--(11.0317,-6.473)--(11.0510,-6.478)--(11.0702,-6.482)--(11.0892,-6.487)--(11.1080,-6.491)--(11.1267,-6.495)--(11.1453,-6.499)--(11.1636,-6.504)--(11.1819,-6.508)--(11.2000,-6.512)--(11.4719,-6.570)--(11.7151,-6.619)--(11.9351,-6.661)--(12.1359,-6.697)--(12.3207,-6.729)--(12.4918,-6.757)--(12.6510,-6.782)--(12.8000,-6.805)--(13.0719,-6.844)--(13.3151,-6.877)--(13.5351,-6.904)--(13.7359,-6.929)--(13.9207,-6.950)--(14.0918,-6.968)--(14.2510,-6.985)--(14.4000,-7.000)--(14.6719,-7.026)--(14.9151,-7.047)--(15.1351,-7.066)--(15.3359,-7.081)--(15.5207,-7.095)--(15.6918,-7.107)--(15.8510,-7.118)--(16.0000,-7.128)--(16.2719,-7.145)--(16.5151,-7.159)--(16.7351,-7.171)--(16.9359,-7.181)--(17.1207,-7.190)--(17.2918,-7.198)--(17.4510,-7.205)--(17.6000,-7.212);
\fill (0.0000,-3.3333) circle (0.03cm);
\fill (1.6000,-1.3835) circle (0.03cm);
\fill (2.5359,-2.0334) circle (0.03cm);
\fill (3.2000,-2.3584) circle (0.03cm);
\fill (3.7151,-2.9041) circle (0.03cm);
\fill (4.1359,-3.1445) circle (0.03cm);
\fill (4.4918,-3.3550) circle (0.03cm);
\fill (4.8000,-3.4109) circle (0.03cm);
\fill (5.0719,-3.8834) circle (0.03cm);
\fill (5.3151,-4.0228) circle (0.03cm);
\fill (5.5351,-4.4334) circle (0.03cm);
\fill (5.7359,-4.6032) circle (0.03cm);
\fill (5.9207,-4.7834) circle (0.03cm);
\fill (6.0918,-4.9289) circle (0.03cm);
\fill (6.2510,-4.9670) circle (0.03cm);
\fill (6.4000,-4.9680) circle (0.03cm);
\fill (6.5399,-5.1434) circle (0.03cm);
\fill (6.6719,-5.2397) circle (0.03cm);
\fill (6.7967,-5.3049) circle (0.03cm);
\fill (6.9151,-5.3044) circle (0.03cm);
\fill (7.0277,-5.3147) circle (0.03cm);
\fill (7.1351,-5.2707) circle (0.03cm);
\fill (7.2377,-5.2153) circle (0.03cm);
\fill (7.3359,-5.1154) circle (0.03cm);
\fill (7.4302,-5.3098) circle (0.03cm);
\fill (7.5207,-5.4875) circle (0.03cm);
\fill (7.6078,-5.6520) circle (0.03cm);
\fill (7.6918,-5.7998) circle (0.03cm);
\fill (7.7728,-5.9384) circle (0.03cm);
\fill (7.8510,-6.0562) circle (0.03cm);
\fill (7.9267,-6.1658) circle (0.03cm);
\fill (8.0000,-6.2345) circle (0.03cm);
\fill (8.0710,-6.3296) circle (0.03cm);
\fill (8.1399,-6.4159) circle (0.03cm);
\fill (8.2069,-6.4761) circle (0.03cm);
\fill (8.2719,-6.5025) circle (0.03cm);
\fill (8.3351,-6.6227) circle (0.03cm);
\fill (8.3967,-6.6470) circle (0.03cm);
\fill (8.4566,-6.6759) circle (0.03cm);
\fill (8.5151,-6.6633) circle (0.03cm);
\fill (8.6277,-6.7951) circle (0.03cm);
\fill (8.7351,-6.7437) circle (0.03cm);
\fill (8.9359,-6.4384) circle (0.03cm);
\fill (9.6000,-7.0515) circle (0.03cm);
\fill (9.8719,-6.9876) circle (0.03cm);
\fill (11.2000,-7.7598) circle (0.03cm);
\end{tikzpicture}
\end{center}
\caption{The implied kissing number in high dimensions,
compared with the record kissing number \cite{SPLAG, 1608.07270} and the excluded volume lower bound for the kissing number from \cite{Wyner}.}
\label{figure:impliedkissinghigh}
\end{figure}

Figure~\ref{figure:impliedkissinghigh} shows how the implied kissing number
grows in high dimensions. Comparing it with upper bounds turns out to be
surprisingly subtle, and we will do so in
Figure~\ref{figure:impliedkissingratio} once we have explained more about the
needed bounds. Aside from low dimensions,
Figure~\ref{figure:impliedkissinghigh} looks similar to
Figure~\ref{figure:loglog}, and that is not a coincidence:
Table~\ref{table:LPIK} indicates that the implied kissing number is
$2^{d+o(d)}$ times the linear programming bound as $d \to \infty$. This
relationship is easily explained using the $\textup{U}(1)^c$ Cardy formula, because
the sphere packing density is $\rho_c(\Delta_1) \Delta_1/(c 4^c)$ in terms of
the spectral gap $\Delta_1$ and $c=d/2$. We can approximate the implied
kissing number $d_1^\textup{LP}(c)$ by $\rho_c(\Delta_1^\textup{LP}(c))$
using the Cardy formula; because of the size of $\Delta_1^\textup{LP}(c)$ we
expect some error, but the error factor should be subexponential in $c$, in fact roughly
$\Delta_2^\textup{LP}(c)-\Delta_1^\textup{LP}(c)$. We conclude that the linear
programming bound for the packing density is $d_1^\textup{LP}(c)/(4+o(1))^c$
as $c \to \infty$, which is the desired relationship.

\begin{table}
\caption{Numerical comparison of the linear programming bound for the sphere packing density in $\R^d$ (denoted LP) with the
implied kissing number (denoted IK), both in the form $\log_2(\cdot)/d$.}
\label{table:LPIK}
\begin{center}
\begin{tabular}{rccc}
\toprule
$d$ & $\phantom{-}\textup{LP}$ & $\phantom{-}\textup{IK}$ & $\phantom{-}1+\textup{LP}-\textup{IK}$\\
\midrule
$1$ & $\phantom{-}0.00000$ & $\phantom{-}1.00000$ & $\phantom{-}0.00000$\\
$2$ & $-0.07049$ & $\phantom{-}1.29248$ & $-0.36297$\\
$4$ & $-0.15665$ & $\phantom{-}1.18103$ & $-0.33768$\\
$8$ & $-0.24737$ & $\phantom{-}0.98836$ & $-0.23573$\\
$16$ & $-0.33192$ & $\phantom{-}0.81510$ & $-0.14702$\\
$32$ & $-0.40382$ & $\phantom{-}0.68279$ & $-0.08661$\\
$64$ & $-0.46101$ & $\phantom{-}0.58837$ & $-0.04937$\\
$128$ & $-0.50432$ & $\phantom{-}0.52325$ & $-0.02757$\\
$256$ & $-0.53589$ & $\phantom{-}0.47929$ & $-0.01518$\\
$512$ & $-0.55824$ & $\phantom{-}0.45003$ & $-0.00827$\\
$1024$ & $-0.57370$ & $\phantom{-}0.43077$ & $-0.00447$\\
$2048$ & $-0.58418$ & $\phantom{-}0.41822$ & $-0.00240$\\
$4096$ & $-0.59120$ & $\phantom{-}0.41009$ & $-0.00128$\\
\midrule
$\infty$ & $-0.6044\phantom{0}$ & $\phantom{-}0.3956\phantom{0}$ & $\phantom{-}0.00000$\\
\bottomrule
\end{tabular}
\end{center}
\end{table}

To prove bounds for kissing numbers, the relevant optimization problem is the
spherical code problem, which is a compact analogue of the sphere packing
problem. In dimension $d$ and with minimal angle $\theta$ this problem asks
how large a subset $\mathcal{C}$ of the unit sphere in $\R^d$ can be if
$\langle x, y\rangle \geq s$ for all distinct $x,y \in \mathcal{C}$, where $s
= \cos\theta$. In other words, all points in $\mathcal{C}$ must be separated
by at least a distance of $\theta$ along the surface of the sphere, and so
$\mathcal{C}$ yields a packing with spherical caps of radius $\theta/2$. Such
a set is called a \emph{spherical code} with minimal angle $\theta$. The
kissing problem amounts to the case $\theta = \pi/3$; note that here we are
considering the kissing problem for a single sphere, rather than averaged
over a packing in Euclidean space.

Let $A(d,s)$ be the largest possible size of such a code. Delsarte, Goethals,
and Seidel \cite{DGS77} introduced a linear programming bound for $A(d,s)$,
which Kabatyanskii and Levenshtein \cite{KL78} used to obtain the best sphere
packing density bounds known in Euclidean space.

After briefly reviewing this linear programming bound, we will discuss two
applications of spherical codes to the sphere packing problem, followed by a
new average kissing bound. First we discuss the Kabatyanskii-Levenshtein
bound, using the approach from \cite{Lev98}. Then we discuss a strengthening
of the linear programming bound for the sphere packing problem through bounds
for spherical codes, and its implications for when the bound can be tight. We
conclude with a linear programming bound for the average kissing number.

The results of this section rely on the geometry of the sphere packing
problem and do not appear to have any direct application to CFTs. On the
other hand, conceptually, upper bounds on the average kissing number are
similar to upper bounds on operator-product coefficients often considered in
the bootstrap literature (e.g., \cite{Hellerman:2010qd,El-Showk:2014dwa}). It
is also our hope that the methods in this section (in particular, the
application of the Christoffel-Darboux formula to produce positive auxiliary
functions) will inspire new analytic approaches to the conformal bootstrap.

\subsection{The linear programming bound}

The analogue of the radial Fourier transform on the surface of a sphere is
the expansion in terms of zonal spherical harmonics, which uses the following
orthogonal polynomials.\footnote{One key conceptual difference between
spheres and Euclidean space is that $\R^d$ is its own Pontryagin dual, which
means Fourier eigenfunctions make sense, while that concept does not apply to
spheres.} Let $d$ be the dimension of the spherical code, and let $a$ and $b$
be nonnegative integers. (We can take $a=b=0$ for now, but we will make use
of $a$ and $b$ in the Kabatyanskii-Levenshtein bound.)

Let
\begin{equation}
w^{a,b}(t) = \frac{\Gamma(a+b+d-1)}{2^{a+b+d-2}\Gamma(a+\frac{d-1}{2})\Gamma(b+\frac{d-1}{2})}  (1-t)^a (1+t)^b (1-t^2)^{(d-3)/2},
\end{equation}
where the normalization is chosen so that $\int_{-1}^1 dt \, w^{a,b}(t) = 1$.
Define the orthogonal polynomials $\smash{Q_i^{a,b}(t)}$ with
$\deg(\smash{Q_i^{a,b}}) = i$ and positive leading coefficients by
\begin{equation}
\int_{-1}^1dt\,Q_i^{a,b}(t) Q_j^{a,b}(t) w^{a,b}(t) =  \delta_{i,j}
\end{equation}
for $i,j \ge 0$. Up to normalization, these polynomials are the Jacobi
polynomials with parameters $(d-3)/2+a$ and $(d-3)/2+b$.

The parameters $a=b=0$ are particularly important, and the polynomials $Q_i
:= Q_i^{0,0}$ are known as the ultraspherical polynomials in dimension $d$.
For a continuous function $f \colon [-1,1] \to \R$,
\begin{equation}
f_0 := \int_{-1}^1 dt \, f(t) w^{0,0}(t) = \int_{S^{d-1}} d\mu(x) \, f(\langle x, e\rangle),
 \end{equation}
where $e \in S^{d-1}$ is an arbitrary point and $\mu$ is the surface measure
on the sphere $S^{d-1}$, normalized so that $\mu(S^{d-1})=1$. Therefore the
polynomials $Q_i$ are orthogonal if we think of them as zonal functions on
$S^{d-1}$, i.e., functions $x \mapsto Q_i(\langle x,e \rangle)$ invariant
under the stabilizer subgroup of $O(d)$ with respect to $e$. Moreover, these
polynomials are of positive type: for all finite $\mathcal{C} \subseteq
S^{d-1}$ and all coefficients $c_x \in \R$ for $x \in \mathcal{C}$,
\begin{equation} \label{eq:positivetype}
\sum_{x,y \in \mathcal{C}} c_x c_y Q_i(\langle x, y\rangle) \geq 0.
\end{equation}
This inequality follows from the addition formula
\begin{equation}
\frac{Q_i(\langle x,y\rangle)}{Q_i(1)} = \frac{1}{r_i}\sum_{j=1}^{r_i} v_{i,j}(x) v_{i,j}(y),
\end{equation}
where the functions $v_{i,j} \colon \R^d \to \R$ for $j=1,\dots,r_i$ are an
orthonormal basis of the spherical harmonics of degree $i$ (see
\cite[Theorem~9.6.3]{AAR}); specifically, the addition formula shows that the
left side of \eqref{eq:positivetype} is a square and therefore nonnegative.

The linear programming bound for spherical codes converts an auxiliary
function into an upper bound on the greatest size of a spherical code:

\begin{theorem}[\cite{DGS77}]\label{theorem:delsarte}
Let $f \colon [-1,1] \to \R$ be a continuous function and $s \in [-1,1]$. If
$f$ is of positive type as a zonal function on $S^{d-1}$, $f(t) \leq 0$ for
$t \in [-1,s]$, and $f_0 \neq 0$, then $A(d,s)$ is at most $f(1)/f_0$.
\end{theorem}

By Schoenberg's theorem \cite{Schoenberg}, every continuous function $f
\colon [-1,1] \to \R$ of positive type is of the form $ f(x) =
\sum_{n=0}^\infty f_n Q_n(x) $ with $f_n \ge 0$, where convergence is uniform
and absolute. Therefore the linear programming bound can be approximated
arbitrarily well by the minimum of $f(1)/f_0$ over functions $f$ of the form
\begin{equation}
f(t) = \sum_{n=0}^N f_n Q_n(t)
\end{equation}
with $f_0 >0$, $f_1,\dots,f_N \geq 0$, and $f(t) \leq 0$ for $t \in [-1,s]$.

One can optimize this bound numerically for given $d$ and $N$ using
semidefinite programming. Specifically, we can create a semidefinite program
where $f_0,\dots,f_N$ are one-by-one positive semidefinite matrices, and the
inequality constraint is modeled as
\begin{equation}
f(t) = \begin{cases} -v_0(t)^{\sf T} X v_0(t) - (t+1)(s - t) v_{1}(t)^{\sf T} Y v_{1}(t) & \text{if $N$ is even, and}\\
-(t+1) v_0(t)^{\sf T} X v_0(t) - (s - t) v_0(t)^{\sf T} Y v_0(t) & \text{if $N$ is odd,}\\
\end{cases}
\end{equation}
where $X$ and $Y$ are positive semidefinite matrices and
\begin{equation}
v_k(t) = (Q_0(t),\dots,Q_{\lfloor N/2 \rfloor - k}(t)).
\end{equation}
If we additionally require $f_0 = 1$, then the objective is $f(1)$, which
means we are minimizing a linear functional over positive semidefinite
matrices with linear constraints. This semidefinite program can be solved
numerically on a computer.

\begin{figure}
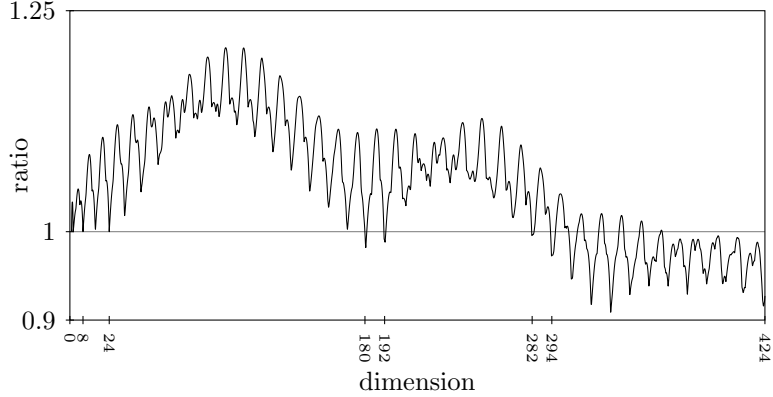

\begin{center}

\end{center}
\caption{The ratio of the implied kissing number to the linear programming
bound for the kissing number. It dips below $1$ at dimensions 180--181, 192, 282--283, 294--296,
and many beyond that.}
\label{figure:impliedkissingratio}
\end{figure}

Shtrom \cite{Shtrom} computed the exact linear programming bound for the
kissing number $A(d,1/2)$ when $d \le 146$, by determining the optimal value
of $N$, beyond which there is no improvement. We have extended these
computations to $d \le 424$ using $N=95$, which appears to be high enough in
this range of dimensions and in any case should closely approximate the
optimum. Figure~\ref{figure:impliedkissingratio} shows the ratio of the
implied kissing number to this upper bound. They are very close to each other
in size, but their precise ratio seems difficult to predict, and we do not
know what happens as $d \to \infty$. No sphere packing can match the linear
programming bound for density when this ratio is strictly greater than $1$.
Our initial hope was that this condition would rule out every dimension
$d>24$, but it does not. Instead, further progress may depend on more
powerful bounds for the kissing number, such as semidefinite programming
bounds \cite{BV,1609.05167}.

This approach can rule out exact equality in the linear programming bound for
packing density, but it does not give a quantitative improvement. We will
return to the problem of improving the density bound, once we explain
Levenshtein's universal bound and the Kabatyanskii-Levenshtein bound.

\subsection{Levenshtein's universal bound} \label{ss:lub}

Levenshtein's universal bound is the best bound that has been analytically
derived from Theorem~\ref{theorem:delsarte}; our explanation follows
Section~5.4 in \cite{Lev98}. Let $t_k^{a,b} \in [-1,1)$ be the largest root
of $Q_k^{a,b}(t)$, where we set $t_0^{1,1} = -1$. One can show that
\begin{equation}
t_{k-1}^{1,1} < t_k^{1,0} < t_k^{1,1}
\end{equation}
for all $k \ge 1$ (see (5.89) in \cite{Lev98}).  For $s \in [-1,1)$, define the function
\begin{equation}
f^{(s)}(t) = \begin{cases}(t-s) \big(K_{k-1}^{1,0}(t,s)\big)^2 & \text{if $t_{k-1}^{1,1} \le s< t_k^{1,0}$, and}\\
 (t+1)(t-s) \big(K_{k-1}^{1,1}(t,s)\big)^2& \text{if $t_k^{1,0} \le s< t_k^{1,1}$,}\end{cases}
\end{equation}
where
\begin{equation}
K_{k-1}^{a,b}(t,s) = \sum_{i=0}^{k-1} Q_i^{a,b}(t) Q_i^{a,b}(s).
\end{equation}

This function $f^{(s)}$ will prove a bound in Theorem~\ref{theorem:delsarte},
once we verify the hypotheses of the theorem (see Theorem~5.42 in
\cite{Lev98} for more details than we give here). By construction,
$f^{(s)}(t) \leq 0$ for $-1 \leq t \leq s$. To show $f^{(s)}(t)$ is of
positive type for $t_{k-1}^{1,1} \leq s < t_k^{1,0}$, one can check that for
such $s$,
\begin{equation}
Q_0^{1,0}(s) > 0, \quad \dots, \quad Q_{k-1}^{1,0}(s) >0, \quad\text{and}\quad {Q_k^{1,0}(s)} < 0.
\end{equation}
Since $Q_k^{1,0}(t)$ is of positive type, these inequalities show
$K_{k-1}^{1,0}(t,s)$ is of positive type as a function of $t$. Moreover, they
also show $(t-s)K_{k-1}^{1,0}(t,s)$ is of positive type as a function of $t$
by using the Christoffel-Darboux formula, which says
\begin{equation}
(t-s)K_{k-1}^{1,0}(t,s) = (c_{k-1}/c_k) \big(Q_k^{1,0}(t)Q_{k-1}^{1,0}(s) - Q_k^{1,0}(s)Q_{k-1}^{1,0}(t)\big),
\end{equation}
where $c_k>0$ is the leading coefficient of $Q_k^{1,0}$. Since the product of
functions of positive type is also of positive type, it follows that
$f^{(s)}(t)$ is of positive type for $t_{k-1}^{1,1} \leq s < t_k^{1,0}$. By
using the  property that
\begin{equation}
(t+1)Q_i^{1,1}(t)Q_j^{1,1}(t)
\end{equation}
is of positive type, this argument can be extended to show
$\smash{f^{(s)}(t)}$ is of positive type for all $s \in [-1,1]$.

Thus, these function can be used as auxiliary functions for
Theorem~\ref{theorem:delsarte}, which gives Levenshtein's universal bound for
the sphere. In terms of the normalized polynomials $\mathcal{Q}_i(s) =
Q_i(s)/Q_i(1)$, we arrive at the bound
\begin{equation}
A(d,s) \leq  \begin{cases}
\binom{k+d-3}{k-1}\left(\frac{2k+d-3}{d-1} - \frac{\mathcal{Q}_{k-1}(s)-\mathcal{Q}_k(s)}{(1-s)\mathcal{Q}_k(s)}\right) & \text{if $t_{k-1}^{1,1} \le s< t_k^{1,0}$, and}\\
\\ 
\binom{k+d-2}{k}\left(\frac{2k+d-1}{d-1} - \frac{(1+s)(\mathcal{Q}_k(s)-\mathcal{Q}_{k+1}(s))}{(1-s)(\mathcal{Q}_k(s)+\mathcal{Q}_{k+1}(s))}\right) & \text{if $t_k^{1,0} \le s< t_k^{1,1}$}.\end{cases}
\end{equation}
In certain cases this bound is the best that can be obtained from
Theorem~\ref{theorem:delsarte}, but in general it does not fully optimize the
choice of auxiliary function.

\subsection{The Kabatyanskii-Levenshtein bound} \label{ss:kl}

The following geometric inequality shows how the sphere packing density
$\Delta_{\R^d}$ in $\R^d$ can be bounded using $A(d,s)$:
\begin{equation} \label{eq:Deltabd}
\Delta_{\R^d} \leq \min_{\pi/3 \leq \theta \leq \pi} \sin^d(\theta/2) A(d,\cos\theta) = \min_{-1 \leq s \leq 1/2} \left(\frac{1-s}{2}\right)^{d/2} A(d,s)
\end{equation}
(see (6.9) in \cite{Lev98} or Proposition~2.1 in \cite{CZ}).

To obtain a good bound for fixed $d$, this inequality can be combined with
Levenshtein's universal bound for $A(d,s)$, where the best value of $s$ can
be found by optimizing a piecewise differentiable function. The resulting
bound is the one shown as the Kabatyanskii-Levenshtein bound in
Figure~\ref{figure:loglog}, although Kabatyanskii and Levenshtein \cite{KL78}
used a slightly worse bound for $A(d,s)$ as well as for $\Delta_{\R^d}$ in
terms of $A(d,s)$.

To obtain an asymptotic bound as $d \to \infty$, we can use the inequality
\begin{equation}
A(d,s) \leq A(d,t_k^{1,1}) \leq \frac{f^{(t_k^{1,1})}(1)}{f^{(t_k^{1,1})}_0} = 2 \binom{d+k-1}{k}
\end{equation}
when $s \le t_k^{1,1}$ (see (6.13) in \cite{Lev98}). If $k,d \to \infty$ with
$k/d \to \alpha$, then
\begin{equation}
t_k^{1,1} \to \frac{2\sqrt{\alpha(1+\alpha)}}{2\alpha+1}
\end{equation}
by Corollary~5.17 in \cite{Lev98}. For $\theta < \pi/2$, taking
\begin{equation}
\alpha = \frac{1-\sin \theta}{2 \sin \theta}
\end{equation}
ensures that $t_k^{1,1} \to \cos \theta$, and applying Stirling's formula
shows that
\begin{equation}
\frac{1}{d} \log_2 A(d,\cos \theta) \le \frac{1+\sin\theta}{2\sin\theta}\log_2\frac{1+\sin\theta}{2\sin\theta} - \frac{1-\sin\theta}{2\sin\theta}\log_2\frac{1-\sin\theta}{2\sin\theta} + o(1)
\end{equation}
as $d \to \infty$. It now follows from \eqref{eq:Deltabd} that the sphere
packing density is at most $2^{-(\kappa+o(1))d}$, where
\begin{equation}
\kappa = -\log_2\sin\frac{\theta}{2} - \frac{1+\sin\theta}{2\sin\theta}\log_2\frac{1+\sin\theta}{2\sin\theta} + \frac{1-\sin\theta}{2\sin\theta}\log_2\frac{1-\sin\theta}{2\sin\theta}.
\end{equation}
Optimizing for the best choice of $\theta$ between $\pi/3$ and $\pi/2$ yields
the root $\theta = 1.09951240\dots$ of
\begin{equation}
\sec \theta + \tan \theta = e^{(\tan \theta + \sin \theta)/2},
\end{equation}
at which point we obtain the Kabatyanskii-Levenshtein bound $\kappa =
0.59905576\dots$.

Cohn and Zhao \cite{CZ} gave a general transformation showing that any bound
obtained from Theorem~\ref{theorem:delsarte} and \eqref{eq:Deltabd} can also
be obtained directly from the Euclidean linear programming bound. Thus, there
is no need to use spherical codes to obtain the Kabatyanskii-Levenshtein
bound. However, the transformation sheds little additional light on this
bound, and it is difficult to see how someone might think of it without using
spherical codes.

\subsection{Implied kissing numbers}

One can strengthen the linear programming bound by taking into account
constraints on spherical codes. The following relaxation of the kissing
number will prove useful in doing so.  Let $\mathcal{C}$ be the set of sphere
centers in a packing.  For $x \in \mathcal{C}$ and $r\ge 0$, let
\begin{equation}
N_x(r) = \#\{ y \in \mathcal{C}: 0 < |x-y| \le r\},
\end{equation}
and let $N(r)$ be the average of $N_x(r)$ over $x \in \mathcal{C}$ (we can
restrict our attention to periodic packings, so that this average is well
defined). If $r_0$ is the minimal distance in $\mathcal{C}$, then $N(r_0)$ is
the average kissing number of $\mathcal{C}$, and for $t>0$ we define the
\emph{average $t$-neighbor number} to be $N(tr_0)$. In other words, the
average kissing number is the average $1$-neighbor number.

The following strengthening of the linear programming bound is a special case
of Theorem~1.4 in \cite{LOV12}, where it was used to give improved bounds in
dimensions $4$ through $7$ and $9$. The proof in \cite{LOV12} amounts to
retaining more terms in the Poisson summation argument from \cite{CE}.

\begin{theorem}[\cite{LOV12}]\label{theorem:LOV}
Let $g \colon \R^d \to \R$ be a radial Schwartz function, and suppose that
$g$ satisfies the following inequalities for some $\eta\ge 0$ and $s>r>0$:
\begin{enumerate}
\item[\textup{(1)}] $g(0)>0$ and $\widehat{g}(0)>0$,
\item[\textup{(2)}] $g(x) \le \eta g(0)$ for $r \le |x| \le s$,
\item[\textup{(3)}] $g(x) \le 0$ for $|x| \ge s$, and
\item[\textup{(4)}] $\widehat{g}(y) \ge 0$ for all $y$.
\end{enumerate}
Suppose furthermore that every sphere packing in $\R^d$ has average
$s/r$-neighbor number at most $M$. Then the sphere packing density in $\R^d$
is at most
\begin{equation}
\frac{\pi^{d/2}}{(d/2)!}
\left(\frac{r}{2}\right)^d
\frac{g(0)}{\widehat{g}(0)}
(1+\eta M).
\end{equation}
\end{theorem}

The linear programming bound is equivalent to taking $\eta=0$. The extra
flexibility of being able to choose $g(0)$ and $\widehat{g}(0)$ is
irrelevant, because we can rescale $g$ and its input variable, but it will be
convenient below.

When the implied kissing number is impossibly large, we can apply
Theorem~\ref{theorem:LOV} to improve on the linear programming bound from
Theorem~\ref{theorem:LP} as follows. Suppose $h$ is a Schwartz function
satisfying the hypotheses of the linear programming bound with radius $r$.
One can check by a rescaling argument that the average kissing number of any
sphere packing that achieves this bound must be
\begin{equation}
K = -\frac{d}{r h'(r)},
\end{equation}
where $h'(r)$ denotes the radial derivative at radius $r$ (see Lemma~5.1 in
\cite{CM}). Thus, if $h$ is the optimal auxiliary function in the bound, then
$K$ must be the implied kissing number.

Suppose furthermore that for some $t>1$ we can prove an upper bound $B$ for
the average $t$-neighbor number in every packing in $\R^d$ with $B<K$. For
example, $B$ could be an upper bound for $A(d,\cos \theta)$ for some $\theta
> \pi/3$, which can be arbitrarily close to $\pi/3$.

Given such a bound $B<K$, Theorem~\ref{theorem:LOV} proves a strictly
stronger density bound than Theorem~\ref{theorem:LP} using $h$, as follows.
Let $s= (1+\varepsilon)r$ for some small $\varepsilon>0$, and define $g$ by
$g(x) = h(x/(1+\varepsilon))$.  Then $g$ satisfies the hypotheses of
Theorem~\ref{theorem:LOV} with $\eta = -r h'(r) \varepsilon +
O(\varepsilon^2)$, $g(0)=1$, and
\begin{equation}
\widehat{g}(0) = (1+\varepsilon)^d = 1+d\varepsilon + O(\varepsilon^2).
\end{equation}
If $\varepsilon$ is small enough, then we can take $M=B$ in
Theorem~\ref{theorem:LOV}, and
\begin{equation}
\frac{g(0)}{\widehat{g}(0)}
(1+\eta M) = \frac{1-r h'(r) B \varepsilon + O(\varepsilon^2)}{1+d\varepsilon+O(\varepsilon^2)}
= 1 + d (B/K-1) \varepsilon + O(\varepsilon^2),
\end{equation}
which is less than $1$ when $\varepsilon$ is sufficiently small, because
$B<K$. The improvement here is not large, but the resulting density bound is
strictly better than that from Theorem~\ref{theorem:LP}. Thus,
Figure~\ref{figure:impliedkissingratio} shows that we can extend the improved
density bound from \cite{LOV12} to $10 \le d \le 23$, $25 \le d \le 179$, and
a number of larger values of $d$. However, we do not know what happens as $d
\to \infty$.

\subsection{Bounds for the average kissing number}
\label{subsec:avgkiss}

The implied kissing number has a concrete geometric meaning, beyond being the
average kissing number of a hypothetical packing. It turns out to be an upper
bound for the average kissing number of any sphere packing, subject to some
conjectures about interpolation. The key tool is the following theorem, which
is the Euclidean analogue of Proposition~4.1 in \cite{BP} by Bourque and
Petri.

\begin{theorem} \label{theorem:kiss}
Let $f \colon \R^d \to \R$ be a radial Schwartz function and $r>0$, and
suppose that $f$ satisfies the following inequalities:
\begin{enumerate}
\item[\textup{(1)}] $f(r)<0$,
\item[\textup{(2)}] $f(x) \le 0$ for $|x| \ge r$, and
\item[\textup{(3)}] $\widehat{f}(y) \ge 0$ for all $y$.
\end{enumerate}
Then the average kissing number of any $d$-dimensional sphere
packing is at most \begin{equation}-\frac{f(0)}{f(r)}.\end{equation}
\end{theorem}

Here $f(r)$ denotes the value of $f(x)$ when $|x|=r$.

\begin{proof}
It suffices to prove the inequality for finite packings and take a limit.
Let $\mathcal{C}$ be any finite subset of $\R^d$ with minimal distance $r$,
and let
\begin{equation}
N = \#\{(x,y) \in \mathcal{C}^2 : |x-y|=r\}/|\mathcal{C}|
\end{equation}
be its average kissing number. Then Fourier inversion implies that
\begin{equation}
\sum_{x_1, x_2 \in \mathcal{C}} f(|x_1-x_2|) = \int_{\R^d} dy\, \widehat{f}(y) \left| \sum_{x \in \mathcal{C}} e^{2\pi i \langle x,y \rangle} \right|^2 \ge 0,
\end{equation}
while
\begin{equation}
\sum_{x_1, x_2 \in \mathcal{C}} f(|x_1-x_2|) \le |\mathcal{C}| f(0) + N |\mathcal{C}| f(r)
\end{equation}
thanks to the inequalities for $f$. By combining these two bounds, we
conclude that $N \le -f(0)/f(r)$.
\end{proof}

One can also prove this theorem using Poisson summation, along the lines of
\cite{CE} or \cite{BP}. The conditions for equality are similar to those for
the linear programming bound for packing density, if we assume
self-duality.\footnote{This is the same issue as the conjectured agreement
between the linear programming bound and the $-1$ eigenfunction uncertainty
principle.} Specifically, equality holds iff $f$ vanishes at radius $r_n :=
\sqrt{2\Delta_n}$ for $n \ge 2$, $\widehat{f}$ vanishes at $r_n$ for $n \ge
0$ (with $r_0=0$), $r=r_1$, and $f'(r_1)=0$ (because otherwise shifting $r$
would improve the bound). For comparison, the equality conditions for $h$ in
Theorem~\ref{theorem:LP} are identical, except that the conditions
$f'(r_1)=0$ and $\widehat{f}(0)=0$ are replaced with $h(r_1)=0$.

Suppose $r_n = \sqrt{2\Delta^\textup{LP}_n(d/2)}$ and $d_n =
d^\textup{LP}_n(d/2)$ come from the optimal solution to the linear
programming bound in $\R^d$. The crossing equation says that
\begin{equation}
\sum_{n \ge 0} d_n f(r_n) = \sum_{n \ge 0} d_n \widehat{f}(r_n)
\end{equation}
for every radial Schwartz function $f \colon \R^d \to \R$. If $f$ satisfies
the equality conditions for Theorem~\ref{theorem:kiss}, then this equation
reduces to
\begin{equation}
f(0) + d_1 f(r_1) = 0.
\end{equation}
In other words, the bound $-f(0)/f(r_1)$ for the average kissing number is
$d_1$, as desired.

When should such a function $f$ exist and satisfy the hypotheses of
Theorem~\ref{theorem:kiss}? First, note that the condition $\widehat{f}(0)=0$
is redundant, for the following reason. If we let $F(x) = |x| f'(x)$, then
$\widehat{F}(y) = -d \widehat{f}(y) - |y| \widehat{f}\,'(y)$. The other
conditions on $f$ guarantee that $F(0)=F(r_n)=\widehat{F}(r_n)=0$ for $n \ge
1$, and then the crossing equation implies that $\widehat{F}(0)=0$ and hence
$\widehat{f}(0)=0$. What we need is for $f$ to satisfy the same equality
conditions as $h$, except for changing $h(r_1)=0$ to $f'(r_1)=0$.

These conditions arise naturally in interpolation problems \cite{RV, CKMRV2}.
Specifically, Open Problem~7.3 from \cite{CKMRV2} raised the question of
whether radial Schwartz functions $g \colon \R^d \to \R$ are uniquely
determined by the values and radial derivatives of $g$ and $\widehat{g}$ at
the radii $r_n$ for $n \ge 1$. While this assertion fails for $d \le 2$ and
is difficult to test for $d=3$, it seems to hold numerically for $d \ge 4$.
Proving or disproving it would be an important step forward in our
understanding of the modular bootstrap.

The conditions on $f$ and $h$ mean they are part of an interpolation basis
for reconstructing $g$ from these values, since all but one of the values
must vanish for $f$ and $h$. Thus, Theorem~\ref{theorem:kiss} gives a natural
geometric interpretation for one of the basis functions, just as
Theorem~\ref{theorem:LP} does.

Aside from $d=8$ or $24$ (in which case \cite{CKMRV2} proves an interpolation
theorem), we do not know how to prove that an interpolation basis exists, or
that the basis functions satisfy the right sign conditions for these
theorems. However, the numerical evidence indicates that both are true. If
so, the implied kissing number is an upper bound for the average kissing
number of every sphere packing.

This relationship has a pleasing consequence: in each dimension, either the
implied kissing number is the best bound known for the average kissing
number, or we can use a better bound in Theorem~\ref{theorem:LOV} to improve
on the packing density bound.\footnote{Technically we need a bound for the
average $t$-neighbor number for some $t>1$, but that is practically the same
as a bound for $t=1$.} In other words, if we fail to improve on the linear
programming bound for density, it can only be because we have obtained an
excellent bound for the average kissing number.

\appendix

\section{The limiting case of the spinless modular bootstrap}
\label{appendix:vanishat0}

In this appendix, we explain why $\Delta_{\textup{gap}}$ is an upper bound
for the spectral gap in the spinless modular bootstrap even if
$\omega(\Phi_0) = 0$ (in the notation of
Section~\ref{subsec:modularbootstrap}). We expect that any such functional
$\omega$ is the limit of functionals with $\omega(\Phi_0)
> 0$, but it is not clear how to justify that expectation. Instead, we can
use essentially the same argument as the proof of Proposition~2.4 in
\cite{CG}. We will translate it into modular bootstrap terms for the
convenience of the reader.

First, we note that the spectrum $0 = \Delta_0 < \Delta_1 < \Delta_2 <
\dotsb$ must satisfy
\begin{equation}
\lim_{j \to \infty} \frac{\Delta_{j+1}}{\Delta_j} = 1,
\end{equation}
since otherwise \eqref{eq:Karamata} could not hold for $\Delta_j < A <
\Delta_{j+1}$ with $j$ large.

Now suppose $\omega$ is a linear functional such that $\omega(\Phi_0) \ge 0$,
$\omega(\Phi_\Delta) \ge 0$ whenever $\Delta \ge \Delta_{\textup{gap}}$, and
$\omega(\Phi_\Delta)$ is not identically zero. We wish to obtain a
contradiction if $\Delta_1 > \Delta_{\textup{gap}}$. The crossing equation
\begin{equation}
\omega(\Phi_0) + \sum_{\Delta > 0} d_\Delta \omega(\Phi_\Delta) = 0
\end{equation}
shows that $\omega(\Phi_{\Delta_j})$ must vanish for each $j$, but that does
not directly yield a contradiction.

Instead, for each constant $\lambda \ge 1$ we will replace $\omega$ with a
modified functional $\omega_\lambda$ such that
$\omega_{\lambda}(\Phi_{\Delta}) = \omega(\Phi_{\lambda \Delta})+\lambda^{-c}
\omega(\Phi_{\Delta/\lambda})$. To see why this is possible, let $f(r) =
\omega(\Phi_{r^2/2})$. As explained in Section~\ref{sec:prelim}, $f$ is a
$-1$ eigenfunction for the radial Fourier transform in $2c$ dimensions, and
conversely any such function arises for a suitable choice of $\omega$. Let
$f_\lambda(r) = f(\lambda^{1/2} r)$, so that $\widehat{f_\lambda}(r) =
\lambda^{-c} \widehat{f}(r/\lambda^{1/2}) = -\lambda^{-c}
f(r/\lambda^{1/2})$, and define $\omega_\lambda$ by
$\omega_{\lambda}(\Phi_{r^2/2}) = f_\lambda(r) - \widehat{f_\lambda}(r) =
f(\lambda^{1/2} r) + \lambda^{-c} f(r/\lambda^{1/2})$, which works because
$f_\lambda - \widehat{f_\lambda}$ is again a $-1$ eigenfunction.

This new functional satisfies $\omega_\lambda(\Phi_0) \ge 0$ and
$\omega_\lambda(\Phi_\Delta) \ge 0$ whenever $\Delta \ge
\lambda\Delta_{\textup{gap}}$. Furthermore, the only way
$\omega_{\lambda}(\Phi_{\Delta})$ can vanish at a point $\Delta \ge
\lambda\Delta_{\textup{gap}}$ is if $\omega(\Phi_{\lambda \Delta}) =
\omega(\Phi_{\Delta/\lambda}) = 0$.

Now suppose  $\Delta_1 > \Delta_{\textup{gap}}$, and let
$\lambda_{\textup{gap}} = \Delta_1/\Delta_{\textup{gap}}$.  Then we conclude
that $\omega(\Phi_{\Delta})=0$ whenever $\Delta$ is in one of the intervals
$(\lambda_{\textup{gap}}^{-1}\Delta_j, \lambda_{\textup{gap}} \Delta_j)$.
Because $\lim_{j \to \infty} \Delta_{j+1}/\Delta_j = 1$, these intervals
cover an entire half-line $[R,\infty)$ for some $R>0$. However, an
eigenfunction of the Fourier transform cannot have compact support unless it
vanishes identically, because it must be an entire function if it
(equivalently, its Fourier transform) has compact support. Thus, because
$\omega(\Phi_{\Delta})$ does not vanish identically, we conclude that
$\Delta_1 \le \Delta_{\textup{gap}}$, as desired.

\section{Convergence of the spinless bootstrap}
\label{appendix:convergence}

In this appendix, we examine the convergence rate of the spinless modular
bootstrap as a function of the truncation order, and in particular explain
why we are confident that the numerical calculations in
Sections~\ref{sec:numerics} and~\ref{sec:properties} have been fully
optimized.

\begin{figure}
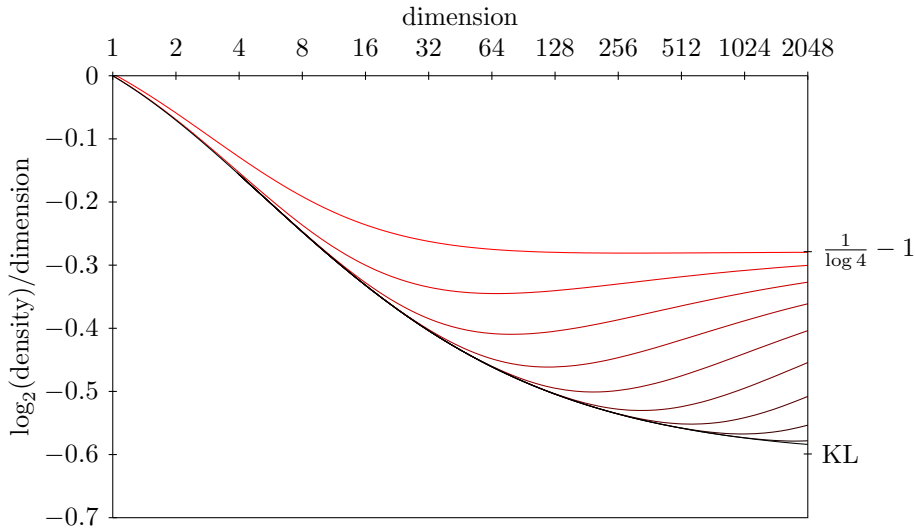

\begin{center}

\end{center}
\caption{Density bounds based on $N=1$, $2$, $4$, $8$, \dots, $512$
(colored red through black).}
\label{figure:densitybounds512}
\end{figure}

Figure~\ref{figure:densitybounds512} shows the density bounds obtained using
truncation orders $N = 1$, $2$, $4$, \dots, $512$ for dimensions $d \le
2048$, in the same format as
Figure~\ref{figure:loglog}.\footnote{Technically, when $d$ is small we plot
only the smaller values of $N$, to avoid the failure of our numerical method
for $d=3$ and $N>32$, which we noted in Section~\ref{subsec:numerics}.} Each
fixed $N$ seems to lead to the same limit as $d \to \infty$, analogously to
\cite[Section~3.2]{Friedan:2013cba}, but they closely approximate the optimal
linear programming bound over increasingly large ranges of $d$. In
particular, doubling $N$ more or less doubles the range of dimensions over
which we obtain a close approximation.

\begin{figure}
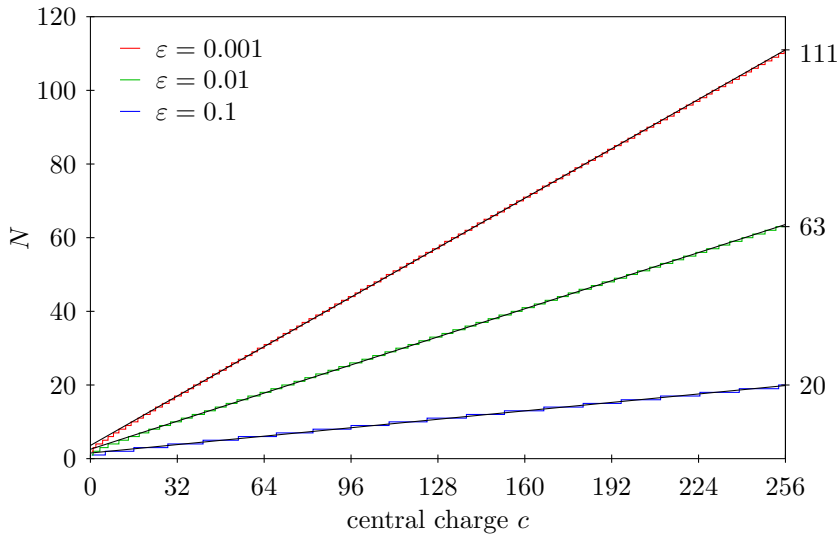

\begin{center}

\end{center}
\caption{Minimal $N$ such that $\Delta_1^{\textup{LP},N}(c) \le (1+\varepsilon) \Delta^\textup{LP}_1(c)$
for $\varepsilon = 0.1$, $0.01$, and $0.001$,
together with the least squares regression lines.}
\label{figure:regression}
\end{figure}

What Figure~\ref{figure:densitybounds512} indicates is that $N$ should be
chosen proportionally to $d$ if we wish to obtain comparably accurate
results. Figure~\ref{figure:regression} makes this assertion more precise,
and shows that the required truncation order is remarkably close to linear in
$d$. We obtained the limiting values $\Delta_1^\textup{LP}(c)$ in
Figure~\ref{figure:regression} by taking $N$ quite large, in particular more
than twice as large as needed to make the plotted values stop changing.

\begin{figure}
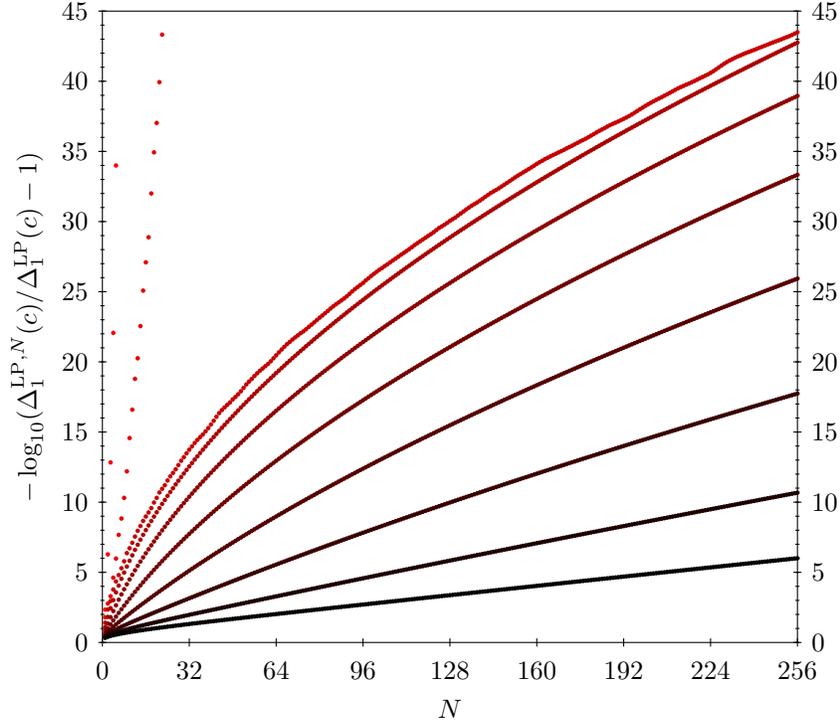

\begin{center}

\end{center}
\caption{Convergence of $\Delta_1^{\textup{LP},N}(c)$ as a function of $N$
for $c=1/2$, $1$, $2$, $4$, \dots, $256$ (colored red through black).}
\label{figure:lotsofdigits}
\end{figure}

We do not know a formula for the slopes in Figure~\ref{figure:regression}.
When we need to estimate the convergence rate in high dimensions, we
extrapolate from lower dimensions and then make a conservative underestimate.
What makes this procedure reliable is how close to linear
Figure~\ref{figure:regression} is. By contrast,
Figure~\ref{figure:lotsofdigits} fixes $c$ and examines how many digits of
$\Delta_1^{\textup{LP},N}(c)$ have converged as a function of $N$. Here, the
behavior is not nearly as linear, and it is more difficult to extrapolate.

\begin{table}
\caption{The truncation order $N_k$ required to approximate
$(\log_2 \textup{density})/d$ to within $0.5\cdot 10^{-k}$,
so that rounding to $k$ decimal places leads to error at most $10^{-k}$.}
\label{table:Ni}
\begin{center}
\begin{tabular}{rrrrrrr}
\toprule
$d$ & $N$ & $N_1$ & $N_2$ & $N_3$ & $N_4$ & $N_5$\\
\midrule
$1$ & $\infty$ & $1$& $1$& $2$& $2$& $2$\\
$2$ & $33$ & $1$& $2$& $3$& $4$& $5$\\
$4$ & $308$ & $1$& $2$& $4$& $5$& $7$\\
$8$ & $\infty$ & $2$& $3$& $5$& $7$& $9$\\
$16$ & $933$ & $2$& $4$& $7$& $10$& $13$\\
$32$ & $997$ & $3$& $7$& $11$& $15$& $19$\\
$64$ & $1025$ & $5$& $11$& $18$& $25$& $33$\\
$128$ & $780$ & $8$& $20$& $33$& $46$& $59$\\
$256$ & $844$ & $14$& $37$& $61$& $86$& $111$\\
$512$ & $972$ & $25$& $71$& $118$& $167$& $215$\\
$1024$ & $852$ & $48$& $138$& $232$& $327$& $422$\\
$2048$ & $1700$ & $93$& $271$& $458$& $647$& $836$\\
$4096$ & $1750$ & $\phantom{1}183$& $\phantom{1}536$& \textcolor{red}{$\phantom{1}916$}& \textcolor{red}{$1294$}& \textcolor{red}{$1672$}\\
\bottomrule
\end{tabular}
\end{center}
\end{table}

The most delicate numerical estimation in this paper occurs in obtaining the
number $-0.6044$ as the infinite-dimensional limit of the LP column in
Table~\ref{table:KLLP}. Table~\ref{table:Ni} gives evidence that the values
in Table~\ref{table:KLLP} are correctly extrapolated to infinite truncation
order. Specifically, for each dimension Table~\ref{table:Ni} lists the
largest truncation order $N$ we have computed, together with the smallest
order $N_k$ that agrees with order $N$ to $k$ decimal places. The numbers in
black are exact, meaning that truncation order $N_k-1$ is not enough. In each
such case $N$ is at least $2 N_k$, and often much larger than that; this
margin of safety gives us confidence that these values do reflect the limit
as $N \to \infty$. The red numbers are obtained by doubling the numbers above
them, which seems to produce an overestimate and would work in every other
case with $d>2$. Even for the red numbers, $N_5 < N$, and therefore we
believe that our truncation orders are high enough for all the numbers in
Table~\ref{table:KLLP} to have stabilized.

\section*{Acknowledgements}

We are grateful to Ganesh Ajjanagadde, Matthew de Courcy-Ireland,
Abhinav Kumar, Dalimil Maz\'a\v{c}, Stephen D.\ Miller, Danylo Radchenko,
Leonardo Rastelli, Peter Sarnak, and Maryna Viazovska for helpful conversations.
The work of TH and AT is supported by the Simons Foundation (Simons Collaboration
on the Nonperturbative Bootstrap). NA is supported by the Leo Kadanoff Fellowship.

\end{spacing}

\providecommand{\href}[2]{#2}\begingroup\raggedright\endgroup

\end{document}